\documentclass[a4paper,traditabstract,fleqn]{aa}
\usepackage{graphicx}
\usepackage{natbib,amssymb,amsmath,txfonts,astro}
\usepackage[version=3]{mhchem}
\usepackage[normalem]{ulem}
\usepackage{longtable,booktabs}
\begin{document}
\setlength{\mathindent}{0pt}
\def\reac#1#2{\begin{equation}\label{r:r#1}\ce{#2}\end{equation}}
\def\reacnn#1{\[\ce{#1}\]}
\def\refreac#1{(\ref{r:r#1})}
\def\abS{\ensuremath{[\ce{S}]_{\rm tot}}}
\def\ab#1{\ensuremath{[\ce{#1}]}}
\def\abtot#1{\ensuremath{[\ce{#1}]_{\rm tot}}}
\def\nn#1{\ensuremath{n(\ce{#1})}}
\def\fch2{\ensuremath{f_{\ce{CH2}}}}
\def\fhcn{\ensuremath{f_{\rm HCN}}}
\def\fhnc{\ensuremath{f_{\rm HNC}}}
\def\fcn{\ensuremath{f_{\rm CN}}}
\def\nhp{\ensuremath{n({\rm H^+})}}
\def\opr{\ensuremath{{\rm o/p}}}
\def\tkin{\ensuremath{T_{\rm kin}}}
\def\tspin{\ensuremath{T_{\rm spin}}}
\def\hp{\ce{H+}}
\def\ohhp{\ensuremath{{\rm o\ce{H2+}}}}
\def\phhp{\ensuremath{{\rm p\ce{H2+}}}}
\def\phhhp{\ce{p-H_3+}}
\def\ohhhp{\ce{o-H_3+}}
\def\ec{\ce{e-}}
\def\ohh{\ensuremath{{\ce{H2}(\rm o)}}}
\def\phh{\ensuremath{{\ce{H2}(\rm p)}}}
\def\onhh{\ensuremath{{\ce{NH2}(\rm o)}}}
\def\pnhh{\ensuremath{{\ce{NH2}(\rm p)}}}
\def\onhhh{\ensuremath{{\ce{NH3}(\rm o)}}}
\def\pnhhh{\ensuremath{{\ce{NH3}(\rm p)}}}

\def\ntot{\ensuremath{n_{\rm tot}}} 
\def\nhh{\ensuremath{n_\hh}} 
\def\nhp{\ensuremath{n_\hp}} 
\def\no{\ensuremath{n_{\rm o}}} 
\def\np{\ensuremath{n_{\rm p}}} 
\def\htrop{\ensuremath{{\rm (\opr)_{HT}}}}
\def\tcrit{\ensuremath{T_{\rm crit}}}
\def\kform{\ensuremath{k_{\rm form-\hh}}}
\title{Interstellar chemistry of nitrogen hydrides in dark clouds}

\author{%
  R.~Le Gal      \inst{1}  \and
  P.~Hily-Blant  \inst{1,2}  \and
  A.~Faure       \inst{1}  \and
  G. Pineau des For\^ets \inst{3,4} \and
  C.~Rist       \inst{1} \and
  S.~Maret      \inst{1}
}

\institute{$^1$ Universit\'e Joseph Fourier/CNRS, Institut de
  Plan\'etologie et d'Astrophysique de Grenoble (IPAG) UMR 5274,
  Grenoble,
  France\\\email{romane.legal@obs.ujf-grenoble.fr;pierre.hily-blant@obs.ujf-grenoble.fr;\\
    alexandre.faure@obs.ujf-grenoble.fr}\\ $^2$ Institut Universitaire
  de France\\ $^3$ Universit\'e de Paris Sud/CNRS, IAS (UMR 8617),
  Orsay, France\\$^4$ LERMA / CNRS (UMR 8112) / Observatoire de Paris, France}

\date{Received 29-07-2013; Accepted 18-11-2013}


\abstract{ Nitrogen, amongst the most abundant metals in the
  interstellar medium, has a peculiar chemistry which differs from
  those of carbon and oxygen. Recent observations of several
  nitrogen-bearing species in the interstellar medium suggest
  abundances in sharp disagreement with current chemical
  models. Although some of these observations show that some
  gas-grains processes are at work, gas-phase chemistry needs first to
  be revisited. Strong constraints are provided by recent Herschel
  observations of nitrogen hydrides in cold gas. The aim of the
  present work is to perform a comprehensive analysis of the
  interstellar chemistry of nitrogen, focussing on the gas-phase
  formation of the smallest polyatomic species and in particular
  nitrogen hydrides. We present a new chemical network in which the
  kinetic rates of critical reactions have been updated based on
  recent experimental and theoretical studies, including nuclear spin
  branching ratios. Our network thus treats the different spin
  symmetries of the nitrogen hydrides self-consistently together with
  the ortho and para forms of molecular hydrogen. This new network is
  used to model the time evolution of the chemical abundances in dark
  cloud conditions. The steady-state results are analysed, with
  special emphasis on the influence of the overall amounts of carbon,
  oxygen, and sulphur. Our calculations are also compared with
  Herschel/HIFI observations of NH, \ce{NH2}, and \ce{NH3} detected
  towards the external envelope of the protostar IRAS~16293-2422. The
  observed abundances and abundance ratios are reproduced for a C/O
  gas-phase elemental abundance ratio of $\sim0.8$, provided that the
  sulphur abundance is depleted by a factor larger than 2. The
  ortho-to-para ratio of \hh\ in these models is $\sim\dix{-3}$. Our models
  also provide predictions for the ortho-to-para ratios of \ce{NH2}
  and \ce{NH3} of $\sim2.3$ and $\sim0.7$ respectively. We
  conclude that the abundances of nitrogen hydrides in dark cloud
  conditions are consistent with the gas-phase synthesis predicted
  with our new chemical network.}

   
\maketitle
%

\section{Introduction}

Nitrogen-bearing species are commonly observed in the interstellar
medium (ISM), since the discovery of ammonia, the first polyatomic
interstellar species, by \cite{cheung1968}. Nitrogenated species are
useful probes of the physics and chemistry of the ISM over a broad
range of conditions. Inversion lines of ammonia serve as temperature
probes in molecular clouds \citep{ho1983, maret2009}, and rotational
lines of diazenylium (\ce{N2H+}) and its deuterated isotopologue,
\ce{N2D+}, may be used at much higher densities
\citep[$n\sim\dix{5}\ccc$, ][]{pagani2007, crapsi2007}. In the diffuse
ISM, CN absorption lines allowed the first estimate of the CMB
temperature \citep{thaddeus1972}. Due to its sensitivity to Zeeman
splitting and to its hyperfine structure, CN is also a powerful tool
to measure the line-of-sight magnetic field intensity in dense regions
\citep{crutcher2012}.

The reservoir of nitrogen in molecular clouds is still controversial,
but is expected to be gaseous, either in atomic or molecular
forms. Atomic nitrogen in the diffuse ISM is observed through
absorption lines in the UV \citep{nieva2012}. Searches for the \ce{N2}
molecule in interstellar space had been unfruitful until its first
detection in the far-ultraviolet by \cite{knauth2004}, in absorption
against the background star HD~124314. The derived column density of
\ce{N2}, 4.6\tdix{13}\cc, is several orders of magnitude lower than
that of atomic nitrogen (2.0\tdix{17}\cc), indicating that nitrogen is
mainly atomic. The total visual extinction is 1.5 magnitude or
$N_\h=2.8\tdix{21}\cc $ (assuming standard grain properties), leading
to abundances, with respect to hydrogen nuclei, of 7.2\tdix{-5} and
1.6\tdix{-8} for atomic and molecular nitrogen respectively. The
column density of \ce{N2} is also about one order of magnitude higher
than the predictions of \cite{li2013} for translucent clouds which
take into account the photodissociation of \ce{N2}. The strong
discrepancy between observations and model predictions suggests that
our understanding of nitrogen chemistry in such diffuse to translucent
environments remains poor. In dense molecular clouds, where hydrogen
is molecular, the situation is even worse, because NI and \ce{N2} are
not observable directly. Constraints on their abundances are thus only
indirect. In dense molecular clouds, \ce{N2H+}, a direct chemical
product of \ce{N2}, was observed by \cite{womack1992b} and
\cite{maret2006} who concluded that atomic N is likely the dominant
reservoir of nitrogen. In addition, in prestellar cores with gas densities
$\sim\dix{5}\ccc$, \cite{hilyblant2010n} derived an upper
limit on the gas-phase abundance of atomic nitrogen which suggested
that nitrogen may be predominantly hidden in ices coating dust
grains. More recently, \cite{daranlot2012} found that gaseous nitrogen
is mostly atomic in dense clouds, but that the dominant reservoir of
nitrogen is indeed in the form of ammonia ices at the surface of dust
grains. Yet, as stressed by these authors, the predicted amount of icy
ammonia is larger than what is observed in dark clouds and much larger
than what is observed in comets. From the above, one can safely
consider that the question of the reservoir of gaseous (and solid)
nitrogen in dark clouds still remains an open issue.

In recent years, additional observations have challenged our
understanding of the first steps of the chemistry of nitrogen
\citep[e.g.][]{hilyblant2010n, persson2012}. By first steps we here
refer to the synthesis of the smallest N-bearing molecules. One such
discrepancy between observations and models is evidenced by the CN:HCN
abundance ratio towards several starless dark clouds, the value of
which is underpredicted in the models of
\citet{hilyblant2010n}. Another fundamental issue concerns the HCN:HNC
abundance ratio. These two isomers are the products of the
dissociative recombination (hereafter noted DR) of \ce{HCNH+}, with
equal measured branching ratio \citep{mendes2012}. The predicted
abundance ratio HCN:HNC is so expected close to unity
\citep{herbst2000}. Observed ratios however show a large scatter
around unity \citep{hirota1998} which may reflect different chemical
routes to these molecules. Another fundamental question is the
formation of ammonia in dense clouds. \cite{lebourlot1991} suggested
that the gas-phase synthesis through the \ce{N+ + H2} reaction,
followed by hydrogen abstractions and DR reactions, was efficient
enough to reproduce the observed amounts. Yet, very recently,
\cite{dislaire2012} revisited the experimental data available for the
\ce{N+ + H2} reaction. The new rate is significantly smaller and falls
below the critical value inferred by \cite{herbst1987} to explain the
observed abundances of ammonia. The efficiency of gas-phase synthesis
of ammonia versus hydrogenation of atomic nitrogen at the surfaces of
dust grains remains an open issue \citep[e.g.][]{tielens1982,
  dhendecourt1985, charnley2002, hidaka2011}. Observational
constraints on the amount of ammonia locked into ices coating dust
grains are rare, because the N-H vibrational feature at 2.95\micr\ is
heavily obscured by the deep 3\micr\ water ice bands. However,
observations of \ce{NH3} ices in young star formation regions indicate
that an abundance of 5\% relative to water seems to be a reasonable
value \citep{bottinelli2010}. Up to now, N$_2$ ices have not been
detected in dense regions \citep{sandford2001}. Perhaps related to the
ammonia issue is a new constraint based on the abundance ratios of
nitrogen hydrides NH:\ce{NH2}:\ce{NH3} towards the Class 0 protostar
IRAS~16293-2422, obtained with the HIFI (Heterodyne Instrument for the
Far-Infrared) instrument onboard the Herschel satellite in the
framework of the CHESS key program \citep{ceccarelli2010}. The
absorption lines arising from the low-lying rotational levels of these
hydrides lead to abundance ratios NH:\ce{NH2}:\ce{NH3} = 5:1:300. These
abundance ratios could not be reproduced by model calculations in dark
gas at a temperature of 10~K and a gas density of $\dix{4}\ccc$
\citep{hilyblant2010nh}. Last, in diffuse to translucent
environments, Herschel/HIFI observations of the ortho and para forms
of ammonia indicate an ortho-to-para ratio of $\sim 0.7$ which could
not be explained with the standard nitrogen chemistry
\citep{persson2012}.

The present paper is devoted to the chemistry of nitrogen in dense
regions of the ISM which are efficiently shielded from ultraviolet
photons by the dust and molecular hydrogen. One major difficulty to
make progress on the issue of nitrogen chemistry in the dense ISM is
that both N and \ce{N2} are not observable. The determination of their
abundances thus relies on observations of trace nitrogen-containing
molecules. Chemical models are then essential. Such carriers usually
include \ce{NH3}, CN, HCN, HNC, \ce{N2H+}, together with \thc, D, and
\fifn\ isotopologues, and to a lesser extent NO. The typical
abundances -- with respect to the total H nuclei -- for the major
isotopologues are $\approx$ \dix{-10}--\dix{-9}, except for NO whose
abundance may be as high as \dix{-8} \citep{suzuki1992, gerin1992,
  akyilmaz2007, hilyblant2010n, padovani2011}. Chemical models allow
to make predictions for the abundances of chemical species under
specified physical conditions \citep[for a review, see
][]{wakelam2010}. Time-dependent models follow the abundances with
time, until the steady-state is eventually reached. In the
interstellar medium, however, the steady-state does not coincide with
the thermodynamical equilibrium, and models should thus solve the
time-dependent chemical and physical equations in a self-consistent
fashion \cite[e.g. ][]{tassis2012}. However, a full coupling,
e.g. including also the impact of the chemical abundances on the
thermodynamical state of the gas through radiative transfer, remains
beyond the current numerical capabilities. Simplifications must be
made, such as adopting analytical prescriptions for the time evolution
of the gas physical conditions \citep{bergin1997,
  flower2005}. Regarding the chemical processes, the cornerstone of
any chemical model is really the network of chemical reactions that
describe the formation and destruction of the chemical species. The
determination, either theoretically or experimentally, of the kinetic
rates of these reactions at the low temperatures prevailing in the
diffuse, translucent, and dense ISM ($5-80$~K) is extremely
time-demanding. This task is simply out of reach for the few thousands
reactions involved in current chemical networks. It is thus crucial to
identify the key reactions whose rates have a major effect on the
chemistry \citep{wakelam2010,wakelam2012}.

In recent years, motivated by the above challenges, rate constants for
several key reactions involved in the chemistry of interstellar
nitrogen have been computed \cite[e.g.][]{jorfi2009a,jorfi2009b} and
measured \citep[e.g.][]{bergeat2009,daranlot2011,daranlot2012} down to
$\sim$10 and $\sim$50~K, respectively. These challenges also triggered
theoretical investigations. Separate collisional rates were computed
for HCN and HNC with \hh\ by \citet{sarrasin2010}. They show in
particular very different rates for the (1-0) rotational
transition. Using those new rates, \cite{padovani2011} derived an
abundance ratio HCN:HNC$\approx1$ in three starless cores, in
agreement with model predictions, suggesting that the current chemical
route to HCN and HNC is consistent with observations. Further
observations are however needed to put this result on a firmer
basis. A theoretical investigation of the ortho-para chemistry of
ammonia in the cold interstellar medium was done by
\cite{rist2013}. They calculated separate branching ratios for the
hydrogen abstractions and DR reactions leading to the ortho and para
forms of nitrogen hydrides, taking into account nuclear spin selection
rules. With these new rates and the new chemical network fully
  described in this paper, \cite{faure2013} showed that the
ortho-to-para ratio of $\sim 0.7$ for ammonia, observed by
\cite{persson2012}, is in fact consistent with gas-phase chemistry in
a para-enriched \hh\ gas. In this model the anomalous values of the
ortho-to-para ratios of nitrogen hydrides are a consequence of the low
ortho-to-para ratio of \hh, and result from the conservation of
nuclear spin in chemical reactions. In a similar fashion,
\cite{dislaire2012} showed that the NH:\ce{NH2} abundance ratio in the
envelope of IRAS~16923-2422 can be reproduced only for an
ortho-to-para ratio of H$_2$ $\approx \dix{-3}$.

These recent experimental and theoretical results motivated
the present work aimed at presenting a new network of the nitrogen
chemistry in dense clouds, thus superseding the classical networks of
\cite{herbst1973} and \cite{pineau1990}. This work puts special
emphasis on nitrogen hydrides for which the ortho and para forms are
treated self-consistently together with the ortho and para forms of
\hh. The outline of the paper is as follows. In Section 2, we present
our new nitrogen network. Section 3 describes our model calculations,
with in particular a discussion on initial abundances. Results and
comparisons with Herschel observations of nitrogen hydrides are the
subject of Section~4. Section~5 summarises our new findings and
suggests further lines of investigation.

\begin{figure*}[t] 
  \centering
  \includegraphics[width=0.7\hsize]{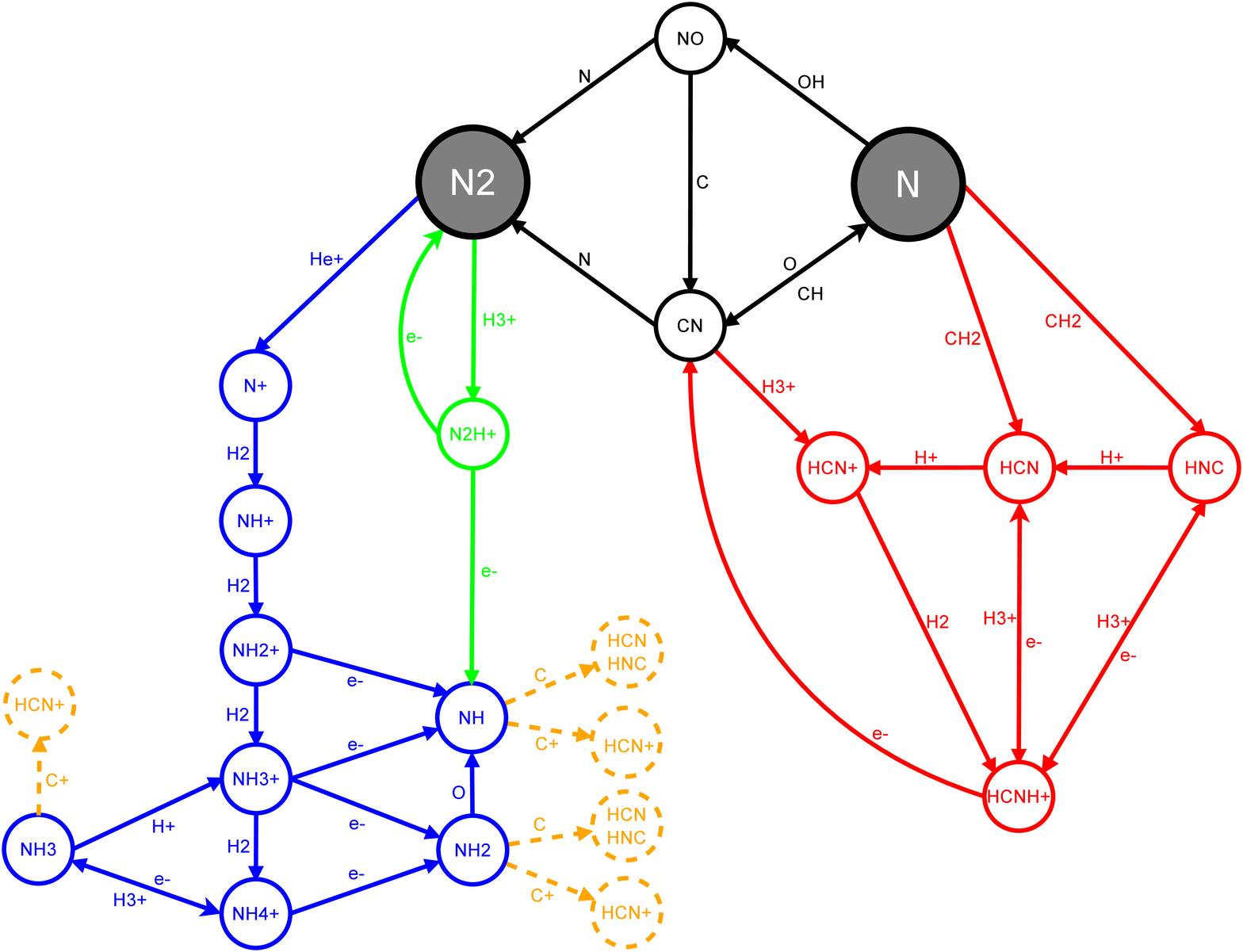}
  \caption{Schema adapted from \cite{hilyblant2013} representing the
    principal gas-phase reactions involved at first stages of nitrogen
    interstellar chemistry in dark clouds. Nitrogen hydrides (blue, left)
    and nitriles (red, right) have been clearly separated. The main
    formation route of NH is highlighted (green). Chemical pathways from
    nitrogen hydrides to nitriles, efficient at specific C/O ratio ranges,
    are also represented (dashed yellow).}
  \label{fig:network}
\end{figure*}

\section{Chemical network}
\label{sec:chem_network}

\subsection{General considerations}
\label{subsec:gl_considerations}

The formation of nitrogen hydrides (NH, \ce{NH2}, and \ce{NH3})
  is similar to that of carbon and oxygen hydrides, but has a
  different origin. First, on the contrary to carbon, \ce{N+} is not
available directly because the ionization potential of N (14.53~eV) is
larger than that of \ce{H}. In addition, the reaction \ce{N + H3+
  $\to$ NH+ + H2} -- whose analogue is responsible for the production
of \ce{CH+} and \ce{OH+} in dark gas -- is endothermic because the
proton affinity of N (3.55~eV) is lower than that of \ce{H2}
(4.38~eV). The alternative exothermic pathway, \ce{N + H3+ $\to$ NH2+
  + H}, shows a high activation energy \citep{herbst1987}. Thereby in
the current understanding of nitrogen chemistry, the production of
\ce{N+} results from molecular nitrogen reacting with \ce{He+}. This
reaction is found to be more efficient than the direct cosmic-ray
ionization of atomic nitrogen. Once \ce{N+} is formed, rapid
  hydrogen-abstraction reactions initiated with \reac{np-h2}{N+ + H2
    -> NH+ + H} lead to the \ce{NH+}, \ce{NH2+}, \ce{NH3+}, and
  \ce{NH4+} ions which, by dissociative recombinations, generate the
  formation of neutral nitrogen hydrides (see
  Fig.~\ref{fig:network}). In addition, and because molecular nitrogen
  has a proton affinity of 5.12~eV, the reaction of \ce{N2} with
  \ce{H3+} yielding \ce{N2H+} is exothermic, and proceeds at a
  significant rate. The DR of \ce{N2H+} then leads to \ce{NH}.  Hence,
  the formation of nitrogen hydrides has its origin in \ce{N2}.  On
  the other hand, nitriles (such as HCN) derive from atomic nitrogen,
  through ion-neutral reactions with \ce{CH2+} and
  \ce{CH3+}. Therefore, if nitrogen hydrides are rather late-time
  species, nitriles are anticipated to form at earlier times. This has
  important consequences on the $^{15}$N fractionation processes as
  discussed recently in \cite{hilyblant2013}.

The formation of nitrogen hydrides is therefore a relatively slow
process compared to carbon and oxygen hydrides, precisely because the
first step involves the synthesis of molecular nitrogen which is
thought to proceed through neutral-neutral reactions
\citep{pineau1990, hilyblant2010n}: \reac{n-oh}{N + OH -> NO + H}
\reac{n-no}{NO + N -> N2 + O} and \reac{n-ch}{N + CH -> CN + H}
\reac{no-c}{NO + C -> CN + O} \reac{cn-n}{CN + N -> N2 + C} We
anticipate that in those instances where the C/O gas-phase elemental
abundance ratio is lower than unity, reaction \refreac{no-c} is
expected to dominate the formation of CN over reaction
\refreac{n-ch}.

Reaction \refreac{np-h2} consequently appears to be a key reaction for
the synthesis of nitrogen hydrides. In fact, it was studied in detail
by \cite{lebourlot1991} who derived separate rates for the reaction
with ortho and para \ce{H2} (\ohh\ and \phh\ respectively in the
remainder of the text). The reaction with \phh\ was assumed to occur
with a 170~K activation energy, while being barrierless with \ohh. The
rate of reaction \refreac{np-h2} strongly depends on the ortho-to-para
ratio of \ce{H2} (noted \opr\ (\hh) hereafter). More recently,
\cite{dislaire2012} revised those rates (see
  Table~\ref{tab:newin}) using available experimental data
\citep{marquette1988,gerlich1993} for \phh, normal-\hh\
(i.e. \opr\ (\hh)=3:1) and \phh\ containing admixtures of \ohh,
leading to a rate at 10~K typically two orders of magnitude lower than
in \cite{lebourlot1991}. In particular, for \opr\
(\hh)~$\lesssim$\dix{-2}, the new rate is below the critical value of
\dix{-13}\cccs\ required to account for the observed amounts of
gas-phase ammonia \citep{herbst1987}. Nevertheless,
\cite{dislaire2012} could reproduced the abundance of ammonia observed
towards IRAS~16293-2422, provided the total C/O gas-phase abundance
ratio was lower than 0.4. Recent measurements by
  \cite{zymak2013} have confirmed the rate used by
  \cite{dislaire2012}, although issues related to the fine structure
  relaxation of N$^+$ deserve further experimental and theoretical studies.
\cite{dislaire2012} also showed that a branching ratio $< 10\%$ for
the channel of \ce{N2H+ + e-} leading to NH is sufficient to reproduce
the observed amount of NH. The abundance of NH is thus independent of
the \opr\ (\hh), contrary to \ce{NH2} and \ce{NH3} which proceed from
reaction \refreac{np-h2}. One consequence is that the NH:\ce{NH2}
abundance ratio depends on the \opr\ (\hh). Using the NH:\ce{NH2}
abundance ratio, \cite{dislaire2012} could finally determine an \opr\
(\hh) of $\approx$\dix{-3}. However, these authors did not treat
explicitly the formation of \ohh\ and \phh.

In the following, we discuss the new reaction rates that have been
implemented to model self-consistently the formation of nitrogen
hydrides, of \ohh\ and \phh. This new network is an updated version
of the gas-phase network of \cite{flower2006n}, as detailed below.

\subsection{The ortho-to-para ratio of \hh}

The \opr\ (\hh) plays a significant role in the chemistry of nitrogen
hydrides, and it is therefore important to understand the processes
that may affect its value, in order to self-consistently model the
chemistry. As \hh\ is not directly observable in the cold gas, its
\opr\ (\hh) ratio derives from indirect measurements \citep[e.g. ][and
references therein]{troscompt2009}. An upper limit of 0.01 was
proposed by \cite{pagani2009} to explain the large deuteration
fraction of \ce{N2D+} in cold cores. Values consistent with this upper
limit were predicted by the models of \cite{flower2006op}, with \opr\
(\hh)$\approx \dix{-3}$. Molecular hydrogen is assumed to form on dust
grain surfaces \citep{hollenbach1971}, in a strongly exothermic
reaction ($\approx 4.5$ eV) such that the outcoming \ohh\ and \phh\
should be in the ratio 3:1, imposed by their nuclear spin statistical
weights. This corresponds to the maximum value permitted under thermal
equilibrium. At temperatures lower than 100~K, only the first
rotational levels of \phh\ and \ohh\ (lying 170.5~K above \phh) are
populated significantly, and the equilibrium value of \opr\ (\hh) is
given by the usual \emph{low-temperature} approximation
\begin{equation}
  \opr\ (\hh) = 9 \exp(-170.5/T)
  \label{eq:opr}
\end{equation}
In the interstellar medium, the characteristic timescale for radiative
spontaneous spin flip from $J=1$ of \ohh\ to $J=0$ of \phh\ is
$\approx \dix{13}$~yr \citep{raich1964, pachucki2008}, much longer
than the lifetime of molecular clouds. As a consequence, the ortho and
para forms may be viewed as two different chemical species, and
gas-phase conversion from one form to the other is expected to take
place only through ion-neutral reactions between \ce{H2} and
protonated ions \citep{dalgarno1973, crabtree2011}. In a general
fashion, these reactions can be written \reac{o2p}{\ohh + XH+ -> \phh
  + XH+} \reac{p2o}{\phh + XH+ -> \ohh + XH+}
where \ce{XH+} stands primarily for \ce{H+}, \ce{H3+}, and
\ce{HCO+}. If both exchange reactions proceed much more rapidly than
the formation of \ce{H2} on grains, the \opr\ (\hh) is expected to
tend to a value imposed by the detailed balance relation between the
rates of reactions \refreac{o2p}-\refreac{p2o} which is the thermal
value of Eq.~(7). On the contrary, if the formation rate on dust
grains is much faster than any of the two exchange reactions, the
\opr\ (\hh) value will stay equal to 3:1. As said previously, there
are some evidences that the \opr\ of \hh\ significantly deviates from
both 3:1 and the thermodynamical value of $\approx 3\times \dix{-7}$
at 10~K. This indicates that certain conversion processes do take
place on timescales comparable or even shorter than the formation
process on dust grains.

The gas-phase conversion between \ohh\ and \phh\ has received
particular attention in the last few years. We have therefore adopted
the most recent theoretical results summarised in
Table~\ref{tab:newin}. In particular, the conversion through reaction
with protons has been computed by \cite{honvault2011,honvault2012}
using state-to-state quantum time-independent calculations. Exchange
reactions with \ce{H3+} were studied using a state-to-state
micro-canonical statistical method by \cite{hugo2009}. The H$_3^+$ +
H$_2$ reaction was also studied experimentally below 100~K by
\cite{grussie2012}, confirming the micro-canonical model. The rates of
\cite{hugo2009} are available for temperatures below 50~K, restricting
the validity of our network to this low-temperature range.

In addition to the formation of H$_2$ on grains and to the
ortho-to-para exchange reactions between \hh\ and protonated ions, the
formation and destruction of the H$_3^+$ ion also play an important
role in the \opr\ (\hh). We have therefore updated the rate
coefficients and branching ratios for the reaction \ce{H2+ + H2} and
the dissociative recombination \ce{H3+ + e-}. To this aim, we have
combined the most recent theoretical and experimental values for the
overall rate coefficients with the Oka's formalism \citep{oka2004} to
derive the nuclear spin branching ratios. Briefly, this formalism
accounts for the selection rules which result from the conservation of
the nuclear spin of identical nuclei. The approach of \cite{oka2004}
is based on angular momentum algebra applied to the total nuclear
spins of the reactants. For each allowed nuclear spin of the
intermediate complex, the distribution of product states is deduced
from the nuclear-spin conservation for the reverse reaction. Full
details can be found in \cite{oka2004}. It should be noted that these
pure nuclear spin branching ratios are applicable for processes in
which many rotational states of the products are populated, in
  particular exothermic reactions. In this work, these branching
  ratios were used for strongly exothermic ion-molecule and DR
  reactions where the products are expected to form rotationally hot,
  along with full proton scrambling. The corresponding new rate
  coefficients are listed in Tables~\ref{tab:newin} and
  \ref{tab:newdr}. We note that most measurements have no temperature
  dependence since they were performed at room temperature only.

Very recently, \cite{faure2013} showed that these new rates do not
significantly change the \opr\ (\hh) at steady-state with respect to
the results of \cite{flower2006op}. A constant \opr\ (\hh) of
  $\sim$10$^{-3}$ was thus found below 15~K. Above this temperature,
the \opr\ (\hh) was shown to converge towards the thermal value, as
expected from the competition between the H$_2$ formation on grains
and the exchange reactions with protonated ions. This point will be
further discussed elsewhere.

\subsection{Atomic to molecular nitrogen conversion}

As we previously mentioned in section \ref{subsec:gl_considerations},
the atomic to molecular conversion is the first step towards the
synthesis of nitrogen hydrides in dark clouds. The rates of reactions
\refreac{n-oh}--\refreac{cn-n} have been uncertain at low
temperatures, and in particular the existence of low activation energy
($\approx 20$~K) was suggested by the CN:NO abundance ratio
\citep{akyilmaz2007}. The rates for the \ce{NO + N} and \ce{CN + N}
reactions have been revisited experimentally by
\cite{daranlot2011,daranlot2012} using the CRESU technique down to
temperatures of $\approx 50$~K, and further extrapolated down to
10~K. The new rates implemented in our network are summarised in
Table~\ref{tab:newnn}. We note that these rates present positive and
negative temperature dependences through $\beta$ values (see Table
~\ref{tab:newnn}) and that they all lie between
4\tdix{-12}cm$^3$s$^{-1}$ and 2\tdix{-10}cm$^3$s$^{-1}$ at 10~K. These
radical-neutral reactions are therefore moderately fast but they
provide the main routes in converting N to N$_2$. The typical chemical
timescales of nitrogen chemistry are discussed below.

\subsection{Nitrogen hydrides synthesis}
\label{subsec:N-hyd}

\begin{figure}
  \centering
  \includegraphics[width=0.95\hsize]{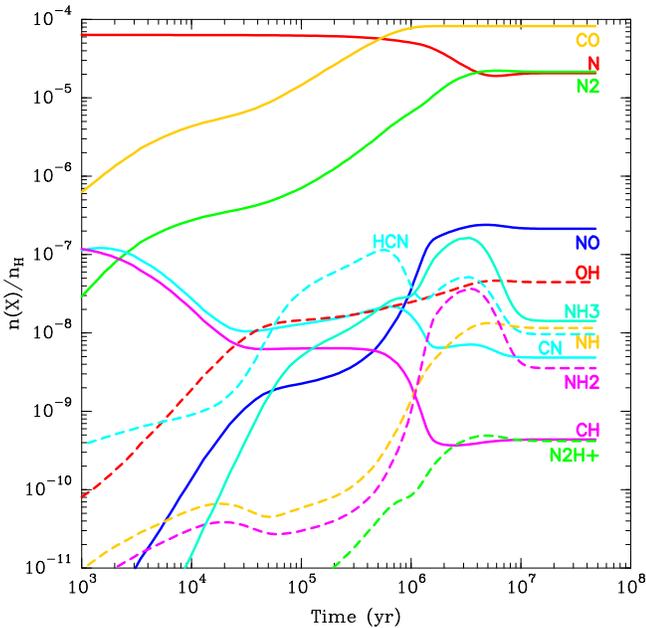}
  \caption{Time evolution of the abundances of the main nitrogenated
    species and species involved in the N-to-\ce{N2} conversion. For
    this particular model (our best model), the total gas-phase
    abundance ratio C/O is 0.8 and the total gas-phase abundance of
    sulphur is 8.0\tdix{-8}.}
  \label{fig:abtime}
\end{figure}

As previously emphasized, the \opr\ (\hh) has a strong effect on the
chemistry of nitrogen hydrides. In order to take into account the
\opr\ (\hh) self-consistently, some care must be taken in the
dissociative recombinations of ionic nitrogen hydrides, NH$_n^+ (n=3,
4) + \rm {e^-}$, since these reactions produce \ohh\ and \phh, in
amounts that depend on the nuclear spin branching ratios. The
associated hydrogen abstractions NH$_n^+ (n=0, 3)$ + H$_2$ are also
crucial in driving the nuclear spin distributions. The corresponding
branching ratios were determined by \cite{rist2013} using the Oka's
formalism discussed above and assuming full scrambling. These
branching ratios were found to be significantly different from those
derived by \cite{flower2006n}. In this pioneering study,
  where the conservation of the total nuclear spin angular momentum
  was included, the branching ratios were indeed derived from simple
  statistical considerations without recourse to the angular momentum
  rules of Oka (2004) (see also \cite{sipila2013} for a recent
  similar work). The new branching ratios derived by
  \cite{rist2013} were employed very recently by \cite{faure2013} in
their work on the \opr\ of ammonia in the cold interstellar gas.

It should be noted that interconversion processes between the nuclear
states of ionic hydrides, of the form \ce{NH$_n$+ +H2 -> NH$_n$+ +H2} ($n=2, 4$), are ignored in this work. Indeed, these proton
scrambling reactions were studied experimentally using deuterated
analogues and H-D exchanges were not observed, with small upper limits
for the rate coefficients (see the discussion in \cite{faure2013}). In
addition, \cite{faure2013} explicitly studied the \ce{NH4+ +H2 -> NH4+
+ H2} thermalization reaction and showed that this reaction
must be negligible in order to explain the observed non-thermal \opr\
of \ce{NH3} ($\sim 0.7$).

Crucial too is the branching ratio of the DR of N$_2$H$^+$. As
emphasized by \cite{dislaire2012}, the channel leading to NH was
pivotal to explain the observed NH:\ce{NH2} ratio higher than
one. These authors showed that a branching ratio lower than 10\% is
sufficient, and recent storage ring measurements by \cite{vigren2012}
confirmed this value, with a branching ratio of 7$\pm ^2
_4$\%. Branching ratios for the DR of the NH$^+_n$ ($n$ = 2-4) ions
are taken from \cite{mitchell1990}, \cite{thomas2005},
\cite{jensen2000}, and \cite{ojekull2004} (see Table~\ref{tab:newdr}),
following \cite{hilyblant2010nh}.

We summarise, in Tables~\ref{tab:newin}, \ref{tab:newdr}, and
\ref{tab:newnn}, all the updates and improvements from the Flower et
al. network, including the ortho- and para-forms of hydrogen and
nitrogen species\footnote{The full chemical network is available on
  request to \texttt{pierre.hily-blant@obs.ujf-grenoble.fr}.}.

\section{Modelling}
\label{sec:modelling}

\subsection{General description}

 \begin{table*}[t]
   \tiny
  \centering
  \caption{The 104 gas-phase chemical species considered in our network$^\dag$.}
  \label{tab:species}
  \renewcommand{\arraystretch}{1.4}
  \begin{tabular}{cccccccccccccc}
    \toprule
    \ce{H} &\ce{H2(p)} &\ce{H2(o)} &\ce{He} &\ce{C} &\ce{CH} &\ce{CH2} &\ce{CH3} 
    &\ce{CH4} &\ce{O} &\ce{O2} &\ce{OH} &\ce{H2O} &\ce{CO} \\
    \ce{CO2} &\ce{C2} &\ce{C2H} &\ce{C2H2} &\ce{C3} &\ce{C3H} &\ce{C3H2}
    &\ce{CH3OH} &\ce{H2CO} &\ce{HCO2H} &\ce{N} &\ce{NH} & \ce{NH2(p)} &\ce{NH2(o)} \\
    \ce{NH3}(p)&\ce{NH3}(o) &\ce{CN} &\ce{HCN} &\ce{HNC} &\ce{N2} &\ce{NO} &\ce{S} &\ce{SH}
    &\ce{H2S} &\ce{CS} &\ce{SO} &\ce{SO2} &\ce{OCS} \\
    \ce{Mg} &\ce{Fe}&\ce{H+} &\ce{H2+(p)}&\ce{H2+(o)} &\ce{H3+(p)} &\ce{H3+(o)}
    &\ce{He+} &\ce{C+} &\ce{CH+} &\ce{CH2+} &\ce{CH3+} &\ce{CH4+}&\ce{CH5+} \\
    \ce{O+} &\ce{O2+} &\ce{OH+} &\ce{OD+} &\ce{H2O+} &\ce{H3O+}
    &\ce{CO+} &\ce{HCO+} &\ce{HCO2+} &\ce{C2+} &\ce{C2H+} &\ce{C2H2+} &\ce{C2H3+}&\ce{C3+} \\
    \ce{C3H+} &\ce{C3H2+} &\ce{C3H3+} &\ce{N+} &\ce{NH+}
    &\ce{NH2+(p)} &\ce{NH2+(o)} &\ce{NH3+(p)} &\ce{NH3+(o)} &\ce{NH4+(p)} &\ce{NH4+(m)} &\ce{NH4+(o)} &\ce{CN+} & \ce{C2N+}\\
   \ce{HCN+} &\ce{H2CN+} &\ce{H2NC+} &\ce{N2+} &\ce{N2H+}
    &\ce{NO+} &\ce{HNO+} &\ce{S+} &\ce{SH+} &\ce{H2S+} &\ce{H3S+} &\ce{CS+} &\ce{HCS+} &\ce{SO+} \\
    \ce{HSO+} &\ce{HSO2+} &\ce{HOCS+} &\ce{Fe+} &\ce{C4H+} &\ce{C6H+} \\

    \bottomrule
  \end{tabular}
  \begin{list}{}{}
  \item $^\dag$ (o), (p), and (m) stand for ortho, para, and meta modifications respectively.
  \end{list}
\end{table*}

The complete network contains 104 species (see
Table~\ref{tab:species}) and 917 reactions. The physical conditions,
which are fixed, are typical of dark clouds: the gas has a constant
number density of hydrogen nuclei $\nh = \nn{H} + 2\nn{H2}$ of
\dix{4}\ccc, shielded from the UV photons by 10 magnitudes of visual
extinction, and the kinetic temperature is kept constant at
10~K. Ionization is driven by cosmic rays, with secondary photons
included, and the ionization rate of hydrogen is
$\zeta=1.3\tdix{-17}\pers$, the standard value usually attributed to
dense regions \citep{prasad1980,caselli1998,wakelam2005}. The
processes involving dust grains are charge exchange, DR reactions, and
the formation of \hh. We adopt a single dust grain radius of 0.1\micr,
following \citet{walmsley2004}, satisfying a dust:gas mass ratio of
1\%. The abundances of all species are followed with time, until a
steady-state is reached. A typical illustrative result is shown in
Fig.~\ref{fig:abtime}.

Even though we compute the time evolution of all the abundances, we
focus our analysis on the steady state, which is reached after
typically $\sim$1~Myr for carbon and oxygen species, $\sim3-5$
  Myr for NO, \ce{N2H+}, and NH, and after $\sim$10~Myr for nitrogen
bearing species (see Fig.~\ref{fig:abtime}). We note that these
timescales are likely upper limits as they correspond to initial
conditions where all elements (except hydrogen) are in atomic form. As
mentioned earlier, nitrogen hydrides are late-time species, a
consequence of the slow neutral-neutral reactions needed to build
\ce{N2}. The longer timescale of 10~Myr is larger than the
gravitational free-fall timescale of $\sim 0.5$~Myr but comparable to
the ambipolar diffusion timescale of $\sim 10$~Myr at a density of
\dix{4}\ccc \citep{tassis2004, bergin2007}. Indeed, in their chemical
models of collapsing cores, \cite{tassis2012} also considered delay
times (before collapse starts) up to 10~Myr. Another assumption of our
model is that freeze-out, which occurs on a typical timescale
  $\sim \dix{10}/\nh$~yr \citep{bergin2007}, is not considered during
  the time evolution of the gas, although observations show depletion
of several species in pre-stellar cores
\citep[e.g. ][]{tafalla2004}. However, observations also show that an
important fraction of molecular species remains in the gas phase, such
that desorption mechanisms are required \citep{hasegawa1993}. Those
are highly uncertain, and depend critically on binding energies of the
adsorbed species. Nevertheless, molecular mantles are observed in the
core of dense dark clouds \citep{whittet1983, boogert2011}, indicating
that molecular condensation on grains occurs efficiently. This effect
is crudely accounted for in our model by adopting initially depleted
elemental abundances in the gas-phase. We also note that although
steady-state is rather unlikely, both steady-state and time-dependent
modelling carry their own limitations. Time-dependent calculations
depend critically on the assumed initial abundances, which are poorly
known, whilst steady-state models do not. In addition, focussing on
the steady-state does not prevent from identifying key reactions and
general trends, while avoiding the uncertainties described above.

\subsection{Elemental abundances}
\label{sec:ref_mod}

The choice of the gas-phase elemental abundances is not a trivial
task, nonetheless, and the abundances of the elements strongly
influence the chemical state of the gas. One example is the relative
amount of carbon and oxygen, which is known to strongly affect the
chemistry \citep{vandishoeck1998, lebourlot1995}. For most of the
elements, including nitrogen, we follow \cite{flower2005},
who have estimated gas-phase elemental abundances by combining
observational constraints from the diffuse ISM and infrared signatures
of ices \citep{gibb2000}, and assuming the composition for the
refractory core of dust grains.  The abundances of carbon, oxygen, and
sulphur, deserve special attention.

The elemental gas-phase abundance of sulphur, noted \abtot{S}, is very
poorly constrained. It can vary by almost three orders of magnitude if
we consider the range delimited by the so-called \emph{low metal
  abundance} and \emph{high metal abundance} cases defined originally
by \cite{graedel1982}. The first category is suggested by the low
electron abundance in dense clouds and may be more representative of
the chemistry in such environments, whilst the second category
reflects the standard $\zeta$ Oph sightline. Further studies of dense
dark clouds have corroborated the \emph{low metal} sulphur gas-phase
abundance, even where elemental carbon, oxygen and nitrogen are
depleted by only factors of a few \citep{ruffle1999}. In the well
studied TMC-1 and L183 dark clouds, \cite{tieftrunk1994} found that
the sum of all detected S-bearing molecules represents only 0.1\% of
the elemental sulphur solar abundance $\abS\approx 1.5\tdix{-5}$
\citep{przybilla2008, asplund2009}. In the following, we have applied
depletion factors from the gas-phase of 200, 20, and 2, or abundances
with respect to H nuclei of 8.0\tdix{-8}, 8.0\tdix{-7}, and
8.0\tdix{-6}, respectively.

The reference cosmic abundance for oxygen is 575~ppm
\citep{przybilla2008}. Based on 56 diffuse sightlines,
  \cite{cartledge2004} evidenced a decreasing gas-phase abundance of
  oxygen with increasing sightline mean density, reflecting the uptake
  of oxygen in silicate and oxydes into grain cores. They found an
  average gas-phase atomic oxygen abundance of 280~ppm for densities
  greater than 1.5\ccc, suggesting that up to 300~ppm of oxygen have
  been removed from the gas phase. This idea was further developped by
  \cite{whittet2010} who performed a comprehensive oxygen budget up to
  densities $\sim 1000$\ccc, and showed that $\sim 100$~ppm and $\sim
  120$~ppm are locked into ices and silicates or oxydes,
  respectively. Taking into account that $\sim$50~ppm is in gaseous
  CO, it must be recognized that up to $\sim$300~ppm of oxygen are in
  unknown carriers. The depletion of oxygen not accounted for by known
  carriers is also demonstrated by \cite{nieva2012} who compared the
  abundances derived by Cartledge et al. with their \emph{Cosmic
    Abundance Standard} based on B-stars.

Depletion of carbon is surprisingly not well constrained \citep[see
  e.g.\ Figure~5 of ][]{jenkins2009}, essentially for observational
limitations. In addition, the column densities derived so far in the
litterature may overestimate the amount of gas-phase carbon, because of
an underestimated oscillator strength for the CII transition at
2325\ang\ \citep{sofia2011}. The error could be up to a factor of
two.

To reflect the large uncertainties in the gas-phase elemental
abundances of carbon and oxygen, we decided to vary the
carbon-to-oxygen gas-phase elemental abundance ratio,
\abtot{C}/\abtot{O}, simply noted C/O hereafter. To do so, we have
assumed a constant elemental abundance of carbon, $\abtot{C} =
8.3\tdix{-5}$, and varied that of oxygen from 50~ppm to 280~ppm. The
lower limit is imposed by gaseous CO, and the upper limit is taken
from \cite{cartledge2004}.  The corresponding range in C/O is 0.3 to
1.5. The upper value of 280~ppm implicitely requires that
$\sim$230~ppm of oxygen has been taken from the various reservoirs
(solid, ices, unknown carriers) and released in the gas phase, by some
unspecified processes. Our strategy is therefore analogous in some
respects to that adopted by \citet{terzieva1998} and
\citet{tassis2012} who considered a similar range of elemental C/O
ratios. However, the main difference with our model is that we do not
consider gas-grain processes apart from charge exchange with grains
and \hh\ formation, such that the total amount of gas-phase
carbon and oxygen remain constant during the time evolution. For each
value of C/O, three models, corresponding to the three sulphur
abundances discussed above, were performed. The intial gas-phase
elemental abundances are summarized in Table~\ref{tab:ab_ini}.

\begin{table}[t]
  \centering
  \caption{\label{tab:ab_ini} Gas-phase elemental abundances$^\dag$
      adopted in our model.}
  \begin{tabular}{lrrr}
    \toprule
    Element & This work & Flower et al 2005 \\ 
    \midrule
    He       & 0.10              & 0.10   \\
    C$^\S$   & 8.3(-5)           & 8.3(-5)\\
    N$^\S$   & 6.4(-5)           & 6.4(-5)\\
    O	     & 5.5(-5) to 2.8(-4)& 1.2(-4)\\
    S$^\ddag$& 8.0(-8) to 8.0(-6)  & 6.0(-7)\\
    Fe$^\S$  & 1.5(-9)           & 1.5(-9)\\
    \bottomrule
  \end{tabular}
  \begin{list}{}{}
  \item $^\dag$ {Fractional abundances relative to total H
      nuclei. Numbers in parentheses are powers of 10.}
  \item $^\ddag$ The lower value of the total gas-phase elemental
    abundance of S is taken from the ``low metal'' case of
    \cite{graedel1982}.
  \item $^\S$ The elemental abundances of gas-phase nitrogen, carbon, and iron,
    are taken from \cite{flower2005}.
  \end{list}
\end{table}

\section{Results}
\label{sec:results}

\subsection{Influence of the C/O ratio}
\label{sec:CsurO}

\begin{figure*}[t]
  \centering
  \includegraphics[height=0.35\hsize]{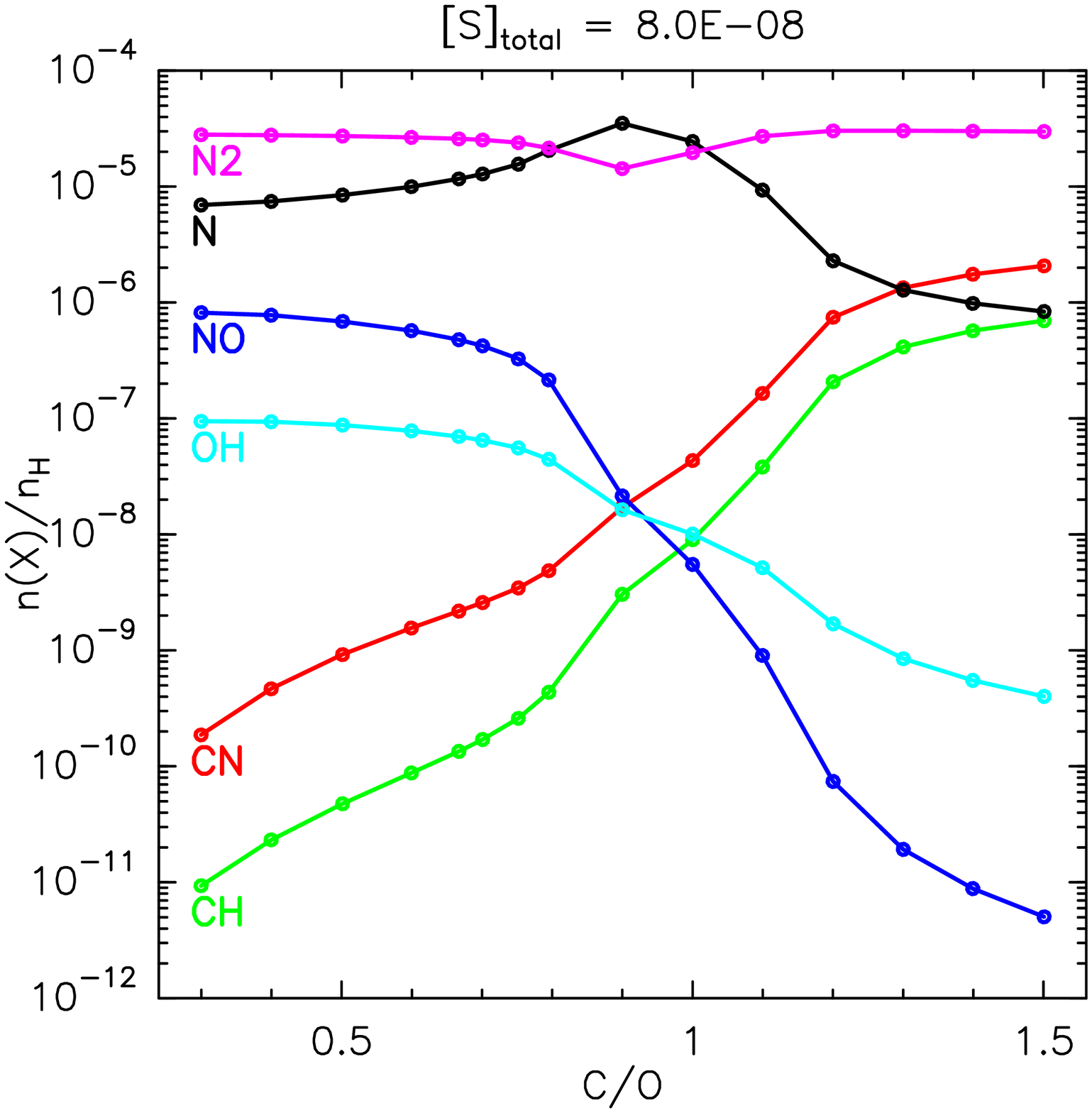}\hfill%
  \includegraphics[height=0.35\hsize]{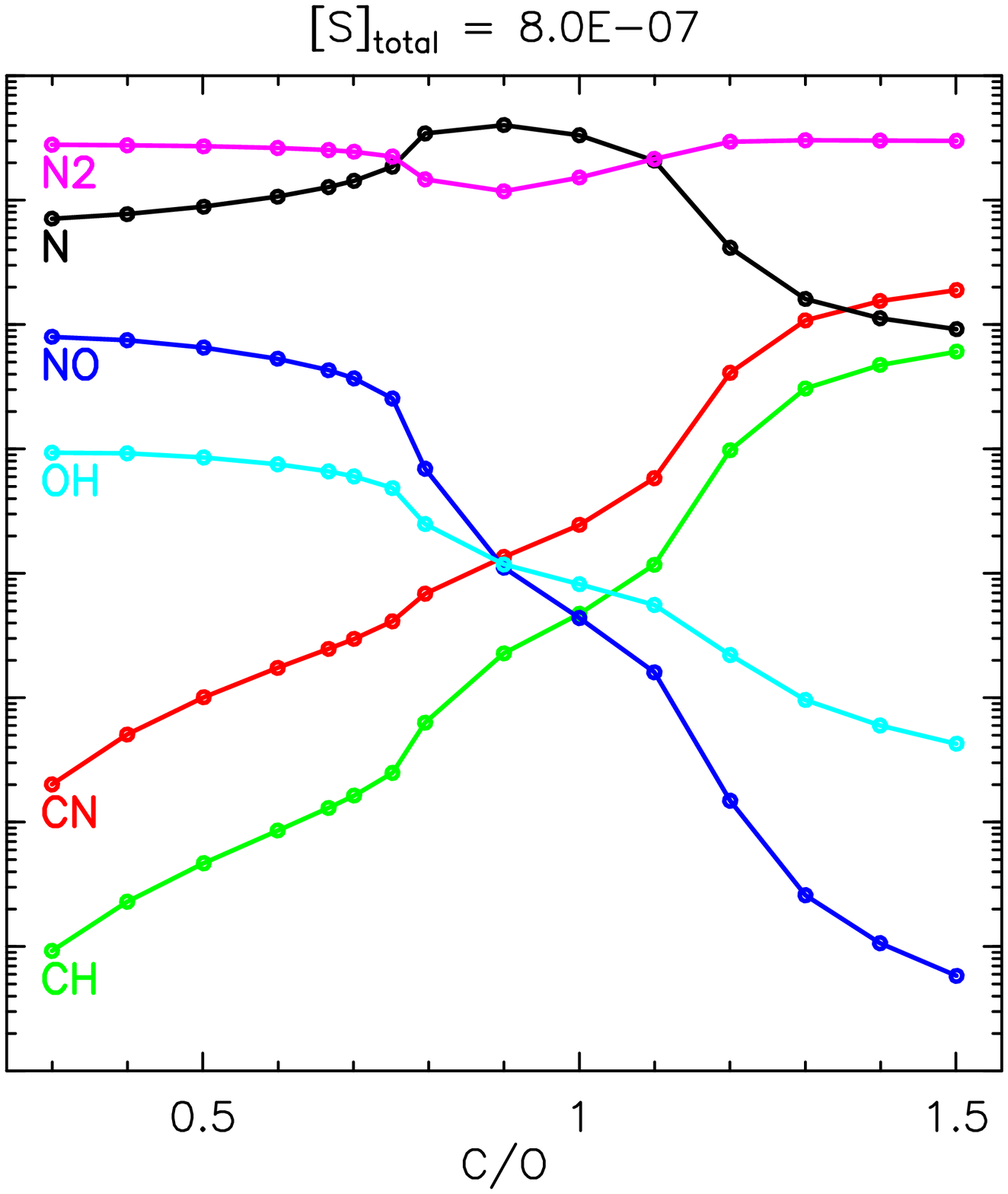}\hfill%
  \includegraphics[height=0.35\hsize]{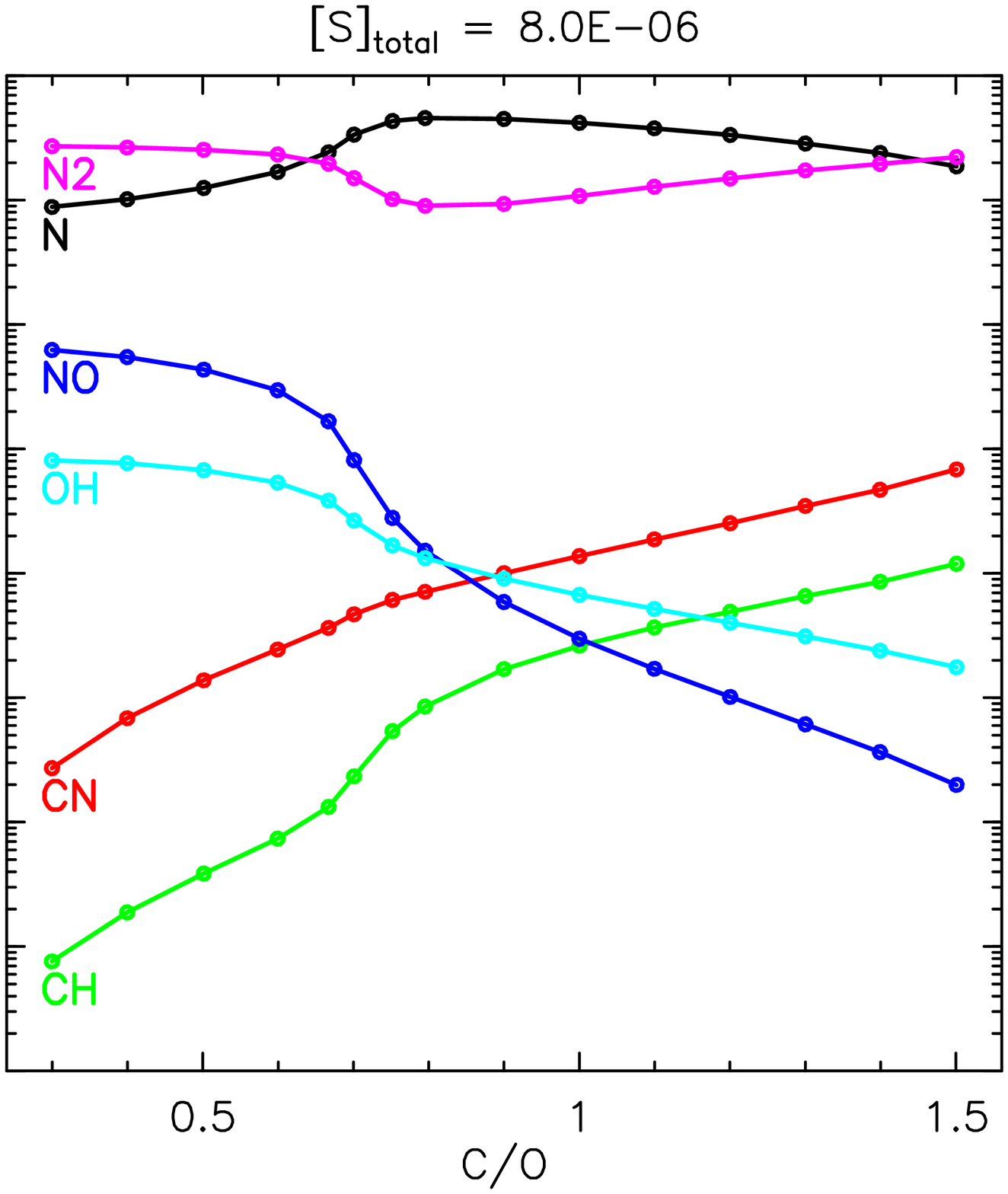}\bigskip\\
  \includegraphics[height=0.35\hsize]{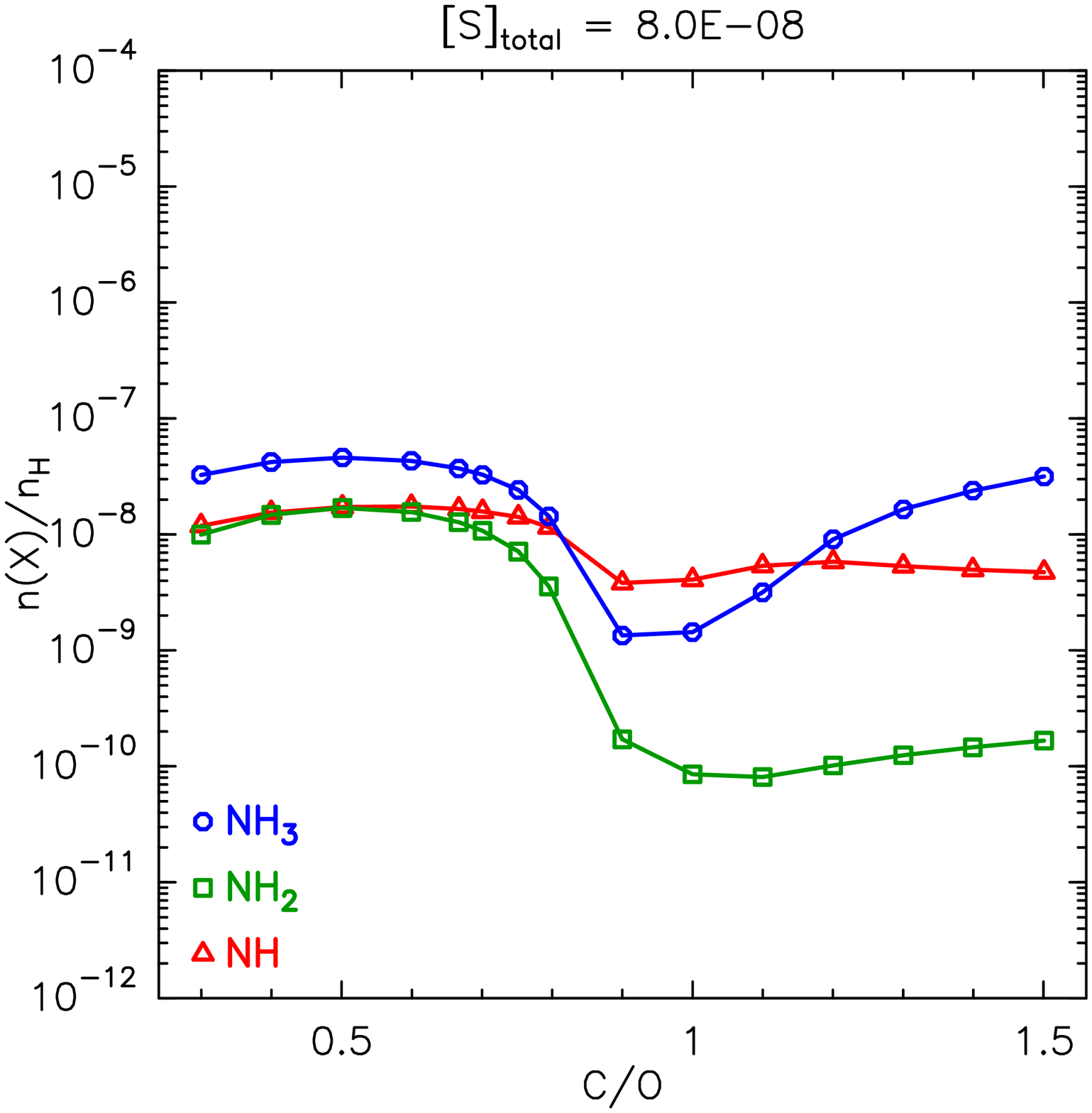}\hfill%
  \includegraphics[height=0.35\hsize]{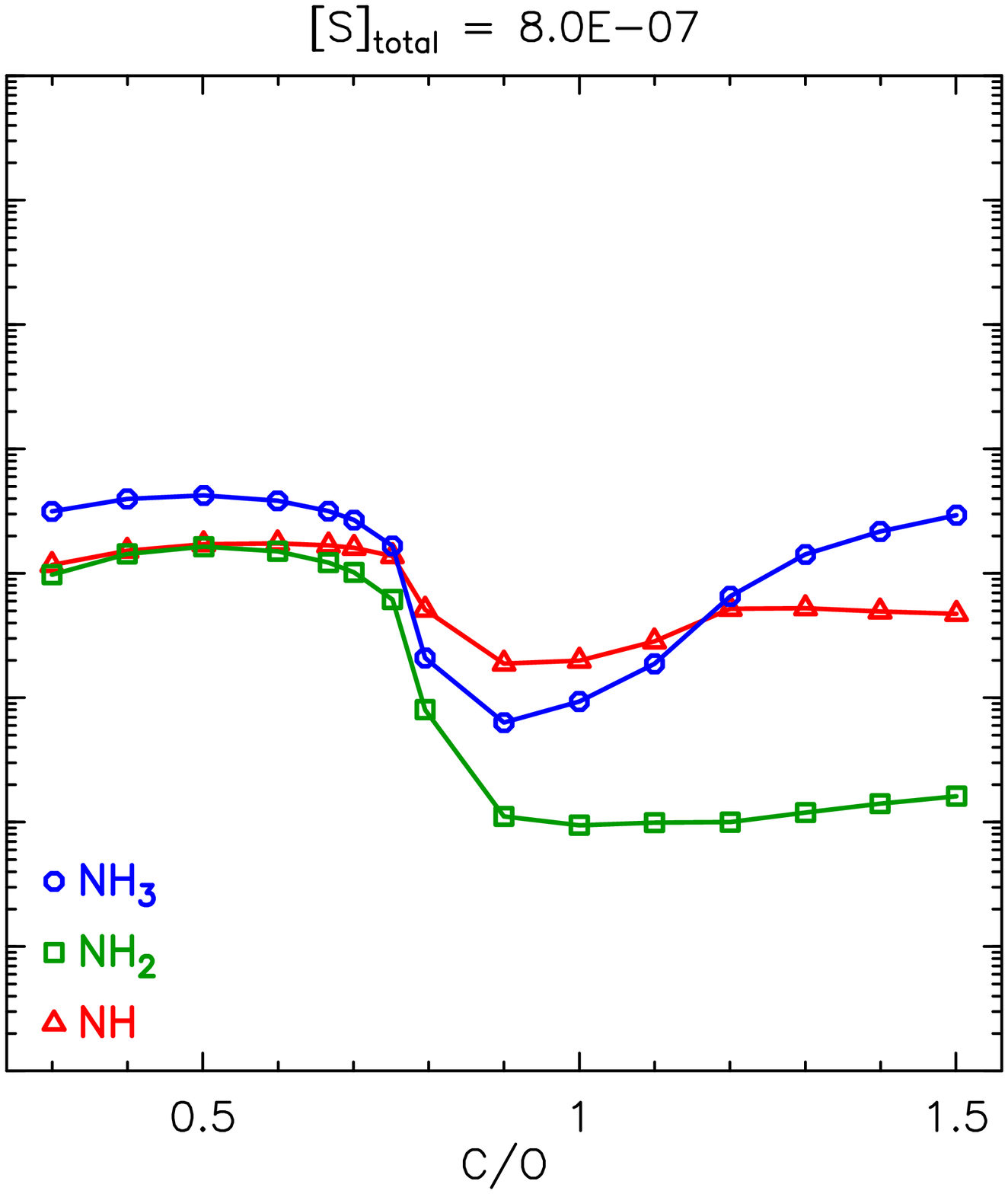}\hfill%
  \includegraphics[height=0.35\hsize]{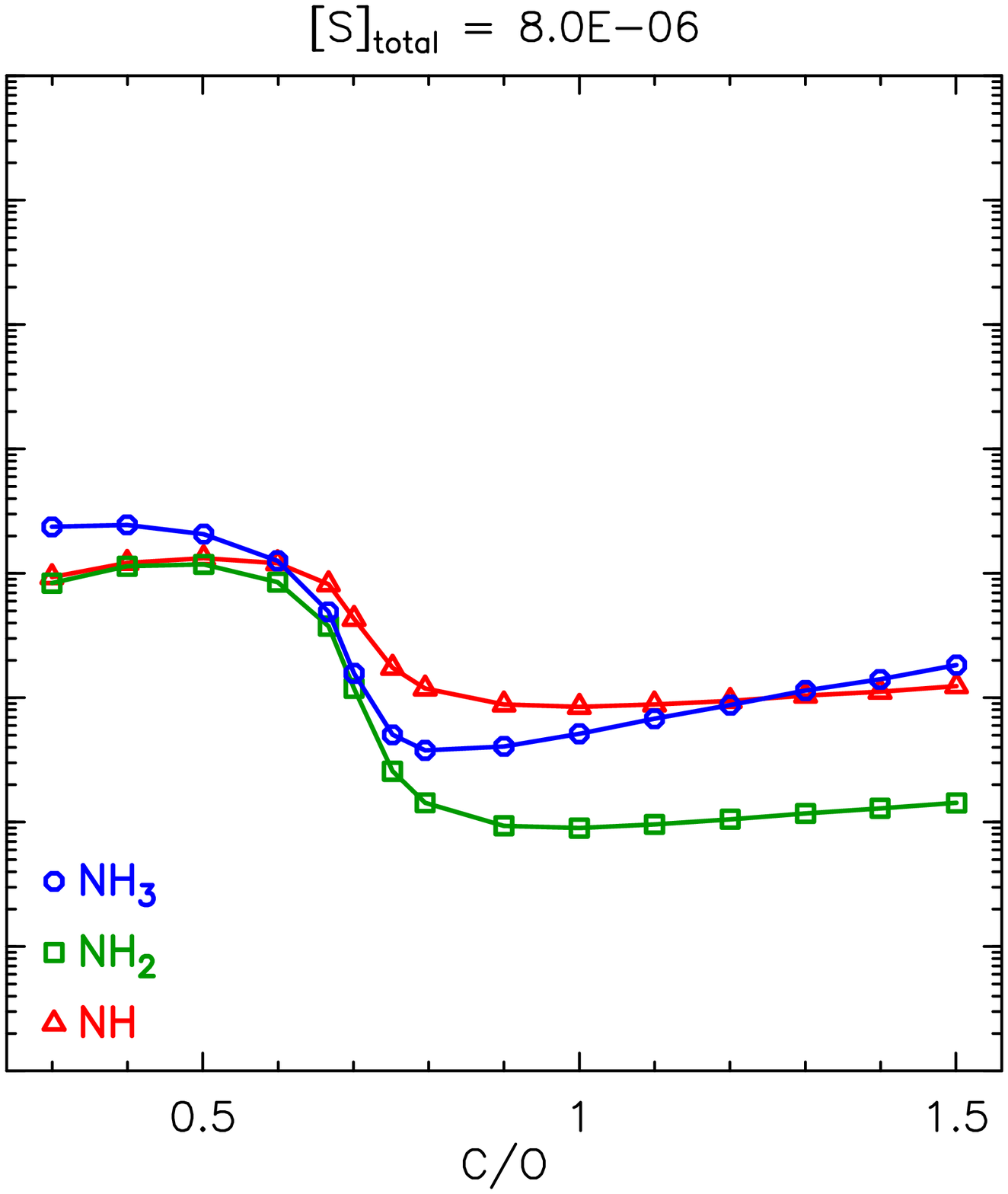}
  \caption{Steady-state abundances of several nitrogenated-bearing and
    nitrogen-chemistry related species, as a function of the gas-phase
    elemental abundance ratio of C/O, for three different values of
    gas-phase elemental abundance of sulphur (from left to right: $\abS
    = 8.0\tdix{-8}$, 8.0\tdix{-7}, 8.0\tdix{-6}). C/O values include
    0.3, 0.4, 0.5, 0.6, 0.7, 0.75, 0.8, 0.9, 1.0, 1.1, 1.2, 1.3,
    1.4, 1.5, and 0.67 which was the ratio employed in the previous studies of
    Flower et al. 2006. \emph{Top panels}: molecules involved in the
    atomic-to-molecular conversion of nitrogen. \emph{Bottom panels}:
    nitrogen hydrides.}
  \label{fig:ab_csuro}
\end{figure*}

The steady-state abundances of several nitrogenated species, together
with chemically related species, are shown in Fig.~\ref{fig:ab_csuro},
as a function of the elemental C/O gas-phase abundance ratio, for
three sulphur abundances. We first focus on the \emph{low metal}
sulphur abundance (left panel). For all values of C/O, atomic and
molecular nitrogen are the most abundant species, except at high C/O,
where CN becomes more abundant than N. Our model predicts that, at low
and high C/O, molecular nitrogen is the dominant carrier of
nitrogen. Oxygen-bearing species such as NO and OH see their abundance
decreasing with increasing C/O, by 2 to 5 orders of magnitude. This is
a consequence of the decreasing total amount of oxygen available in
the gas-phase. On the contrary, the abundances of CN and CH increase
by the same amount. Yet, for the latter species, this is not due to an
increase in the total amount of carbon, but to an increase of the
amount of carbon available in the gas phase, since less carbon is
locked into the very stable CO molecule. The turning point between the
low- and high-C/O regimes occurs around 0.9. In this intermediate
regime ($0.8\le\rm {C/O}\le 1.0$), molecular nitrogen is no longer the
main reservoir of nitrogen because OH and CH, which ensure the
  conversion from N to \ce{N2}, are two orders of magnitude less
  abundant. The above reasoning remains true for a total gas-phase
  sulphur abundance of 8.0\tdix{-7}, although the mid-C/O regime now
  extends up to 1.1. At even higher total gas-phase abundance of
  sulphur, the high-C/O regime in which N:\ce{N2}$<1$ is not
  recovered.

  Concerning nitrogen hydrides their evolution with C/O is not
  uniform.  For sulphur abundances up to 8.0\tdix{-7}, the NH
  abundance varies by less than an order of magnitude over the whole
  range of C/O ratio, in sharp contrast with \ce{NH2} whose abundance
  decreases from \dix{-8} in the low-C/O regime, to \dix{-10} in the
  high-C/O regime. The abundance of ammonia is a few \dix{-8} in both
  the low- and high-C/O regimes, but drops to \dix{-9} in the mid-C/O
  regime. When $\abS=8.0\tdix{-6}$, the behaviour of \ab{NH} is similar
  to that of \ab{NH2}, and drops by more than an order of magnitude
  from the low- to high-C/O regimes. For ammonia, the same trend
  holds, but unlike \ab{NH} or \ab{NH2}, \ab{NH3} manages to increase
  with C/O, although not recovering its value of the low-C/O regime.

We here provide explanations for the observed abundance behaviours,
focussing on the \emph{low metal} sulphur abundance case. The balance
evolution between N and \ce{N2} with C/O is explained by the fact that
as C/O increases, more oxygen and less carbon -- relative to the total
-- are locked into CO. Therefore, there is less oxygen available in
the gas phase, and the abundances of NO and OH drop. The contrary
holds true for carbon: there is more CH and CN. Since the N-to-\ce{N2}
conversion proceeds through NO and/or CN, this explains the peak of
N:\ce{N2} ratio for C/O $\sim$ 1.  As a consequence, the modelled
abundances of all nitrogen hydrides which strongly depend on the
conversion of N into \ce{N2}, decrease at C/O$\sim$1.  However, the
fact that ammonia recovers its low-C/O abundance at high C/O when NH
and \ce{NH2} do not, stems from the fact that these two radicals are
primarily destroyed by reactions with atomic carbon, while ammonia is
not. Reaction of NH with neutral carbon forms CN, and \ce{NH2 + C}
produces HCN and HNC. On the other hand, \ce{NH3} is not destroyed by
neutral carbon but by \ce{H3+} (whose abundance does not strongly
depend on C/O), \ce{HCO+}, and \ce{C+}. As a consequence, in the
high-C/O regime, the large amount of carbon not locked into CO
enhances mostly the destruction of NH and \ce{NH2}.  At the same time,
the N-to-\ce{N2} conversion restarts through CN, and the formation of
ammonia increases.

\subsection{Influence of the sulphur abundance}

We now further analyse the influence of the total sulphur abundance on
the steady-state abundances of nitrogen hydrides. The top panels of
Fig.~\ref{fig:ab_csuro} do not show striking differences. Indeed,
models with $\abS=8.0\tdix{-8}$ and 8.0\tdix{-7} are very similar, and
the only change is the broadening of the mid-C/O regime mentioned
above. For $\abS=8.0\tdix{-6}$, CN and CH abundances drop by more than
an order of magnitude at high C/O, whereas those of NO and OH
increase.

The broadening of the mid-C/O regime as \abS\ increases results from a
combination of several chemical effects. As already mentioned, in
this regime, characterized by N:$\ce{N2}>1$, the conversion from N to
\ce{N2} is weak, because of the low abundances of CH and OH. Comparing
models with $\abS=8.0\tdix{-7}$ and 8.0\tdix{-6}, \ab{CH} drops by about
two orders of magnitude at high C/O while \ab{OH} is only marginally
affected. Although \ab{NO} increases by more than a factor 10, it
remains 100 times less abundant than CN and does not compensate for
the drop of \ab{CN}. The net effect is a global decrease of the
conversion partners. Hence N remains more abundant than \ce{N2} up to
$\rm{C/O}=1.5$. Thus, the effect of \abS\ must be traced back to CH, through
the reaction \reac{sp-ch}{S+ + CH -> CS+ + H} which removes CH. A
similar reaction holds for OH, but its abundance only slightly
decreases. To understand this behaviour, we notice that OH is formed
through the DR of \ce{H3O+}, the abundance of which decreases with
\abS.  Indeed the ionization potential of S (10.36~eV) is lower than
that of H and increasing \abS\ drives up the ionization fraction. The
resulting enhanced DR destruction of \ce{H3O+} thus compensates for
the destruction of OH by \ce{S+}. In the mid-C/O regime at low to
intermediate sulphur abundances, there is such a combination of
effects that both OH and CH remain too rare to ensure an efficient
conversion of N into \ce{N2}, leading to a well-defined C/O range in
which N:$\ce{N2}>1$. In this parameter space, the enhanced destruction
of CH by \ce{S+} dominates over the increase of atomic carbon in the
gas phase, and the increase of \ce{H3O+} does not compensate for the
destruction of OH by \ce{S+}.

Regarding nitrogen hydrides, a variation from low to intermediate
sulphur abundance does not change dramatically the abundances. This is
not true when going from intermediate to high \abS. At high C/O,
\ab{NH} drops by a factor of 5, while \ce{NH2} is unaffected and
remains at a very low abundance. The most dramatic change is that of
\ab{NH3} which drops by a factor of 20. This is due to the
combination of the two reactions \ce{NH3 + S+ -> NH3+ + S} and
\ce{NH3+ + e- -> NH + H + H} which get greatly enhanced at high
\abS. Their net effect is to transfer nitrogen from ammonia to
NH. Nevertherless, the latter sees its amount divided by a factor 5
because the destruction path \ce{NH + N} becomes important in the
N:$\ce{N2}>1$ regime.

\subsection{Comparison with observations}

\begin{figure*}[t]
  \centering
  \includegraphics[height=0.35\hsize]{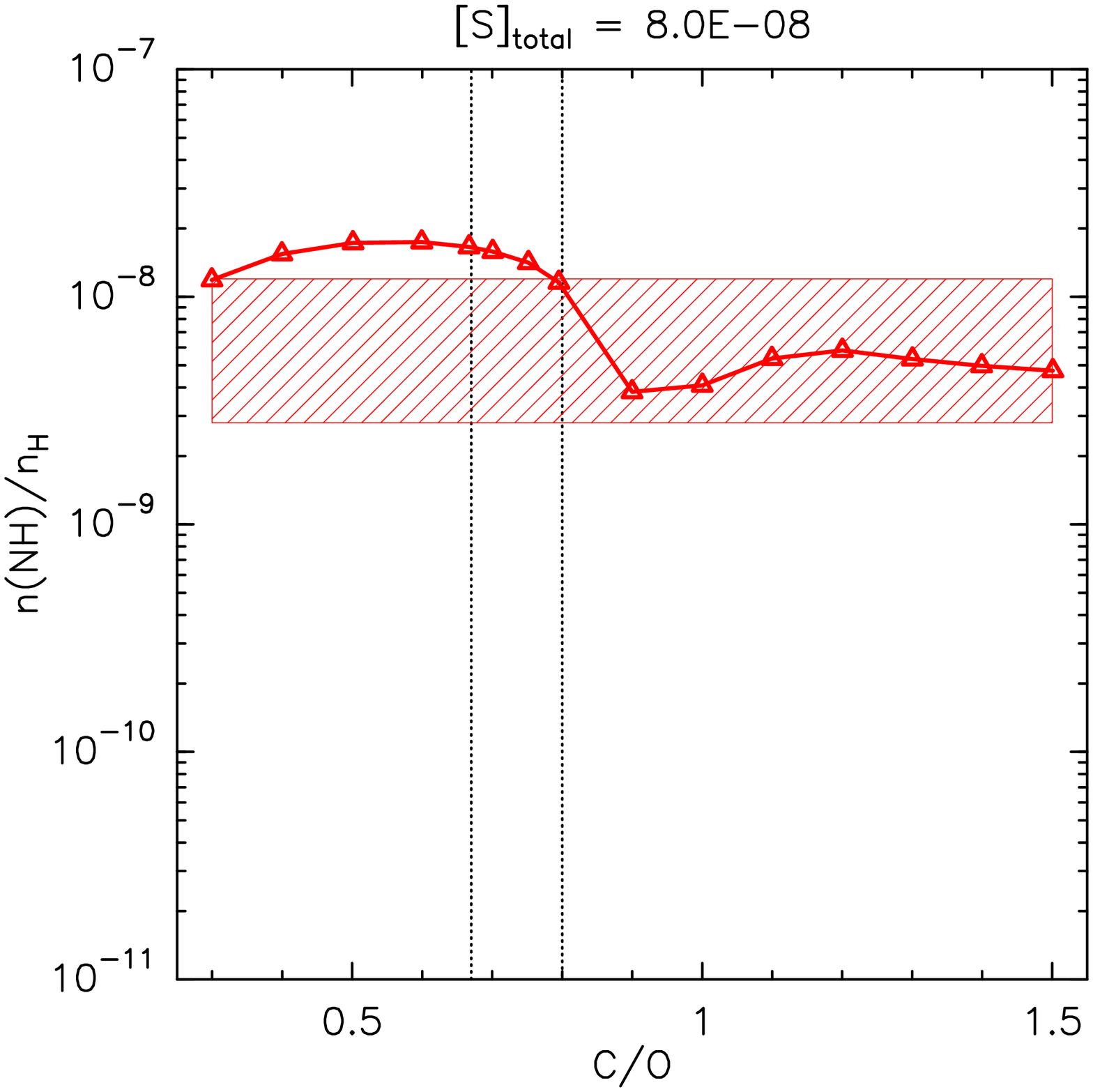}
  \includegraphics[height=0.35\hsize]{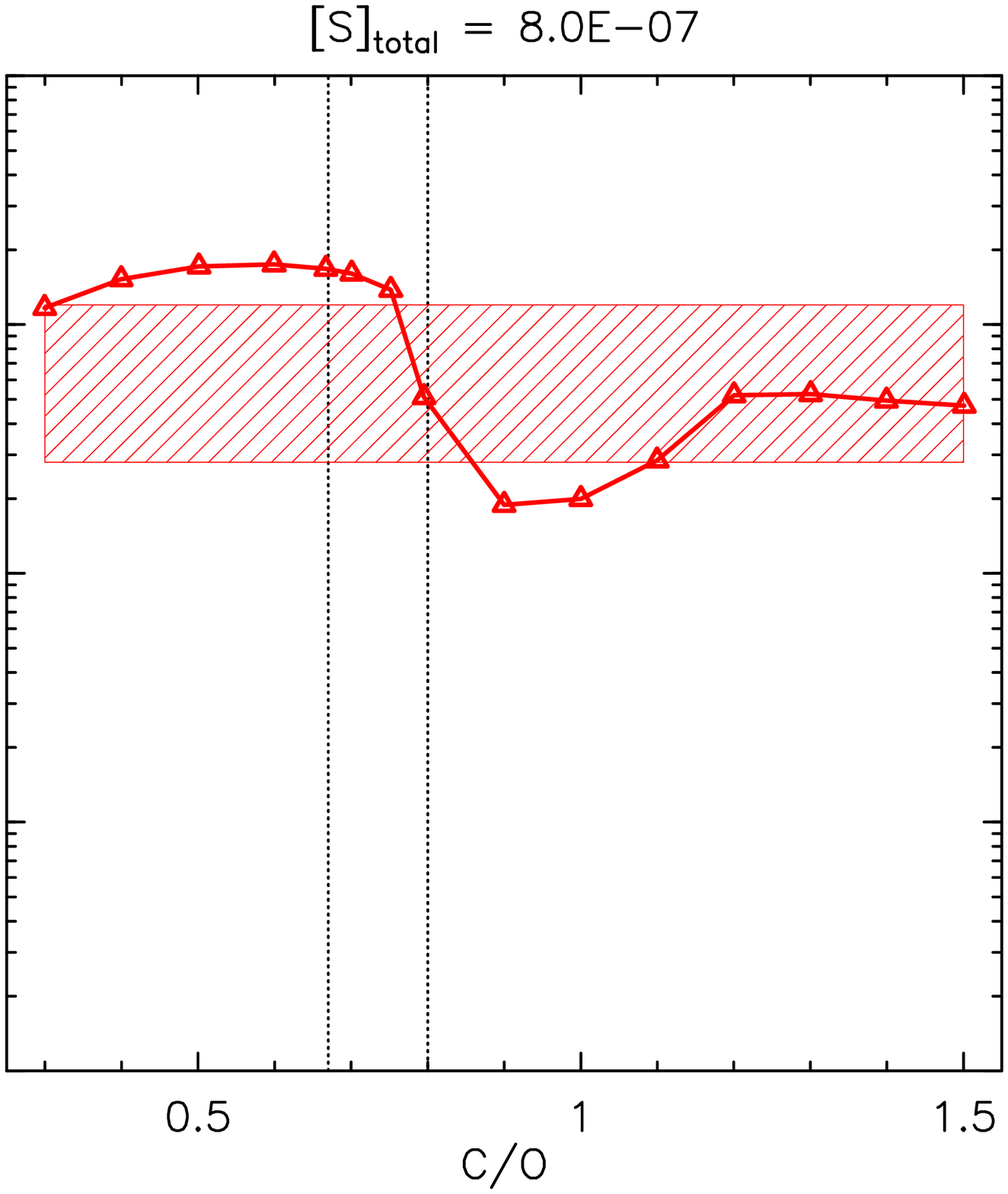}
  \includegraphics[height=0.35\hsize]{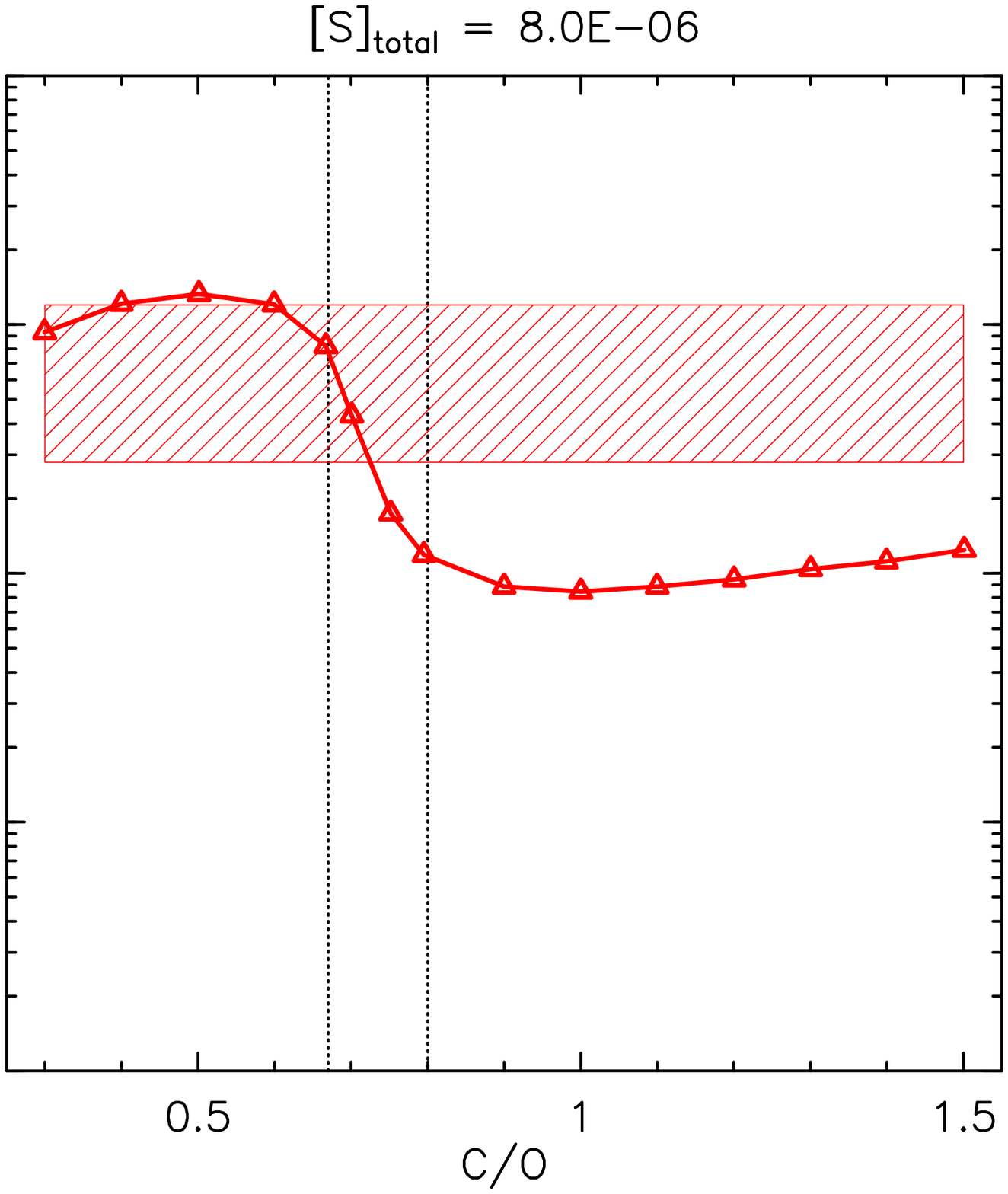}
  \includegraphics[height=0.35\hsize]{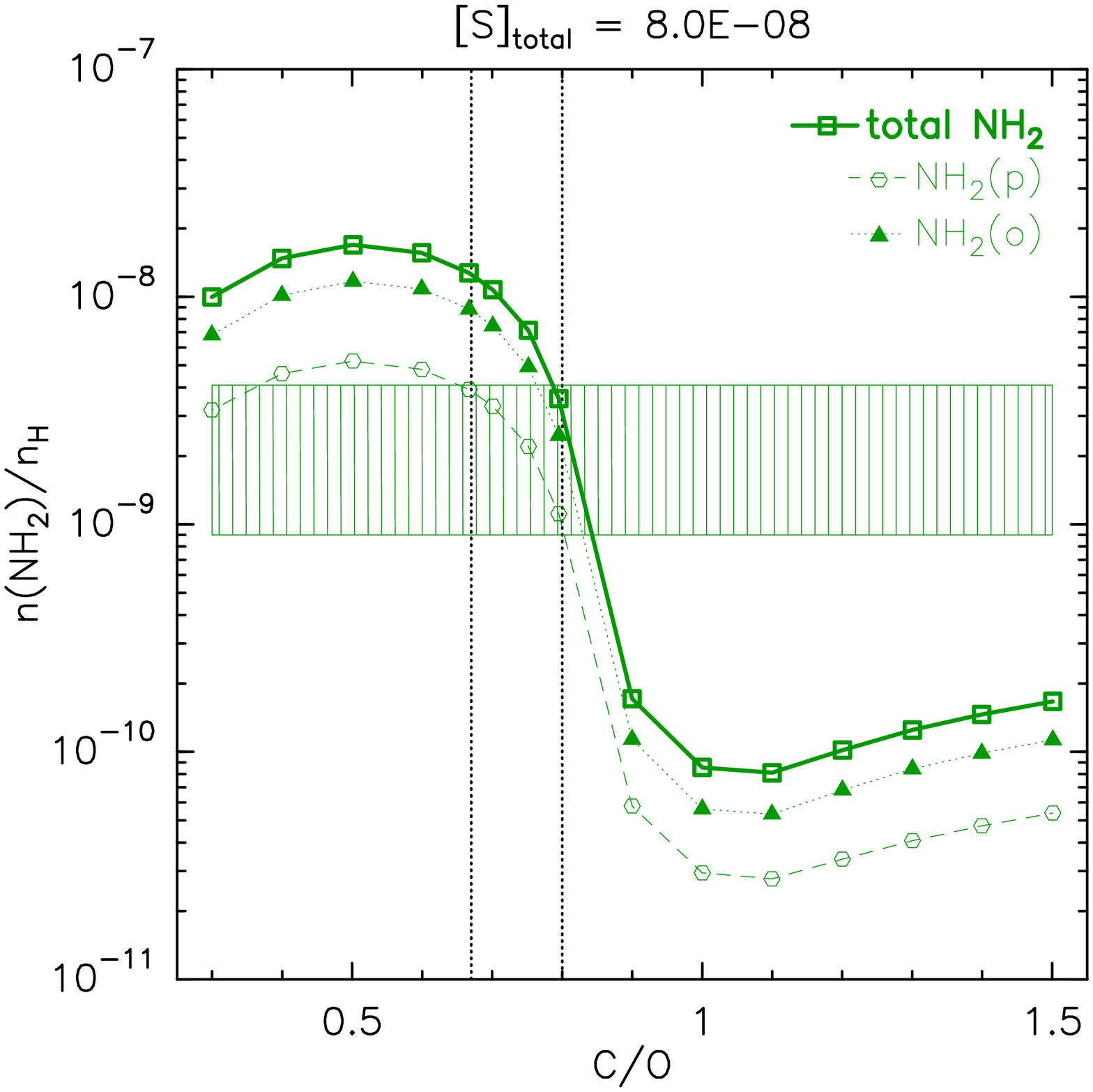}
  \includegraphics[height=0.35\hsize]{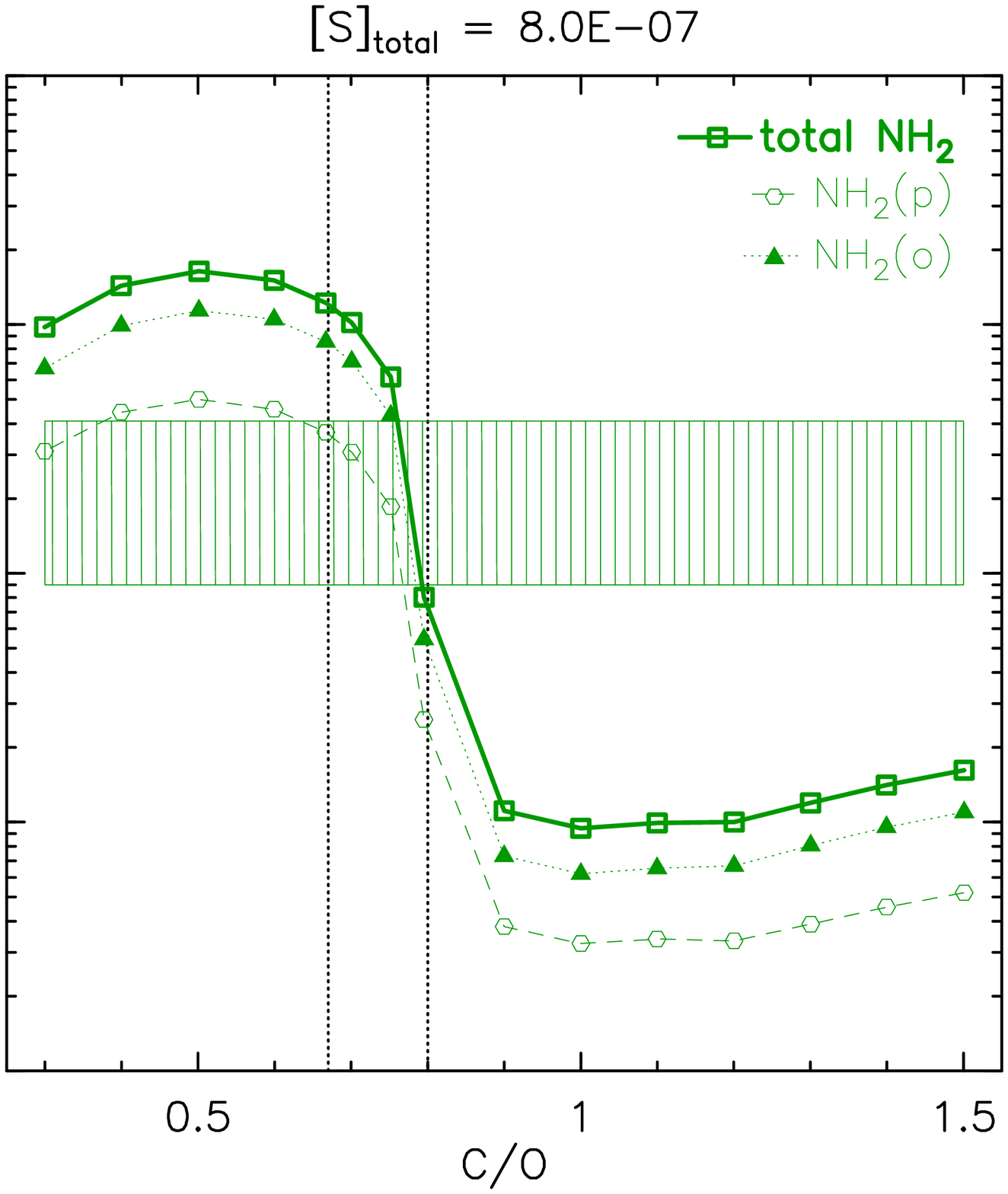}
  \includegraphics[height=0.35\hsize]{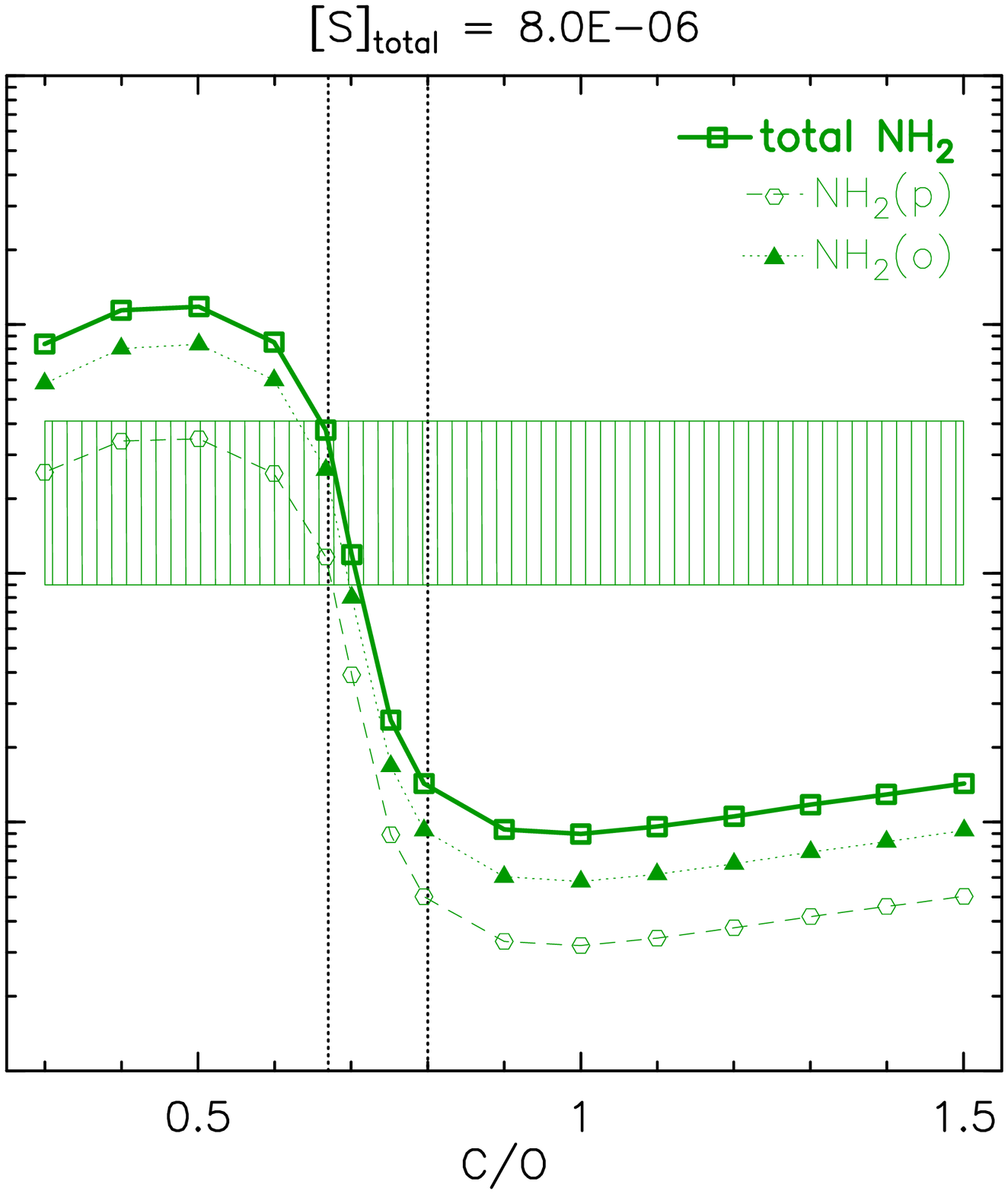}
  \includegraphics[height=0.35\hsize]{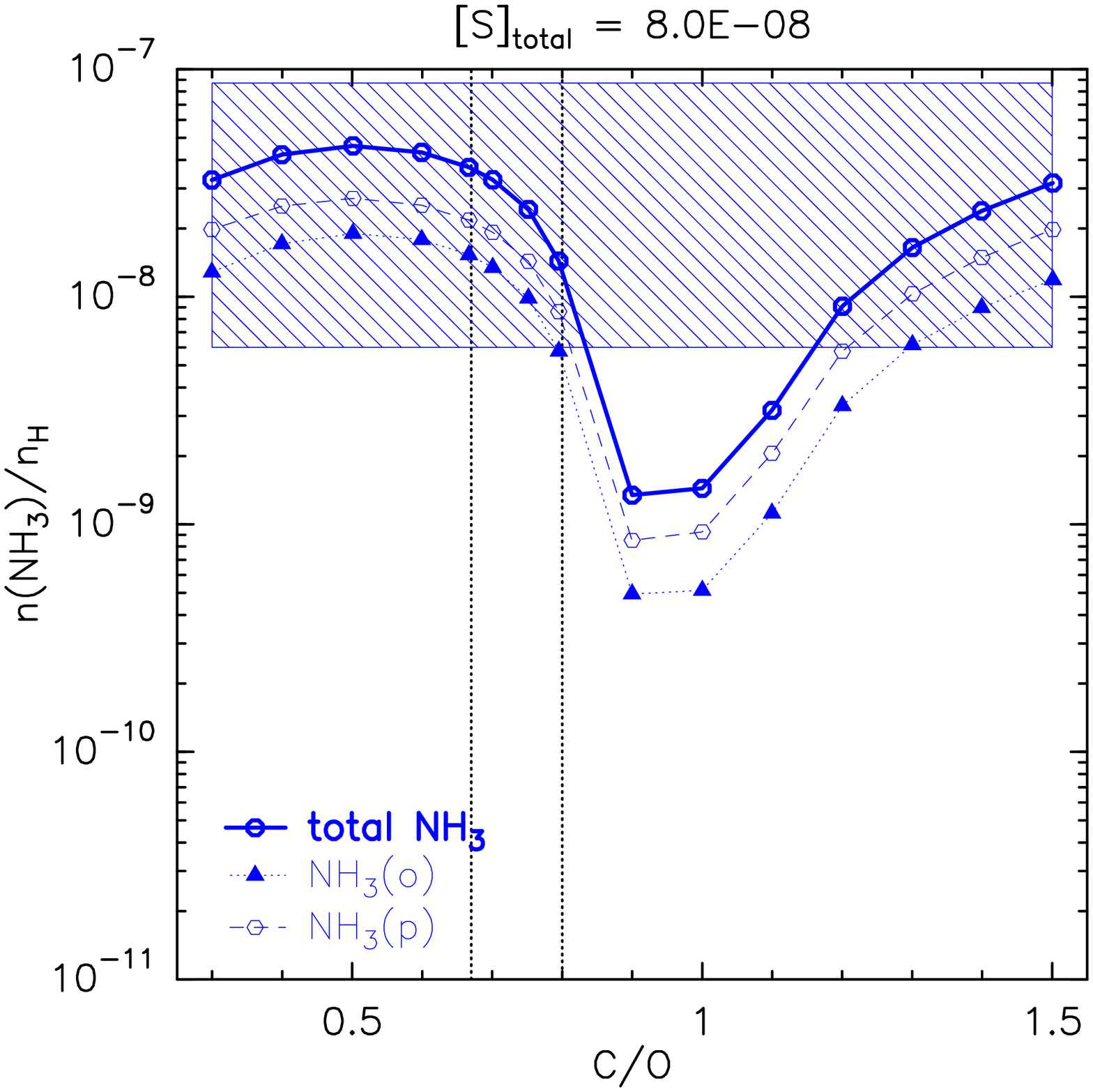}
  \includegraphics[height=0.35\hsize]{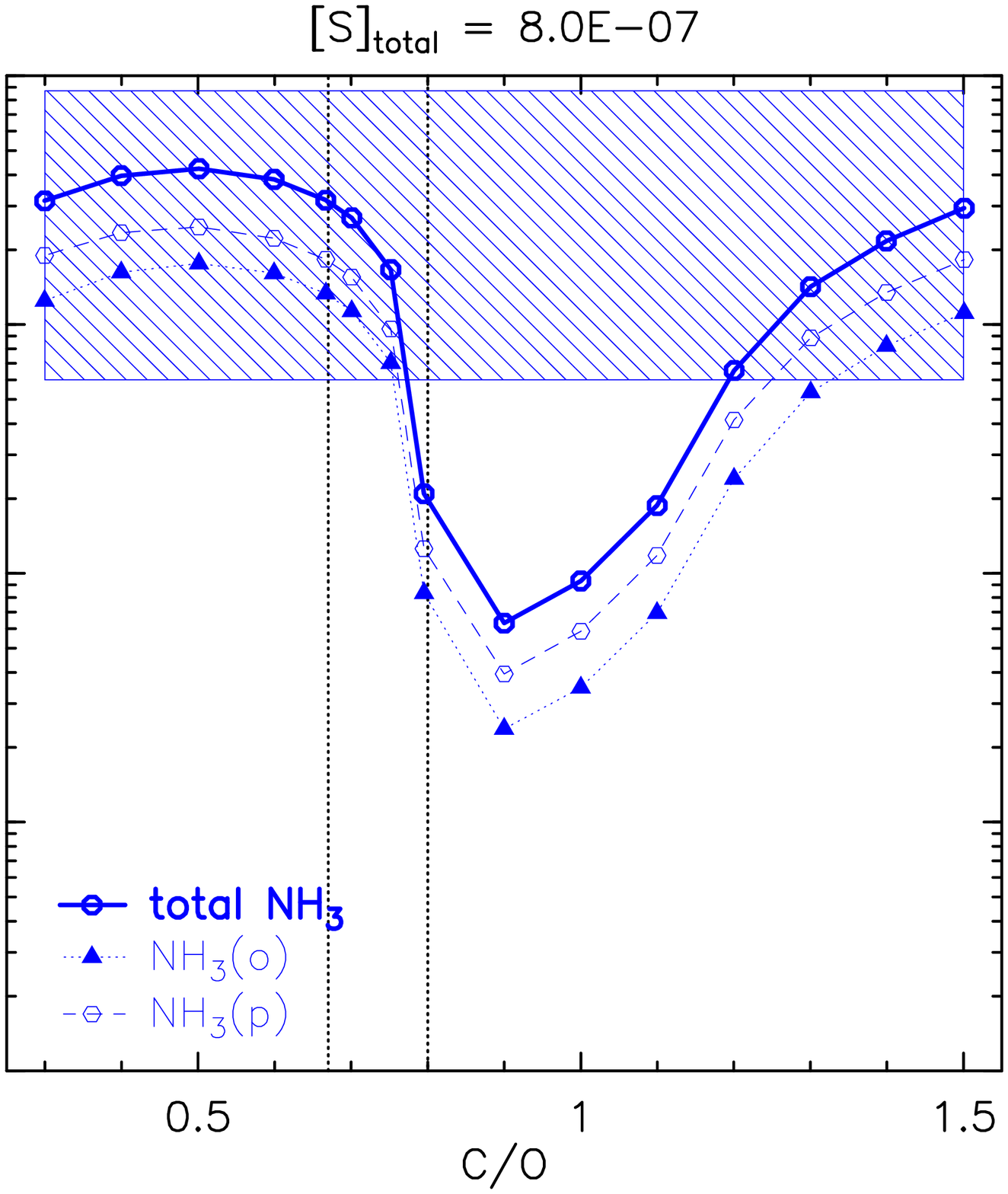}
  \includegraphics[height=0.35\hsize]{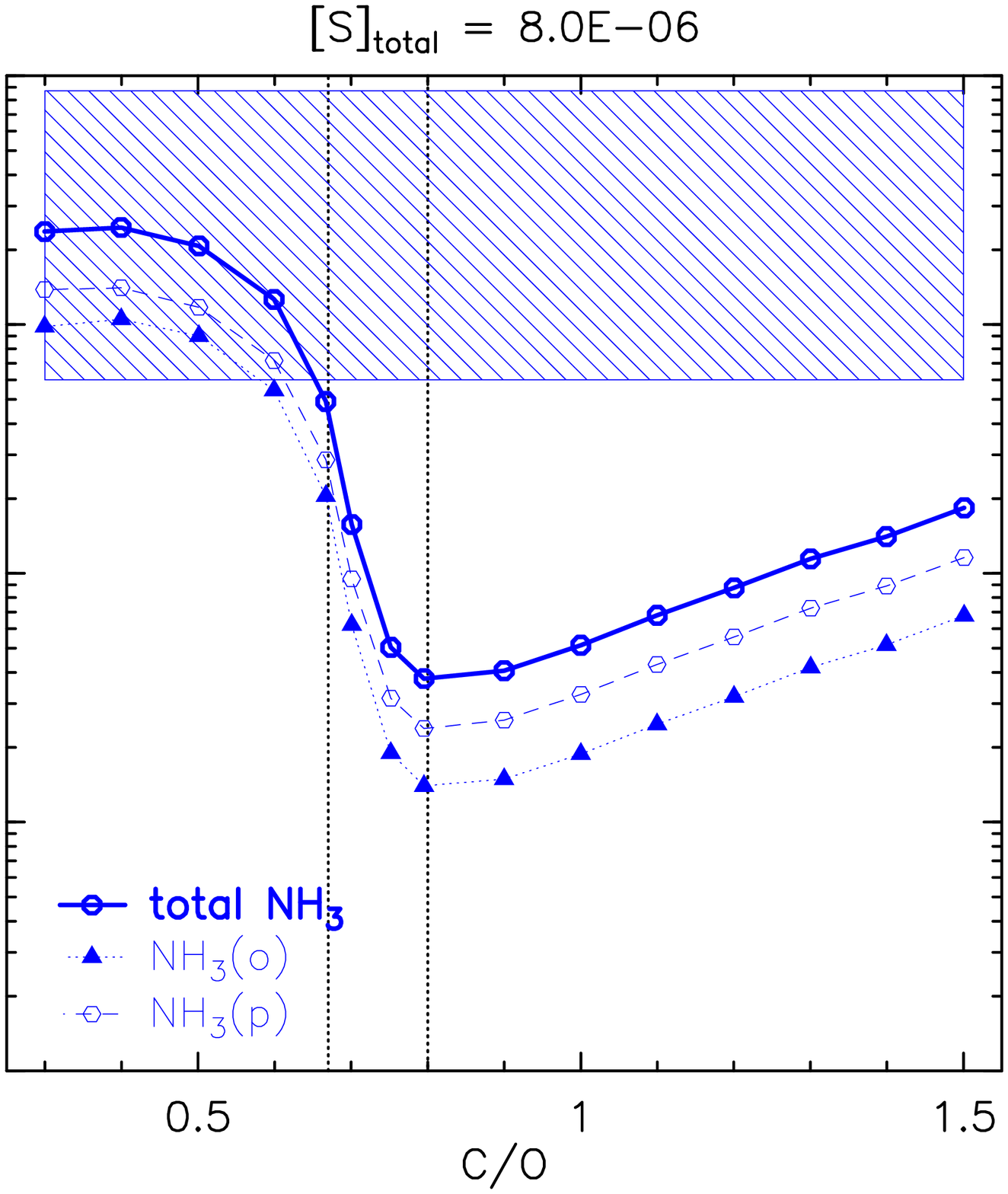}
  \caption{Steady-state abundances of NH, \ce{NH2}, and \ce{NH3} as a
    function of the C/O gas-phase elemental abundance ratio for three
    different values of gas-phase elemental abundance of sulphur (from
    left to right: $\abS = 8.0\tdix{-8}$, 8.0\tdix{-7},
    8.0\tdix{-6}). C/O values include 0.3, 0.4, 0.5, 0.6, 0.67, 0.7,
    0.75, 0.8, 0.9, 1.0, 1.1, 1.2, 1.3, 1.4, 1.5. In each panel,
      the solid line refers to the model predictions and the vertical
      dashed lines locate C/O=0.67 (Flower et al. 2006 value) and 0.8
      (our best model value). For \ce{NH2} and \ce{NH3}, the ortho and
      para abundances are shown separately (dashed and dotted lines
      respectively). The observed abundances towards IRAS~ 16293-2422
      are represented by the hatched boxes, considering an \opr\ ratio
      of \ce{NH2} $\sim 2$, as predicted by our model.}
  \label{fig:abun_hydrides_CsurO}
\end{figure*}

\begin{table*}[t]
\tiny
  \centering
  \caption{Abundances and abundance ratios of nitrogen hydrides at
    steady-state predicted using several chemical networks$^a$.}
  \label{tab:Comparison-of-ratios}
  \renewcommand{\arraystretch}{1.2}
  \begin{tabular}{l c c c c c}
    \toprule
    &\multicolumn{2}{c}{Observations $^b$}
    &\multicolumn{3}{c}{Models}\\
    \cmidrule(lr){2-3}
    \cmidrule(lr){4-6}
    & $N$ (\cc) & $N/\Nh$
    & \cite{dislaire2012} $^c$
    & This work$^d$
    & \cite{flower2006n}$^e$\\
    \midrule
    NH             & 2.2$\pm0.8$(14) & 7.3$\pm4.5$(-9)  & 4.3(-10) & 1.2(-8) & 1.3(-9)\\
    \onhh          &5.0$\pm0.9$(13) & 1.7$\pm0.9$(-9)    & --            & 2.5(-9) & 2.0(-9)\\
    \pnhh          & -- &                    --          &    --         & 1.1(-9) & 2.0(-9)\\
    \ce{NH2}       & 7.5$\pm3.0$(13) & 2.5$\pm1.6$(-9) & 8.2(-11) & 3.6(-9) & 4.0(-9)\\
    \onhhh         & --  &               --              & --           & 5.8(-9) & 2.5(-9)\\
    \pnhhh         & -- &               --              & --            & 8.6(-9) & 7.3(-9)\\
    \ce{NH3}       & 1.4$\pm1.0$(15) & 4.7$\pm4.1$(-8) & 7.5(-10) & 1.4(-8) & 9.8(-9)\\
    \midrule
    \ce{NH2}:NH    & & 0.34$\pm0.18$      & 0.2      & 0.3     & 3.1    \\
    \ce{NH3}:NH    & & 6.4$\pm5.2$        & 1.7      & 1.2     & 7.5\\
    NH:\ce{NH2}:\ce{NH3} & & 3:1:19& 5:1:9 &3:1:3 &0.3:1:2.5\\
    \bottomrule
  \end{tabular}
  \begin{list}{}{}
    \scriptsize
  \item $^a$ {Constant physical conditions were adopted: \tkin=10~K,
      \nh=\dix{4}\ccc, $\zeta=1.3\tdix{-17}\pers$. Numbers in
      parentheses are powers of ten.}
  \item $^b$ {Based on \cite{hilyblant2010nh} and revised as described
      in the Appendix~\ref{chem_updates}. For each species, $N$ is the
      total column density. The adopted value of $\Nh =
      3.0\pm$1.5\tdix{22}\cc.}
  \item $^c$ {Updated version of the osu.09.2008 network used in
      \cite{dislaire2012}, who adopted $\rm{C/O}=0.67$ and
      $\abS=1.5\tdix{-5}$.}
  \item $^d$ {Our network with the parameters of our best model:
      C/O=0.8 (or [O]$_{\rm tot}$=1.04\tdix{-4}) and \abS=8\tdix{-8}.}
  \item $^e$ {Network from \cite{flower2006n} adopting the parameters
      of our best model.}
  \end{list}
\end{table*}

Before the launch of the Herschel space telescope, the only
astronomical source where the three hydrides NH, \ce{NH2} and \ce{NH3}
were detected was Sgr~B2, with the ratios NH:\ce{NH2}:\ce{NH3}
$\sim$1:10:100 \citep{goicoechea2004}. However, this line of
  sight encompasses a variety of physical conditions (e.g. shocks),
and the NH:\ce{NH2}:\ce{NH3} ratios measured in Sgr~B2 may therefore not
be representative of cold dense clouds. Recent Herschel/HIFI
observations towards the solar-type protostar IRAS~16293-2422,
however, allowed to put strong constraints on the nitrogen hydride
chemistry in dark gas \citep{hilyblant2010nh, bacmann2010}.

IRAS~16293-2422 is situated deep within a high column density core in
a filament of the $\rho$Oph cloud complex. From the results of
extensive interferometric and single dish-mapping at centimetre,
millimetre and far-infrared wavelengths, detailed models of the
physical structure of the source have been constructed \cite[e.g.][and
references therein]{crimier2010}. The region can be divided into four
components: the ambient molecular cloud, the circumbinary envelope,
the three protostellar sources A1, A2 and B, and the outflow
components \cite[see][and references therein]{loinard2013}. The total
column density of \hh\ towards IRAS~16293-2422 is $\sim2\tdix{23}\cc$
\citep{vandishoeck1995}. In the most external part of the source where
nitrogen hydrides are seen in absorption, the column density is
necessarily much lower. Radiative transfer computations where we have
coupled non local thermodynamic equilibrium (non-LTE) calculations
with a simple physical model of the source (see
Appendix~\ref{appendice}) have shown that the absorbing region must
have a H$_2$ density $\lesssim \dix{4}\ccc$
(i.e. $\nh<2\tdix{4}\ccc$), which does not correspond to the
circumbinary envelope but to the ambient cloud (although the
distinction is artificial since the density profile is
continuous). For this component, the H$_2$ column density is rather
uncertain but we adopted the value derived from C$^{18}$O observations
by \cite{vandishoeck1995}, $N(\hh) = 1.5\tdix{22}$\cc. We note that
\cite{hilyblant2010nh} employed a different value, $N(\hh) =
5.5\pm2.5\tdix{22}\cc$, based on an extrapolation of the Crimier et
al. profile. The \ce{NH3} column density derived by
\cite{hilyblant2010nh} was highly uncertain because, in contrast to NH
and \ce{NH2}, the hyperfine structure was not resolved. Guessed
excitation temperatures of 8-10~K were therefore employed whithin a
LTE treatment. In the present work, non-LTE radiative transfer
calculations for NH and \ce{NH3} were performed, as described in the
Appendix \ref{appendice}. The best agreement with the observations
(see Fig.~\ref{fig:ratran-nh3}) is obtained for a total ammonia column
density of 1.4\tdix{15}\cc. The associated gas temperature is 11~K,
fully consistent with standard dark cloud conditions. The spectrum of
NH is also well reproduced (see Fig.~\ref{fig:ratran-nh}). The column
density of \ce{NH2} has also been revised, since it had been estimated
by \cite{hilyblant2010nh} based on the ortho-\ce{NH2} only and
assuming that the para-\ce{NH2} abundance was negligible. However, our
models, as well as \cite{faure2013}, show that this should not be the
case, and that the \opr\ ratio of \ce{NH2} is more likely $\sim
2.3$. We have therefore increased the total column density of \ce{NH2}
from 5.0\tdix{13} to 7.5\tdix{13}\cc. Finally, the column densities of
NH, \ce{NH2}, and \ce{NH3} used in this work are $2.2\pm
0.8\tdix{14}$, $7.5\pm3.0\tdix{13}$, and $1.4\pm1.0\tdix{15}\cc$,
respectively, or abundance ratios
NH:\ce{NH2}:$\ce{NH3}\sim3$:1:19. The corresponding abundances
relative to the total H nuclei are then obtained using $\Nh =
2\times N(\hh) = 3.0\pm1.5\tdix{22}\cc$.

\begin{figure*}[t]
  \centering
  \includegraphics[height=0.35\hsize]{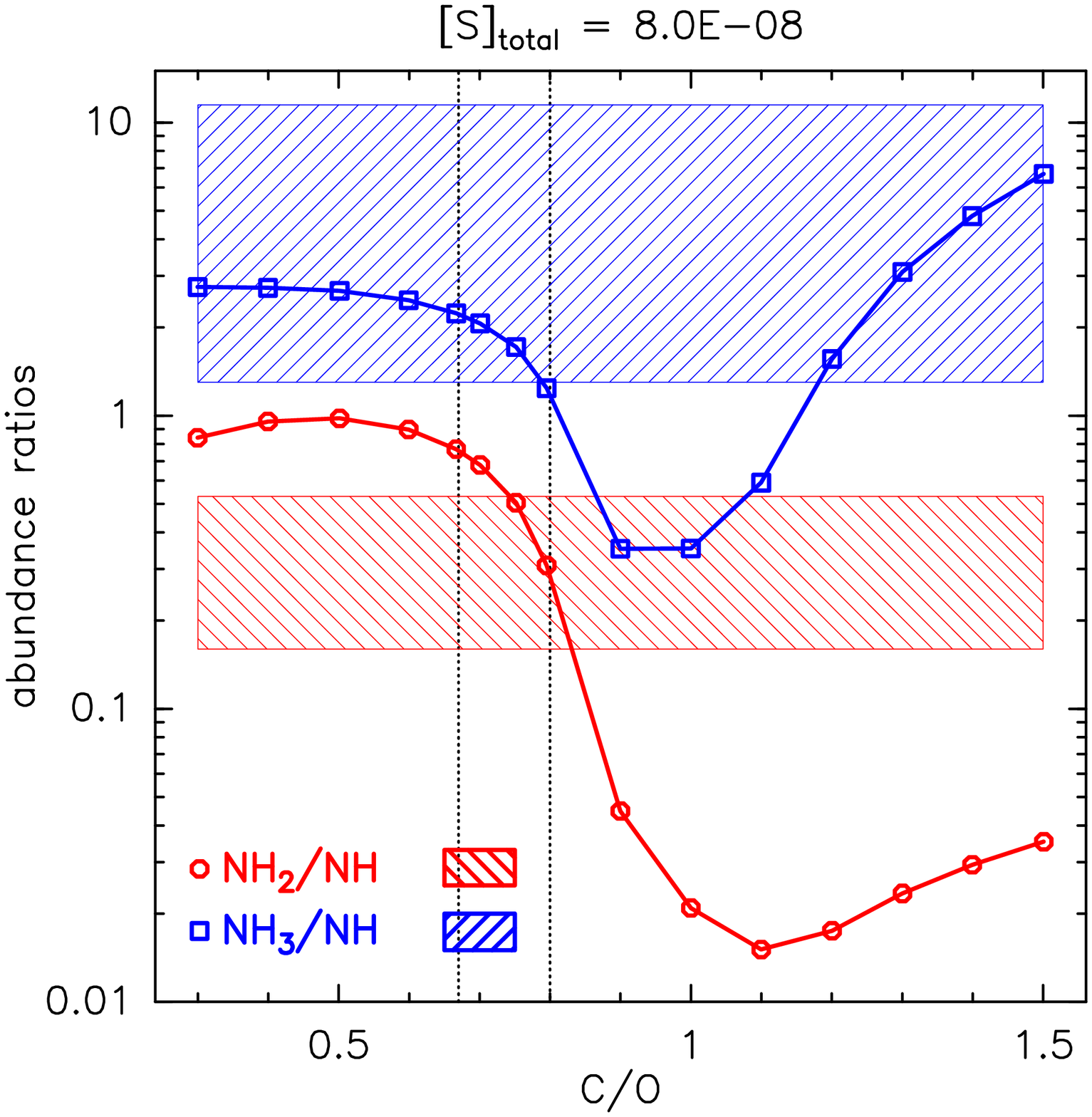}\hfill%
  \includegraphics[height=0.35\hsize]{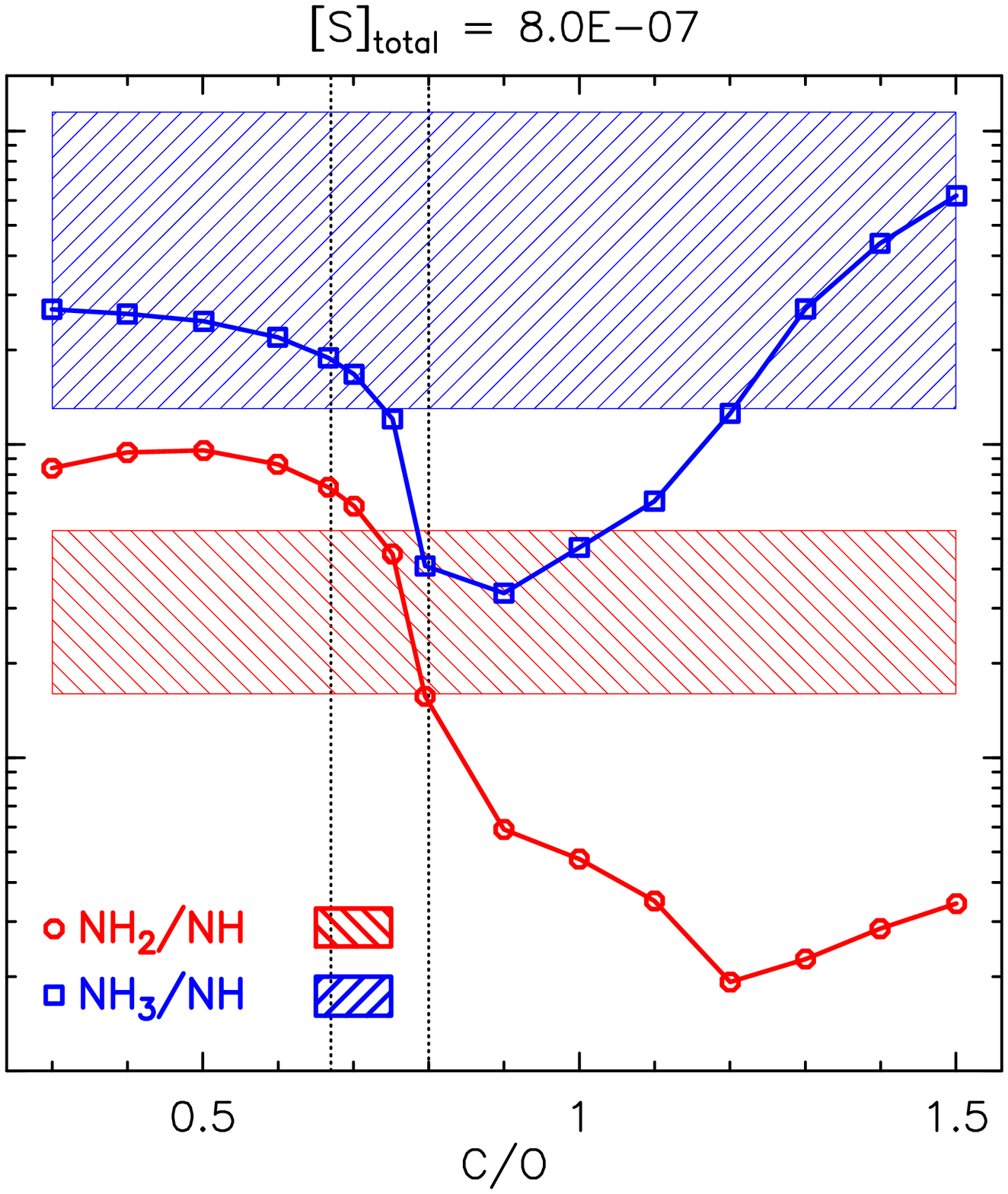}\hfill%
  \includegraphics[height=0.35\hsize]{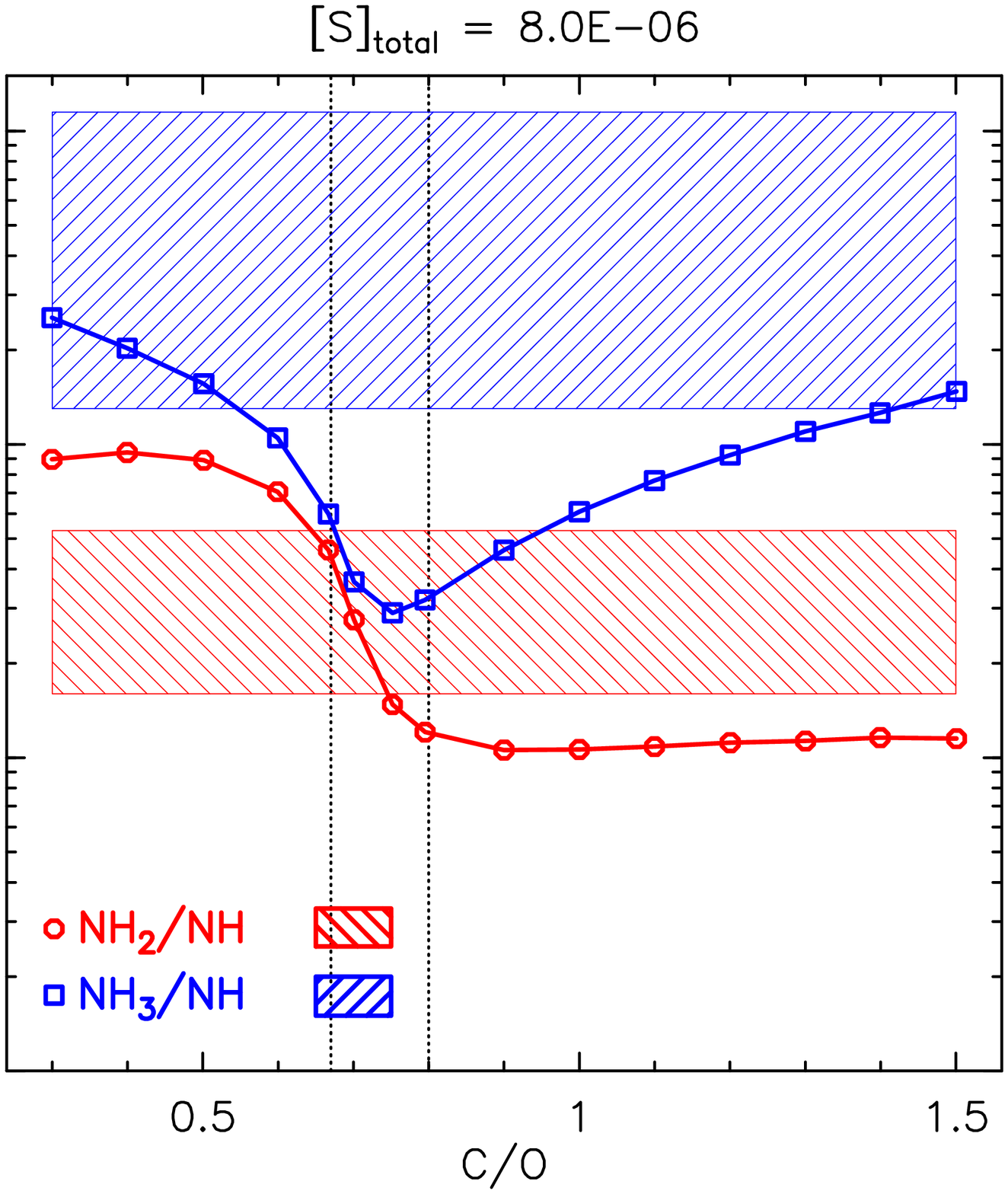}
  \caption{Steady-state abundance ratios of the N-hydrides as a
    function of the gas-phase elemental abundance ratio of C/O for
    three different values of gas-phase elemental abundance of sulphur
    (from left to right: $\abS = 8.0\tdix{-8}$, 8.0\tdix{-7},
    8.0\tdix{-6}). C/O values include 0.3, 0.4, 0.5, 0.6, 0.67, 0.7,
    0.75, 0.8, 0.9, 1.0, 1.1, 1.2, 1.3, 1.4, 1.5. In all panels, the
    vertical dashed lines locate C/O=0.67 (Flower et al. 2006 value)
    and 0.8 (our best model value). The abundances derived from
    Herschel/HIFI observations towards IRAS~ 16293-2422 are represented
    by the hatched boxes, considering an \opr\ ratio of \ce{NH2} $\sim
    2$, as predicted by our model.}
  \label{fig:ratios_hydrides_CsurO}
\end{figure*}

The abundances of NH, \ce{NH2}, and \ce{NH3} towards IRAS~16293-2422
are compared in Fig.~\ref{fig:abun_hydrides_CsurO} to the set of
models described previously. Overall, the predicted abundances
  match the observations only within a small range of C/O ratios,
  which is found to be $0.7-0.8$. However, as discussed above, the
uncertainties on the abundances are rather large due to the difficulty
to estimate the H$_2$ column density in the absorbing layers. In
contrast, abundances (or column density) ratios circumvent this
caveat, and are thus more robust. The ratios \ce{NH2}:NH and
\ce{NH3}:NH are plotted in Fig.\ref{fig:ratios_hydrides_CsurO}. At low
and intermediate sulphur abundance, the \ce{NH2}:NH ratio delineates a
narrow range of C/O values of $0.75-0.80$ consistent with the above
constraints. At intermediate sulphur abundance the \ce{NH3}:NH ratio
further constrains the C/O ratio to 0.75. For a high sulphur
abundance, however, there is no C/O which allows to reproduce
simultaneously both abundance ratios. From Fig.~\ref{fig:abtime},
which shows the evolution with time of the abundances, we note that
the observed abundance ratios NH:\ce{NH2}:\ce{NH3} are also well
reproduced at very early times $\approx 4\tdix{4}$~yr. However, at
such early times, the fractional abundances of the N-hydrides are more
than an order of magnitude smaller than observed. Although there
  was no minimization attempt, the best agreement with the
  observations is found for the set of parameters $\abS =
  8.0\tdix{-8}$ and $\rm{C/O}=0.8$, which will be referred to as our
  best model.

We conclude that the steady-state abundances predicted by our
gas-phase chemical model are consistent with the observational
constraints on the abundances and abundance ratios of NH, \ce{NH2},
and \ce{NH3}, provided that the C/O ratio is $\sim 0.8$. The
  elemental abundance of sulphur is less constrained, but must be
  depleted by more than a factor of 2.

\subsection{Impact of the new rates}

To evidence the influence of the new rates adopted in our network, we
have performed a calculation using the Flower network combined with
our best model parameters. The abundances of \ce{NH2} and \ce{NH3} are
similar to those found with our updated network, with differences less
than 30\%. However, this is not true for NH abundance, which is one
order of magnitude lower with the old network. This large increase in
our model is due to the dissociative recombination channel of
\ce{N2H+} forming NH, as already noted in \cite{dislaire2012}. When
looking into more details, the relatively small change of the
abundances of \ce{NH2} and \ce{NH3} in fact results from a combination
of three effects. First, the updated rate of the reaction \ce{N+ + H2}
lowers by more than a factor of 10 the abundances of \ce{NH2} and
\ce{NH3}. For ammonia, this effect is partially compensated by the
updated branching ratios of the \ce{NH_n+ + e-} dissociative
recombinations. The abundance of \ce{NH2} and that of ammonia are
finally recovered when updating the rate and branching ratios of the
\ce{H3+ + O}. This reaction produces more \ce{OH+} and \ce{H2O+}, thus
enhancing the amount of OH and therefore the N to \ce{N2}
conversion. This demonstrates that updating all those reaction rates
is essential to reproduce the abundances of nitrogen hydrides. We note
that the \cite{dislaire2012} model, based on the \texttt{osu.09.2008}
chemical network
\footnote{www.physics.ohio-state.edu/~eric/research.html},
predicts too small abundances of all three species, which may
  result from the large sulphur abundance assumed by these authors,
although the comparison is not direct because the two chemical
  networks may contain different rate coefficients for non-negligible
  reactions.

The impact of the new rates is further evidenced by the \opr\ of
\ce{NH2} and \ce{NH3} predicted by our models which are 2.3 and 0.7,
respectively (see Table~\ref{tab:Comparison-of-ratios}). These values
are significantly larger than those obtained with the
\cite{flower2006n} network (1 and 0.3 respectively, see
Table~\ref{tab:Comparison-of-ratios}). It should be noted that both
\opr\ ratios are below the statistical values of 3 and 1,
respectively, which are lower limits under thermal equilibrium. In
addition, these \opr\ ratios were found to depend only very weakly on
the C/O value, as expected since the \opr\ of nitrogen hydrides are
driven by the \opr\ ratio of \hh\ (see section \ref{subsec:N-hyd} and
Fig.~\ref{fig:abun_hydrides_CsurO}).

\section{Discussion and conclusions}

We have presented a new gas-phase, ortho/para, chemical network
devoted to the interstellar chemistry of nitrogen under typical
  dark cloud physical conditions. More specifically, this work
  focusses on nitrogen hydrides because these species, which are
  linked by a small number of chemical reactions, have been observed
  with high accuracy with the Herschel/HIFI instrument. The
  absorption lines of NH, \ce{NH2}, and \ce{NH3} trace the cold,
  UV-shielded, and moderately dense, envelope of the IRAS~16293-2422
  protostar, and therefore provide direct observational tests of the
  chemistry of nitrogen in such environments. Although the comparison
  deals with a single object, the present work aims at understanding
  general processes of nitrogen chemistry in dark cloud conditions.

Our network is based on the gas-phase network of
\cite{flower2006n} where the following major improvements have been
implemented: {\it i)} the nitrogen chemistry was revised using the
most recent experimental results, in particular for the conversion of
N to N$_2$ through radical-radical reactions, and for the dissociative
recombinations of the NH$_n^+$($n$=2-4) ions; {\it ii)} we adopted
recent theoretical results for the ortho-to-para conversion of H$_2$
by H$^+$ and H$_3^+$; {\it iii)} we employed the formalism of
\cite{oka2004}, as detailed in \cite{rist2013} and \cite{faure2013},
to derive the nuclear spin branching ratios for all primary reactions
involved in the formation and destruction of \phh, \ohh, NH, \pnhh,
\onhh, \pnhhh\ and \onhhh.

Our findings may be summarized as follows. First, we found that the
abundances of the nitrogen hydrides strongly depend on the gas-phase
elemental C/O ratio adopted in the model. A similar result was noted
in earlier works \citep[e.g. ][]{terzieva1998, tassis2012}, although
we here provide a detailed analysis of the influence of the C/O ratio,
by focussing on a smaller set of species. To summarize, the influence
of the C/O ratio stems from the fact that nitrogen hydrides derive
from N$_2$ which is formed either {\it via} \ce{N + CH} or \ce{N +
  OH}. Two regimes corresponding to $\rm{C/O}<0.8$ and $\rm{C/O}>1$
are clearly identified where the N$_2$/N ratio is larger than 1 and
the abundance of ammonia is a few 10$^{-8}$. In contrast, the
abundances of NH and \ce{NH2} are much lower at high C/O because these
radicals are efficiently destroyed by atomic carbon. Second, we have
shown that increasing the elemental gas-phase abundance of sulphur
\abS\ significantly reduces the efficiency of the N to N$_2$
conversion. Thus a high abundance of sulphur (i.e. 8.0\tdix{-6}) was
found to reduce the abundance of several N-bearing species by up to
two orders of magnitude.

The steady-state abundances predicted by our model were compared to
the Herschel/HIFI observations of NH, \ce{NH2}, and \ce{NH3}, towards
the low-mass protostar IRAS~16293-2422. Our chemical model reproduces
well both the observed abundances and abundance ratios of the three
hydrides. A key point is that our network produces more NH than
\ce{NH2}, as observed. Our best model, which leads to
NH:\ce{NH2}:\ce{NH3} abundance ratios of 3:1:3 (see
Table~\ref{tab:Comparison-of-ratios}), is obtained for a C/O ratio of
0.8 and a low sulphur abundance $\abtot{S}$=8.0\tdix{-8}. Although it
is consistent with the 3:1:19 observed ratios, the predicted abundance
of ammonia is at the lower end of the allowed range. Higher values for
$\abtot{S}$ are also possible but a high abundance of 8.0\tdix{-6} is
clearly excluded by our models, as well as C/O ratios outside the
range $0.7-0.8$. We note that at earlier times ($\sim2$-5~Myr), the
observations can be reproduced with a similar \abS\
(i.e. $<8.0\tdix{-6}$) and slightly higher C/O ratios, in the range
0.9 to 1.1.

Our model calculations thus show that chemical reactive processes on
dust grains are not needed to explain the gas-phase abundances of
nitrogen hydrides in typical dark cloud conditions. This of course
does not preclude the possibility that a fraction of atomic nitrogen
can also form ammonia ices by hydrogenation on the grain mantles. This
process is in fact certainly necessary to account for the high
abundance of ammonia ($\sim \dix{-6}$) detected in e.g. shock regions,
where grain mantles are released in the gas-phase
\citep{umemoto1999}. We note that our best model corresponds to a
regime where N/N$_2\approx 1$. The exact amount of gaseous N$_2$
however depends on the competition between its formation in the gas
and the depletion of atomic nitrogen onto grains, which was neglected
in the present work. Several nitrogenated species were shown
observationally to resist depletion in cold prestellar cores, in
contrast to CO and many other species. This is the case for \ce{NH3}
and N$_2$H$^+$, CN \citep{tafalla2004, crapsi2007, hilyblant2008}, and
also for HCN (and HNC), although for the latter there seems to exist
some variability amongst sources \citep{hilyblant2010n, padovani2011,
  pagani2012}. This observational result indicates that enough atomic
N is always available in the gas phase which is also slowly converted
into N$_2$. However, in similar objects, \cite{akyilmaz2007} observed
that NO disappears from the gas-phase at high densities where CN does
not, which was interpreted as freeze-out onto grains. The different
behaviours among N-bearing species with respect to depletion could
result from a combination of several factors, such as different
molecular properties (e.g. the binding energy), and/or the timescales
of the gas-phase chemical reactions. Further attempts to reproduce, in
a comprehensive fashion, the behaviours of the observed species, may
need to consider all these aspects taking into account the
time-dependent competition between gas-phase and gas-grain processes.

Gas-phase synthesis of N-hydrides is further supported by the
  ortho-to-para ratios of \ce{NH3} and \ce{NH2} predicted by our
  model. \cite{persson2012} measured the ortho- and para-ammonia
column densities in diffuse molecular clouds along the sight-lines
towards the high-mass star-forming regions W49N and G10.6-0.4, and
found $\opr(\ce{NH3}) \approx$ 0.5 -- 0.7. This ratio is in excellent
agreement with our prediction. For \ce{NH2}, preliminary Herschel
measurements indicate ortho-to-para values in the range 1.5 -- 3.5 in
diffuse gas along the line of sight towards G10.6-0.4 (Persson et al.,
private comm.), again in very good agreement with our prediction. The
NH, \ce{NH2}, and \ce{NH3} abundances (relative to total hydrogen)
derived by \cite{persson2010} are 5.6\tdix{-9}, 3.0\tdix{-9}, and
3.2\tdix{-9}, respectively, or abundance ratios
NH:\ce{NH2}:\ce{NH3}=2:1:1. The NH and \ce{NH2} abundances are thus
similar to those in the envelope surrounding IRAS~16293-2422 (see
Table~\ref{tab:Comparison-of-ratios}), although that of ammonia is one
order of magnitude lower. Their abundance ratios are in fact
  similar to the prediction of our best model (see
  Table~\ref{tab:Comparison-of-ratios}). However, our calculations
unlikely apply to a warmer and more tenuous diffuse gas which requires
dedicated models (gas density of $\sim\dix{3}\ccc$ and
  temperature of $\sim$30~K). In particular, the C/O ratio and \abS\
abundance may take different values in the diffuse
ISM. Photodissociation of ammonia may also play a role, as
discussed by \cite{persson2010}, although we do not expect the
UV to alter significantly the \opr\ of \ce{NH2} and \ce{NH3} which are
driven by proton exchange reactions.

The present work gives rise to several questions. First, additional
observations of NH, \ce{NH2}, and \ce{NH3} in various objects are
needed to confirm the gas-phase chemical processes mentioned
above. The determination of the ortho-to-para ratio towards others
Class 0 protostars would also be extremely useful. Another critical
issue is related to the abundances of other N-bearing species such as
N$_2$H$^+$, CN, and nitriles (e.g. HCN) in dark clouds. Observations
in dark clouds are already available for these species, but were not
discussed in this paper. The reason is that we believe that such a
comprehensive comparison with all the observations requires first to
assess the completeness of the current chemical network with respect
to all those simple nitrogen-bearing species. Here, we have focussed
on nitrogen hydrides, and this allowed us to understand several
effects and explain various observational facts. Of course, chemistry
is highly non-linear \citep[e.g. ][]{lebourlot1995}, and we can not
exclude that in some peculiar range of parameters, new rates, e.g. in
the formation of HCN, could affect the above results. However, by
exploring such a broad range of initial abundances, we have been able
to emphasize robust trends and processes. Our approach is really
complementary to other strategies which quantify the sensitivity of
chemical networks to rate uncertainties
\citep[e.g. ][]{wakelam2010}. Time-dependent aspects must also be
explored, taking into account gas-grain processes, and/or including
the effect of gas dynamics \citep[e.g. ][]{brown1990, flower2006n,
  tassis2012}. Concerning future improvements of our network, we are
currently implementing new reaction rates and nuclear spin branching
ratios for the deuterated isotopologues of several nitrogen-bearing
species. Deuterated as well as $^{15}$N isotopologues were not
considered in the present work. However, both are detected in the cold
ISM with significant abundances. The molecular D/H and
$^{15}$N/$^{14}$N ratios are sensitive probes of the chemistry and
future works will explore gas-phase fractionation processes and the
possible link with the D and $^{15}$N-enrichments observed in Solar
System objects \cite[see e.g.][and references therein]{wirstrom2012,
  hilyblant2013}. We note in this context the recent detection of
C$^{15}$N and \ce{NH3}D$^+$ in cold prestellar cores by
\citet{hilyblant2013b} and \cite{cernicharo2013} respectively.

\begin{acknowledgements}
  We thank our anonymous referee for a careful reading and useful
  comments that improved the manuscript.  This work has been supported
  by ``R\'egion Rh\^one-Alpes'' (CIBLE program), the Agence Nationale
  de la Recherche (ANR-HYDRIDES), contract ANR-12-BS05-0011-01, and by
  the CNRS national program ``Physico-Chimie du Milieu
  Interstellaire''.
\end{acknowledgements}

\bibliographystyle{aa}
\bibliography{paper06}

\begin{thebibliography}{116}
\expandafter\ifx\csname natexlab\endcsname\relax\def\natexlab#1{#1}\fi

\bibitem[{{Akyilmaz} {et~al.}(2007){Akyilmaz}, {Flower}, {Hily-Blant}, {Pineau
  des For{\^e}ts}, \& {Walmsley}}]{akyilmaz2007}
{Akyilmaz}, M., {Flower}, D.~R., {Hily-Blant}, P., {Pineau des For{\^e}ts}, G.,
  \& {Walmsley}, C.~M. 2007, \aap, 462, 221

\bibitem[{{Anicich} \& {Huntress}(1986)}]{anicich1986}
{Anicich}, V.~G. \& {Huntress}, Jr., W.~T. 1986, \apjs, 62, 553

\bibitem[{{Asplund} {et~al.}(2009){Asplund}, {Grevesse}, {Sauval}, \&
  {Scott}}]{asplund2009}
{Asplund}, M., {Grevesse}, N., {Sauval}, A.~J., \& {Scott}, P. 2009, Annu. Rev.
  Astron. Astrophys., 47, 481

\bibitem[{{Bacmann} {et~al.}(2010){Bacmann}, {Caux}, {Hily-Blant}, {Parise},
  {Pagani}, {Bottinelli}, {Maret}, {Vastel}, {Ceccarelli}, {Cernicharo},
  {Henning}, {Castets}, {Coutens}, {Bergin}, {Blake}, {Crimier}, {Demyk},
  {Dominik}, {Gerin}, {Hennebelle}, {Kahane}, {Klotz}, {Melnick}, {Schilke},
  {Wakelam}, {Walters}, {Baudry}, {Bell}, {Benedettini}, {Boogert}, {Cabrit},
  {Caselli}, {Codella}, {Comito}, {Encrenaz}, {Falgarone}, {Fuente},
  {Goldsmith}, {Helmich}, {Herbst}, {Jacq}, {Kama}, {Langer}, {Lefloch}, {Lis},
  {Lord}, {Lorenzani}, {Neufeld}, {Nisini}, {Pacheco}, {Pearson}, {Phillips},
  {Salez}, {Saraceno}, {Schuster}, {Tielens}, {van der Tak}, {van der Wiel},
  {Viti}, {Wyrowski}, {Yorke}, {Faure}, {Benz}, {Coeur-Joly}, {Cros},
  {G{\"u}sten}, \& {Ravera}}]{bacmann2010}
{Bacmann}, A., {Caux}, E., {Hily-Blant}, P., {et~al.} 2010, \aap, 521, L42

\bibitem[{{Bergeat}(1999)}]{bergeat1999}
{Bergeat}, A. 1999, Chemical Physics Letters, 308, 7

\bibitem[{Bergeat {et~al.}(2009)Bergeat, Hickson, Daugey, Caubet, \&
  Costes}]{bergeat2009}
Bergeat, A., Hickson, K.~M., Daugey, N., Caubet, P., \& Costes, M. 2009,
  Physical Chemistry Chemical Physics, 11, 8149

\bibitem[{{Bergin} \& {Langer}(1997)}]{bergin1997}
{Bergin}, E.~A. \& {Langer}, W.~D. 1997, \apj, 486, 316

\bibitem[{{Bergin} \& {Tafalla}(2007)}]{bergin2007}
{Bergin}, E.~A. \& {Tafalla}, M. 2007, \araa, 45, 339

\bibitem[{{Boogert} {et~al.}(2011){Boogert}, {Huard}, {Cook}, {Chiar}, {Knez},
  {Decin}, {Blake}, {Tielens}, \& {van Dishoeck}}]{boogert2011}
{Boogert}, A.~C.~A., {Huard}, T.~L., {Cook}, A.~M., {et~al.} 2011, \apj, 729,
  92

\bibitem[{{Bottinelli} {et~al.}(2010){Bottinelli}, {Boogert}, {Bouwman},
  {Beckwith}, {van Dishoeck}, {{\"O}berg}, {Pontoppidan}, {Linnartz}, {Blake},
  {Evans}, \& {Lahuis}}]{bottinelli2010}
{Bottinelli}, S., {Boogert}, A.~C.~A., {Bouwman}, J., {et~al.} 2010, \apj, 718,
  1100

\bibitem[{{Brinch} \& {Hogerheijde}(2010)}]{brinch2010}
{Brinch}, C. \& {Hogerheijde}, M.~R. 2010, \aap, 523, A25

\bibitem[{{Brown} \& {Charnley}(1990)}]{brown1990}
{Brown}, P.~D. \& {Charnley}, S.~B. 1990, \mnras, 244, 432

\bibitem[{{Cartledge} {et~al.}(2004){Cartledge}, {Lauroesch}, {Meyer}, \&
  {Sofia}}]{cartledge2004}
{Cartledge}, S.~I.~B., {Lauroesch}, J.~T., {Meyer}, D.~M., \& {Sofia}, U.~J.
  2004, \apj, 613, 1037

\bibitem[{{Caselli} {et~al.}(1998){Caselli}, {Walmsley}, {Terzieva}, \&
  {Herbst}}]{caselli1998}
{Caselli}, P., {Walmsley}, C.~M., {Terzieva}, R., \& {Herbst}, E. 1998, \apj,
  499, 234

\bibitem[{{Ceccarelli} {et~al.}(2010){Ceccarelli}, {Bacmann}, {Boogert},
  {Caux}, {Dominik}, {Lefloch}, {Lis}, {Schilke}, {van der Tak}, {Caselli},
  {Cernicharo}, {Codella}, {Comito}, {Fuente}, {Baudry}, {Bell}, {Benedettini},
  {Bergin}, {Blake}, {Bottinelli}, {Cabrit}, {Castets}, {Coutens}, {Crimier},
  {Demyk}, {Encrenaz}, {Falgarone}, {Gerin}, {Goldsmith}, {Helmich},
  {Hennebelle}, {Henning}, {Herbst}, {Hily-Blant}, {Jacq}, {Kahane}, {Kama},
  {Klotz}, {Langer}, {Lord}, {Lorenzani}, {Maret}, {Melnick}, {Neufeld},
  {Nisini}, {Pacheco}, {Pagani}, {Parise}, {Pearson}, {Phillips}, {Salez},
  {Saraceno}, {Schuster}, {Tielens}, {van der Wiel}, {Vastel}, {Viti},
  {Wakelam}, {Walters}, {Wyrowski}, {Yorke}, {Liseau}, {Olberg}, {Szczerba},
  {Benz}, \& {Melchior}}]{ceccarelli2010}
{Ceccarelli}, C., {Bacmann}, A., {Boogert}, A., {et~al.} 2010, \aap, 521, L22

\bibitem[{{Cernicharo} {et~al.}(2013){Cernicharo}, {Tercero}, {Fuente},
  {Domenech}, {Cueto}, {Carrasco}, {Herrero}, {Tanarro}, {Marcelino}, {Roueff},
  {Gerin}, \& {Pearson}}]{cernicharo2013}
{Cernicharo}, J., {Tercero}, B., {Fuente}, A., {et~al.} 2013, \apjl, 771, L10

\bibitem[{{Charnley} \& {Rodgers}(2002)}]{charnley2002}
{Charnley}, S.~B. \& {Rodgers}, S.~D. 2002, \apjl, 569, L133

\bibitem[{{Chastaing} {et~al.}(2000){Chastaing}, {Le Picard}, \&
  {Sims}}]{chastaing2000}
{Chastaing}, D., {Le Picard}, S.~D., \& {Sims}, I.~R. 2000, \jcp, 112, 8466

\bibitem[{{Cheung} {et~al.}(1968){Cheung}, {Rank}, {Townes}, {Thornton}, \&
  {Welch}}]{cheung1968}
{Cheung}, A.~C., {Rank}, D.~M., {Townes}, C.~H., {Thornton}, D.~D., \& {Welch},
  W.~J. 1968, Physical Review Letters, 21, 1701

\bibitem[{{Crabtree} {et~al.}(2011){Crabtree}, {Indriolo}, {Kreckel}, {Tom}, \&
  {McCall}}]{crabtree2011}
{Crabtree}, K.~N., {Indriolo}, N., {Kreckel}, H., {Tom}, B.~A., \& {McCall},
  B.~J. 2011, \apj, 729, 15

\bibitem[{{Crapsi} {et~al.}(2007){Crapsi}, {Caselli}, {Walmsley}, \&
  {Tafalla}}]{crapsi2007}
{Crapsi}, A., {Caselli}, P., {Walmsley}, M.~C., \& {Tafalla}, M. 2007, \aap,
  470, 221

\bibitem[{{Crimier} {et~al.}(2010){Crimier}, {Ceccarelli}, {Maret},
  {Bottinelli}, {Caux}, {Kahane}, {Lis}, \& {Olofsson}}]{crimier2010}
{Crimier}, N., {Ceccarelli}, C., {Maret}, S., {et~al.} 2010, ArXiv e-prints

\bibitem[{{Crutcher}(2012)}]{crutcher2012}
{Crutcher}, R.~M. 2012, \araa, 50, 29

\bibitem[{{Dalgarno} {et~al.}(1973){Dalgarno}, {Black}, \&
  {Weisheit}}]{dalgarno1973}
{Dalgarno}, A., {Black}, J.~H., \& {Weisheit}, J.~C. 1973, \aplett, 14, 77

\bibitem[{{Daranlot} {et~al.}(2012){Daranlot}, {Hincelin}, {Bergeat}, {Costes},
  {Loison}, {Wakelam}, \& {Hickson}}]{daranlot2012}
{Daranlot}, J., {Hincelin}, U., {Bergeat}, A., {et~al.} 2012, Proceedings of
  the National Academy of Science, 109, 10233

\bibitem[{{Daranlot} {et~al.}(2011){Daranlot}, {Jorfi}, {Xie}, {Bergeat},
  {Costes}, {Caubet}, {Xie}, {Guo}, {Honvault}, \& {Hickson}}]{daranlot2011}
{Daranlot}, J., {Jorfi}, M., {Xie}, C., {et~al.} 2011, Science, 334, 1538

\bibitem[{{D'Hendecourt} {et~al.}(1985){D'Hendecourt}, {Allamandola}, \&
  {Greenberg}}]{dhendecourt1985}
{D'Hendecourt}, L.~B., {Allamandola}, L.~J., \& {Greenberg}, J.~M. 1985, \aap,
  152, 130

\bibitem[{{Dislaire} {et~al.}(2012){Dislaire}, {Hily-Blant}, {Faure}, {Maret},
  {Bacmann}, \& {Pineau des For{\^e}ts}}]{dislaire2012}
{Dislaire}, V., {Hily-Blant}, P., {Faure}, A., {et~al.} 2012, \aap, 537, A20

\bibitem[{{Dos Santos} {et~al.}(2007){Dos Santos}, {Kokoouline}, \&
  {Greene}}]{dossantos2007}
{Dos Santos}, S.~F., {Kokoouline}, V., \& {Greene}, C.~H. 2007, \jcp, 127,
  124309

\bibitem[{{Dumouchel} {et~al.}(2012){Dumouchel}, {K{\l}os}, {Tobo{\l}a},
  {Bacmann}, {Maret}, {Hily-Blant}, {Faure}, \& {Lique}}]{dumouchel2012}
{Dumouchel}, F., {K{\l}os}, J., {Tobo{\l}a}, R., {et~al.} 2012, \jcp, 137,
  114306

\bibitem[{{Faure} {et~al.}(2013){Faure}, {Hily-Blant}, {Le Gal}, {Rist}, \&
  {Pineau des For{\^e}ts}}]{faure2013}
{Faure}, A., {Hily-Blant}, P., {Le Gal}, R., {Rist}, C., \& {Pineau des
  For{\^e}ts}, G. 2013, \apjl, 770, L2

\bibitem[{{Flower} {et~al.}(2005){Flower}, {Pineau des For{\^e}ts}, \&
  {Walmsley}}]{flower2005}
{Flower}, D.~R., {Pineau des For{\^e}ts}, G., \& {Walmsley}, C.~M. 2005, \aap,
  436, 933

\bibitem[{{Flower} {et~al.}(2006{\natexlab{a}}){Flower}, {Pineau des
  For{\^e}ts}, \& {Walmsley}}]{flower2006n}
{Flower}, D.~R., {Pineau des For{\^e}ts}, G., \& {Walmsley}, C.~M.
  2006{\natexlab{a}}, \aap, 456, 215

\bibitem[{{Flower} {et~al.}(2006{\natexlab{b}}){Flower}, {Pineau des
  For{\^e}ts}, \& {Walmsley}}]{flower2006op}
{Flower}, D.~R., {Pineau des For{\^e}ts}, G., \& {Walmsley}, C.~M.
  2006{\natexlab{b}}, \aap, 449, 621

\bibitem[{{Gerin} {et~al.}(1992){Gerin}, {Viala}, {Pauzat}, \&
  {Ellinger}}]{gerin1992}
{Gerin}, M., {Viala}, Y., {Pauzat}, F., \& {Ellinger}, Y. 1992, \aap, 266, 463

\bibitem[{{Gerlich}(1993)}]{gerlich1993}
{Gerlich}, D. 1993, J. Chem. Soc. Faraday Trans., 89(13), 2199

\bibitem[{{Gibb} {et~al.}(2000){Gibb}, {Whittet}, {Schutte}, {Boogert},
  {Chiar}, {Ehrenfreund}, {Gerakines}, {Keane}, {Tielens}, {van Dishoeck}, \&
  {Kerkhof}}]{gibb2000}
{Gibb}, E.~L., {Whittet}, D.~C.~B., {Schutte}, W.~A., {et~al.} 2000, \apj, 536,
  347

\bibitem[{{Goicoechea} {et~al.}(2004){Goicoechea},
  {Rodr{\'{\i}}guez-Fern{\'a}ndez}, \& {Cernicharo}}]{goicoechea2004}
{Goicoechea}, J.~R., {Rodr{\'{\i}}guez-Fern{\'a}ndez}, N.~J., \& {Cernicharo},
  J. 2004, \apj, 600, 214

\bibitem[{{Graedel} {et~al.}(1982){Graedel}, {Langer}, \&
  {Frerking}}]{graedel1982}
{Graedel}, T.~E., {Langer}, W.~D., \& {Frerking}, M.~A. 1982, \apjs, 48, 321

\bibitem[{{Grussie} {et~al.}(2012){Grussie}, {Berg}, {Crabtree}, {G{\"a}rtner},
  {McCall}, {Schlemmer}, {Wolf}, \& {Kreckel}}]{grussie2012}
{Grussie}, F., {Berg}, M.~H., {Crabtree}, K.~N., {et~al.} 2012, \apj, 759, 21

\bibitem[{{Hasegawa} \& {Herbst}(1993)}]{hasegawa1993}
{Hasegawa}, T.~I. \& {Herbst}, E. 1993, \mnras, 261, 83

\bibitem[{{Herbst} {et~al.}(1987){Herbst}, {Defrees}, \& {McLean}}]{herbst1987}
{Herbst}, E., {Defrees}, D.~J., \& {McLean}, A.~D. 1987, \apj, 321, 898

\bibitem[{{Herbst} \& {Klemperer}(1973)}]{herbst1973}
{Herbst}, E. \& {Klemperer}, W. 1973, \apj, 185, 505

\bibitem[{Herbst {et~al.}(2000)Herbst, Terzieva, \& Talbi}]{herbst2000}
Herbst, E., Terzieva, R., \& Talbi, D. 2000, Monthly Notices of The Royal
  Astronomical Society, 311, 869

\bibitem[{{Hidaka} {et~al.}(2011){Hidaka}, {Watanabe}, {Kouchi}, \&
  {Watanabe}}]{hidaka2011}
{Hidaka}, H., {Watanabe}, M., {Kouchi}, A., \& {Watanabe}, N. 2011, Physical
  Chemistry Chemical Physics (Incorporating Faraday Transactions), 13, 15798

\bibitem[{{Hily-Blant} {et~al.}(2013{\natexlab{a}}){Hily-Blant}, {Bonal},
  {Faure}, \& {Quirico}}]{hilyblant2013}
{Hily-Blant}, P., {Bonal}, L., {Faure}, A., \& {Quirico}, E.
  2013{\natexlab{a}}, \icarus, 223, 582

\bibitem[{{Hily-Blant} {et~al.}(2010{\natexlab{a}}){Hily-Blant}, {Maret},
  {Bacmann}, {Bottinelli}, {Parise}, {Caux}, {Faure}, {Bergin}, {Blake},
  {Castets}, {Ceccarelli}, {Cernicharo}, {Coutens}, {Crimier}, {Demyk},
  {Dominik}, {Gerin}, {Hennebelle}, {Henning}, {Kahane}, {Klotz}, {Melnick},
  {Pagani}, {Schilke}, {Vastel}, {Wakelam}, {Walters}, {Baudry}, {Bell},
  {Benedettini}, {Boogert}, {Cabrit}, {Caselli}, {Codella}, {Comito},
  {Encrenaz}, {Falgarone}, {Fuente}, {Goldsmith}, {Helmich}, {Herbst}, {Jacq},
  {Kama}, {Langer}, {Lefloch}, {Lis}, {Lord}, {Lorenzani}, {Neufeld}, {Nisini},
  {Pacheco}, {Phillips}, {Salez}, {Saraceno}, {Schuster}, {Tielens}, {van der
  Tak}, {van der Wiel}, {Viti}, {Wyrowski}, \& {Yorke}}]{hilyblant2010nh}
{Hily-Blant}, P., {Maret}, S., {Bacmann}, A., {et~al.} 2010{\natexlab{a}},
  \aap, 521, L52

\bibitem[{{Hily-Blant} {et~al.}(2013{\natexlab{b}}){Hily-Blant}, {Pineau des
  For{\^e}ts}, {Faure}, {Le Gal}, \& {Padovani}}]{hilyblant2013b}
{Hily-Blant}, P., {Pineau des For{\^e}ts}, G., {Faure}, A., {Le Gal}, R., \&
  {Padovani}, M. 2013{\natexlab{b}}, \aap, 557, A65

\bibitem[{{Hily-Blant} {et~al.}(2008){Hily-Blant}, {Walmsley}, {Pineau des
  For{\^e}ts}, \& {Flower}}]{hilyblant2008}
{Hily-Blant}, P., {Walmsley}, M., {Pineau des For{\^e}ts}, G., \& {Flower}, D.
  2008, \aap, 480, L5

\bibitem[{{Hily-Blant} {et~al.}(2010{\natexlab{b}}){Hily-Blant}, {Walmsley},
  {Pineau des For{\^e}ts}, \& {Flower}}]{hilyblant2010n}
{Hily-Blant}, P., {Walmsley}, M., {Pineau des For{\^e}ts}, G., \& {Flower}, D.
  2010{\natexlab{b}}, \aap, 513, A41

\bibitem[{{Hirota} {et~al.}(1998){Hirota}, {Yamamoto}, {Mikami}, \&
  {Ohishi}}]{hirota1998}
{Hirota}, T., {Yamamoto}, S., {Mikami}, H., \& {Ohishi}, M. 1998, \apj, 503,
  717

\bibitem[{{Ho} \& {Townes}(1983)}]{ho1983}
{Ho}, P.~T.~P. \& {Townes}, C.~H. 1983, \araa, 21, 239

\bibitem[{{Hogerheijde} \& {van der Tak}(2000)}]{hogerheijde2000}
{Hogerheijde}, M.~R. \& {van der Tak}, F.~F.~S. 2000, \aap, 362, 697

\bibitem[{{Hollenbach} \& {Salpeter}(1971)}]{hollenbach1971}
{Hollenbach}, D. \& {Salpeter}, E.~E. 1971, \apj, 163, 155

\bibitem[{{Honvault} {et~al.}(2011){Honvault}, {Jorfi}, {Gonz{\'a}lez-Lezana},
  {Faure}, \& {Pagani}}]{honvault2011}
{Honvault}, P., {Jorfi}, M., {Gonz{\'a}lez-Lezana}, T., {Faure}, A., \&
  {Pagani}, L. 2011, Physical Chemistry Chemical Physics (Incorporating Faraday
  Transactions), 13, 19089

\bibitem[{{Honvault} {et~al.}(2012){Honvault}, {Jorfi}, {Gonz{\'a}lez-Lezana},
  {Faure}, \& {Pagani}}]{honvault2012}
{Honvault}, P., {Jorfi}, M., {Gonz{\'a}lez-Lezana}, T., {Faure}, A., \&
  {Pagani}, L. 2012, Physical Review Letters, 108, 109903

\bibitem[{{Hugo} {et~al.}(2009){Hugo}, {Asvany}, \& {Schlemmer}}]{hugo2009}
{Hugo}, E., {Asvany}, O., \& {Schlemmer}, S. 2009, \jcp, 130, 164302

\bibitem[{{Jenkins}(2009)}]{jenkins2009}
{Jenkins}, E.~B. 2009, \apj, 700, 1299

\bibitem[{{Jensen} {et~al.}(2000){Jensen}, {Bilodeau}, {Safvan}, {Seiersen},
  {Andersen}, {Pedersen}, \& {Heber}}]{jensen2000}
{Jensen}, M.~J., {Bilodeau}, R.~C., {Safvan}, C.~P., {et~al.} 2000, \apj, 543,
  764

\bibitem[{{Jorfi} \& {Honvault}(2009)}]{jorfi2009b}
{Jorfi}, M. \& {Honvault}, P. 2009, Journal of Physical Chemistry A, 113, 10648

\bibitem[{{Jorfi} {et~al.}(2009){Jorfi}, {Honvault}, \& {Halvick}}]{jorfi2009a}
{Jorfi}, M., {Honvault}, P., \& {Halvick}, P. 2009, Chemical Physics Letters,
  471, 65

\bibitem[{{Knauth} {et~al.}(2004){Knauth}, {Andersson}, {McCandliss}, \&
  {Warren Moos}}]{knauth2004}
{Knauth}, D.~C., {Andersson}, B.-G., {McCandliss}, S.~R., \& {Warren Moos}, H.
  2004, \nat, 429, 636

\bibitem[{{Le Bourlot}(1991)}]{lebourlot1991}
{Le Bourlot}, J. 1991, \aap, 242, 235

\bibitem[{{Le Bourlot} {et~al.}(1995){Le Bourlot}, {Pineau des For\^ets},
  {Roueff}, \& {Flower}}]{lebourlot1995}
{Le Bourlot}, J., {Pineau des For\^ets}, G., {Roueff}, E., \& {Flower}, D.~R.
  1995, \aap, 302, 870

\bibitem[{{Li} {et~al.}(2013){Li}, {Heays}, {Visser}, {Ubachs}, {Lewis},
  {Gibson}, \& {van Dishoeck}}]{li2013}
{Li}, X., {Heays}, A.~N., {Visser}, R., {et~al.} 2013, \aap, 555, A14

\bibitem[{{Loinard} {et~al.}(2013){Loinard}, {Zapata}, {Rodr{\'{\i}}guez},
  {Pech}, {Chandler}, {Brogan}, {Wilner}, {Ho}, {Parise}, {Hartmann}, {Zhu},
  {Takahashi}, \& {Trejo}}]{loinard2013}
{Loinard}, L., {Zapata}, L.~A., {Rodr{\'{\i}}guez}, L.~F., {et~al.} 2013,
  \mnras, 430, L10

\bibitem[{{Maret} {et~al.}(2006){Maret}, {Bergin}, \& {Lada}}]{maret2006}
{Maret}, S., {Bergin}, E.~A., \& {Lada}, C.~J. 2006, \nat, 442, 425

\bibitem[{{Maret} {et~al.}(2009){Maret}, {Faure}, {Scifoni}, \&
  {Wiesenfeld}}]{maret2009}
{Maret}, S., {Faure}, A., {Scifoni}, E., \& {Wiesenfeld}, L. 2009, \mnras, 399,
  425

\bibitem[{{Marquette} {et~al.}(1988){Marquette}, {Rebrion}, \&
  {Rowe}}]{marquette1988}
{Marquette}, J.~B., {Rebrion}, C., \& {Rowe}, B.~R. 1988, \jcp, 89, 2041

\bibitem[{{Marquette} {et~al.}(1989){Marquette}, {Rebrion}, \&
  {Rowe}}]{marquette1989}
{Marquette}, J.~B., {Rebrion}, C., \& {Rowe}, B.~R. 1989, \aap, 213, L29

\bibitem[{{McCall} {et~al.}(2004){McCall}, {Huneycutt}, {Saykally}, {Djuric},
  {Dunn}, {Semaniak}, {Novotny}, {Al-Khalili}, {Ehlerding}, {Hellberg},
  {Kalhori}, {Neau}, {Thomas}, {Paal}, {{\"O}sterdahl}, \&
  {Larsson}}]{mccall2004}
{McCall}, B.~J., {Huneycutt}, A.~J., {Saykally}, R.~J., {et~al.} 2004, \pra,
  70, 052716

\bibitem[{{Mendes} {et~al.}(2012){Mendes}, {Buhr}, {Berg}, {Froese}, {Grieser},
  {Heber}, {Jordon-Thaden}, {Krantz}, {Novotn{\'y}}, {Novotny}, {Orlov},
  {Petrignani}, {Rappaport}, {Repnow}, {Schwalm}, {Shornikov}, {St{\"u}tzel},
  {Zajfman}, \& {Wolf}}]{mendes2012}
{Mendes}, M.~B., {Buhr}, H., {Berg}, M.~H., {et~al.} 2012, \apjl, 746, L8

\bibitem[{{Mitchell}(1990)}]{mitchell1990}
{Mitchell}, B. 1990, Phys. Rep., 186, 215

\bibitem[{{Nieva} \& {Przybilla}(2012)}]{nieva2012}
{Nieva}, M.-F. \& {Przybilla}, N. 2012, \aap, 539, A143

\bibitem[{{{\"O}jekull} {et~al.}(2004){{\"O}jekull}, {Andersson},
  {N\r{a}g\r{a}rd}, {Pettersson}, {Derkatch}, {Neau}, {Ros{\'e}n}, {Thomas},
  {Larsson}, {{\"O}sterdahl}, {Semaniak}, {Danared}, {K{\"a}llberg}, {Ugglas},
  \& {Markovi{\'c}}}]{ojekull2004}
{{\"O}jekull}, J., {Andersson}, P.~U., {N\r{a}g\r{a}rd}, M.~B., {et~al.} 2004,
  \jcp, 120, 7391

\bibitem[{{Oka}(2004)}]{oka2004}
{Oka}, T. 2004, Journal of Molecular Spectroscopy, 228, 635

\bibitem[{{Pachucki} \& {Komasa}(2008)}]{pachucki2008}
{Pachucki}, K. \& {Komasa}, J. 2008, \pra, 77, 030501

\bibitem[{{Padovani} {et~al.}(2011){Padovani}, {Walmsley}, {Tafalla},
  {Hily-Blant}, \& {Pineau des For{\^e}ts}}]{padovani2011}
{Padovani}, M., {Walmsley}, C.~M., {Tafalla}, M., {Hily-Blant}, P., \& {Pineau
  des For{\^e}ts}, G. 2011, \aap, 534, A77

\bibitem[{{Pagani} {et~al.}(2007){Pagani}, {Bacmann}, {Cabrit}, \&
  {Vastel}}]{pagani2007}
{Pagani}, L., {Bacmann}, A., {Cabrit}, S., \& {Vastel}, C. 2007, \aap, 467, 179

\bibitem[{{Pagani} {et~al.}(2012){Pagani}, {Bourgoin}, \& {Lique}}]{pagani2012}
{Pagani}, L., {Bourgoin}, A., \& {Lique}, F. 2012, \aap, 548, L4

\bibitem[{{Pagani} {et~al.}(2009){Pagani}, {Vastel}, {Hugo}, {Kokoouline},
  {Greene}, {Bacmann}, {Bayet}, {Ceccarelli}, {Peng}, \&
  {Schlemmer}}]{pagani2009}
{Pagani}, L., {Vastel}, C., {Hugo}, E., {et~al.} 2009, \aap, 494, 623

\bibitem[{{Persson} {et~al.}(2010){Persson}, {Black}, {Cernicharo},
  {Goicoechea}, {Hassel}, {Herbst}, {Gerin}, {de Luca}, {Bell}, {Coutens},
  {Falgarone}, {Goldsmith}, {Gupta}, {Ka{\'z}mierczak}, {Lis}, {Mookerjea},
  {Neufeld}, {Pearson}, {Phillips}, {Sonnentrucker}, {Stutzki}, {Vastel}, {Yu},
  {Boulanger}, {Dartois}, {Encrenaz}, {Geballe}, {Giesen}, {Godard}, {Gry},
  {Hennebelle}, {Hily-Blant}, {Joblin}, {Ko{\l}os}, {Kre{\l}owski},
  {Mart{\'{\i}}n-Pintado}, {Menten}, {Monje}, {Perault}, {Plume}, {Salez},
  {Schlemmer}, {Schmidt}, {Teyssier}, {P{\'e}ron}, {Cais}, {Gaufre}, {Cros},
  {Ravera}, {Morris}, {Lord}, \& {Planesas}}]{persson2010}
{Persson}, C.~M., {Black}, J.~H., {Cernicharo}, J., {et~al.} 2010, \aap, 521,
  L45+

\bibitem[{{Persson} {et~al.}(2012){Persson}, {De Luca}, {Mookerjea},
  {Olofsson}, {Black}, {Gerin}, {Herbst}, {Bell}, {Coutens}, {Godard},
  {Goicoechea}, {Hassel}, {Hily-Blant}, {Menten}, {M{\"u}ller}, {Pearson}, \&
  {Yu}}]{persson2012}
{Persson}, C.~M., {De Luca}, M., {Mookerjea}, B., {et~al.} 2012, \aap, 543,
  A145

\bibitem[{{Pineau des For{\^e}ts} {et~al.}(1990){Pineau des For{\^e}ts},
  {Roueff}, \& {Flower}}]{pineau1990}
{Pineau des For{\^e}ts}, G., {Roueff}, E., \& {Flower}, D.~R. 1990, \mnras,
  244, 668

\bibitem[{{Prasad} \& {Huntress}(1980)}]{prasad1980}
{Prasad}, S.~S. \& {Huntress}, Jr., W.~T. 1980, \apjs, 43, 1

\bibitem[{{Przybilla} {et~al.}(2008){Przybilla}, {Nieva}, \&
  {Butler}}]{przybilla2008}
{Przybilla}, N., {Nieva}, M.-F., \& {Butler}, K. 2008, \apjl, 688, L103

\bibitem[{{Raich} \& {Good}(1964)}]{raich1964}
{Raich}, J.~C. \& {Good}, Jr., R.~H. 1964, \apj, 139, 1004

\bibitem[{{Rist} {et~al.}(2013){Rist}, {Faure}, {Hily-Blant}, \& {Le
  Gal}}]{rist2013}
{Rist}, C., {Faure}, A., {Hily-Blant}, P., \& {Le Gal}, R. 2013, Journal of
  Physical Chemistry A, 117, 9800

\bibitem[{{Ruffle} {et~al.}(1999){Ruffle}, {Hartquist}, {Caselli}, \&
  {Williams}}]{ruffle1999}
{Ruffle}, D.~P., {Hartquist}, T.~W., {Caselli}, P., \& {Williams}, D.~A. 1999,
  \mnras, 306, 691

\bibitem[{{Sandford} {et~al.}(2001){Sandford}, {Bernstein}, {Allamandola},
  {Goorvitch}, \& {Teixeira}}]{sandford2001}
{Sandford}, S.~A., {Bernstein}, M.~P., {Allamandola}, L.~J., {Goorvitch}, D.,
  \& {Teixeira}, T.~C.~V.~S. 2001, \apj, 548, 836

\bibitem[{{Sarrasin} {et~al.}(2010){Sarrasin}, {Abdallah}, {Wernli}, {Faure},
  {Cernicharo}, \& {Lique}}]{sarrasin2010}
{Sarrasin}, E., {Abdallah}, D.~B., {Wernli}, M., {et~al.} 2010, \mnras, 404,
  518

\bibitem[{{Sipil{\"a}} {et~al.}(2013){Sipil{\"a}}, {Caselli}, \&
  {Harju}}]{sipila2013}
{Sipil{\"a}}, O., {Caselli}, P., \& {Harju}, J. 2013, \aap, 554, A92

\bibitem[{{Sofia} {et~al.}(2011){Sofia}, {Parvathi}, {Babu}, \&
  {Murthy}}]{sofia2011}
{Sofia}, U.~J., {Parvathi}, V.~S., {Babu}, B.~R.~S., \& {Murthy}, J. 2011, \aj,
  141, 22

\bibitem[{{Suzuki} {et~al.}(1992){Suzuki}, {Yamamoto}, {Ohishi}, {Kaifu},
  {Ishikawa}, {Hirahara}, \& {Takano}}]{suzuki1992}
{Suzuki}, H., {Yamamoto}, S., {Ohishi}, M., {et~al.} 1992, \apj, 392, 551

\bibitem[{{Tafalla} {et~al.}(2004){Tafalla}, {Myers}, {Caselli}, \&
  {Walmsley}}]{tafalla2004}
{Tafalla}, M., {Myers}, P.~C., {Caselli}, P., \& {Walmsley}, C.~M. 2004, \aap,
  416, 191

\bibitem[{{Tassis} \& {Mouschovias}(2004)}]{tassis2004}
{Tassis}, K. \& {Mouschovias}, T.~C. 2004, \apj, 616, 283

\bibitem[{{Tassis} {et~al.}(2012){Tassis}, {Willacy}, {Yorke}, \&
  {Turner}}]{tassis2012}
{Tassis}, K., {Willacy}, K., {Yorke}, H.~W., \& {Turner}, N.~J. 2012, \apj,
  753, 29

\bibitem[{{Terzieva} \& {Herbst}(1998)}]{terzieva1998}
{Terzieva}, R. \& {Herbst}, E. 1998, \apj, 501, 207

\bibitem[{{Thaddeus}(1972)}]{thaddeus1972}
{Thaddeus}, P. 1972, \araa, 10, 305

\bibitem[{{Thomas} {et~al.}(2005){Thomas}, {Hellberg}, {Neau}, {Ros{\'e}n},
  {Larsson}, {Vane}, {Bannister}, {Datz}, {Petrignani}, \& {van der
  Zande}}]{thomas2005}
{Thomas}, R.~D., {Hellberg}, F., {Neau}, A., {et~al.} 2005, \pra, 71, 032711

\bibitem[{{Tieftrunk} {et~al.}(1994){Tieftrunk}, {Pineau des For\^ets},
  {Schilke}, \& {Walmsley}}]{tieftrunk1994}
{Tieftrunk}, A., {Pineau des For\^ets}, G., {Schilke}, P., \& {Walmsley}, C.~M.
  1994, \aap, 289, 579

\bibitem[{{Tielens} \& {Hagen}(1982)}]{tielens1982}
{Tielens}, A.~G.~G.~M. \& {Hagen}, W. 1982, \aap, 114, 245

\bibitem[{{Troscompt} {et~al.}(2009){Troscompt}, {Faure}, {Maret},
  {Ceccarelli}, {Hily-Blant}, \& {Wiesenfeld}}]{troscompt2009}
{Troscompt}, N., {Faure}, A., {Maret}, S., {et~al.} 2009, \aap, 506, 1243

\bibitem[{{Umemoto} {et~al.}(1999){Umemoto}, {Mikami}, {Yamamoto}, \&
  {Hirano}}]{umemoto1999}
{Umemoto}, T., {Mikami}, H., {Yamamoto}, S., \& {Hirano}, N. 1999, \apjl, 525,
  L105

\bibitem[{{van Dishoeck} \& {Blake}(1998)}]{vandishoeck1998}
{van Dishoeck}, E.~F. \& {Blake}, G.~A. 1998, \araa, 36, 317

\bibitem[{{van Dishoeck} {et~al.}(1995){van Dishoeck}, {Blake}, {Jansen}, \&
  {Groesbeck}}]{vandishoeck1995}
{van Dishoeck}, E.~F., {Blake}, G.~A., {Jansen}, D.~J., \& {Groesbeck}, T.~D.
  1995, \apj, 447, 760

\bibitem[{{Vigren} {et~al.}(2012){Vigren}, {Zhaunerchyk}, {Hamberg},
  {Kaminska}, {Semaniak}, {Ugglas}, {Larsson}, {Thomas}, \&
  {Geppert}}]{vigren2012}
{Vigren}, E., {Zhaunerchyk}, V., {Hamberg}, M., {et~al.} 2012, \apj, 757, 34

\bibitem[{{Wakelam} {et~al.}(2012){Wakelam}, {Herbst}, {Loison}, {Smith},
  {Chandrasekaran}, {Pavone}, {Adams}, {Bacchus-Montabonel}, {Bergeat},
  {B{\'e}roff}, {Bierbaum}, {Chabot}, {Dalgarno}, {van Dishoeck}, {Faure},
  {Geppert}, {Gerlich}, {Galli}, {H{\'e}brard}, {Hersant}, {Hickson},
  {Honvault}, {Klippenstein}, {Le Picard}, {Nyman}, {Pernot}, {Schlemmer},
  {Selsis}, {Sims}, {Talbi}, {Tennyson}, {Troe}, {Wester}, \&
  {Wiesenfeld}}]{wakelam2012}
{Wakelam}, V., {Herbst}, E., {Loison}, J.-C., {et~al.} 2012, \apjs, 199, 21

\bibitem[{{Wakelam} {et~al.}(2005){Wakelam}, {Selsis}, {Herbst}, \&
  {Caselli}}]{wakelam2005}
{Wakelam}, V., {Selsis}, F., {Herbst}, E., \& {Caselli}, P. 2005, \aap, 444,
  883

\bibitem[{{Wakelam} {et~al.}(2010){Wakelam}, {Smith}, {Herbst}, {Troe},
  {Geppert}, {Linnartz}, {{\"O}berg}, {Roueff}, {Ag{\'u}ndez}, {Pernot},
  {Cuppen}, {Loison}, \& {Talbi}}]{wakelam2010}
{Wakelam}, V., {Smith}, I.~W.~M., {Herbst}, E., {et~al.} 2010, \ssr, 156, 13

\bibitem[{{Walmsley} {et~al.}(2004){Walmsley}, {Flower}, \& {Pineau des
  For{\^e}ts}}]{walmsley2004}
{Walmsley}, C.~M., {Flower}, D.~R., \& {Pineau des For{\^e}ts}, G. 2004, \aap,
  418, 1035

\bibitem[{{Whittet}(2010)}]{whittet2010}
{Whittet}, D.~C.~B. 2010, \apj, 710, 1009

\bibitem[{{Whittet} {et~al.}(1983){Whittet}, {Bode}, {Baines}, {Longmore}, \&
  {Evans}}]{whittet1983}
{Whittet}, D.~C.~B., {Bode}, M.~F., {Baines}, D.~W.~T., {Longmore}, A.~J., \&
  {Evans}, A. 1983, \nat, 303, 218

\bibitem[{{Wirstr{\"o}m} {et~al.}(2012){Wirstr{\"o}m}, {Charnley}, {Cordiner},
  \& {Milam}}]{wirstrom2012}
{Wirstr{\"o}m}, E.~S., {Charnley}, S.~B., {Cordiner}, M.~A., \& {Milam}, S.~N.
  2012, \apjl, 757, L11

\bibitem[{{Womack} {et~al.}(1992){Womack}, {Ziurys}, \&
  {Wyckoff}}]{womack1992b}
{Womack}, M., {Ziurys}, L.~M., \& {Wyckoff}, S. 1992, \apj, 393, 188

\bibitem[{{Zymak} {et~al.}(2013){Zymak}, {Hejduk}, {Mulin}, {Pla{\v s}il},
  {Glos{\'{\i}}k}, \& {Gerlich}}]{zymak2013}
{Zymak}, I., {Hejduk}, M., {Mulin}, D., {et~al.} 2013, \apj, 768, 86

\end{thebibliography}

\appendix
\newpage

\section{Radiative transfer modelling}
\label{appendice}

Our one-dimensional spherical source model consists in two-layers with
uniform density and kinetic temperature. The inner layer, close to the
protostar, has a radius of \dix{4}~AU, a dust temperature of 30~K, and
a total (front + back) column density $N(\hh)=6.6\tdix{22}$\cc\ --
hence, a density $n(\hh)=4.4\tdix{5}\ccc$ -- and we assume this layer
does not contain gas-phase nitrogen hydrides. Furthermore, the dust
opacity is assumed to vary as a power-law of the wavelenght, with a
$\beta$ exponent of 2.83 and a dust opacity at 250~GHz of
1~g/cm$^2$. The values of $N(\hh)$ and $\beta$ were adjusted so that
the continuum observed with HIFI towards that source is well
reproduced by our model.

The external layer has the same \hh\ column density. Hence, the total
column density is 1.3\tdix{23}\cc\ or 140 magnitudes of visual
extinction. We modelled the emergent telescope-convolved spectrum of
all the observed transitions of ammonia, by solving the radiative
transfer with the Monte-Carlo code \texttt{RATRAN}
\citep{hogerheijde2000}. The collisional rates for \ce{NH3}-\phh\ are
taken from \cite{maret2009}. The \opr\ of \ce{NH3} was fixed at 0.7, as
predicted by our chemistry model. The free parameters are the density,
the gas temperature, and the \ce{NH3} column density in the external,
absorbing, layer. Note the radii is set by the \hh\ column density and
the density of that layer. The emergent spectra were then fitted
simultaneously to match the observed ones, and a $\chi^2$ is used to
select to best fit model. We found that the self-absorbed profile of
the $1-0$ transition tightly constraints the \hh\ density in the outer
layer to be $\le \dix{4}$\ccc, as for higher densities the predicted
profile shows no longer absorption. As a result, the absorbing ammonia
molecules reside in the most external layer of the circumbinary
envelope of IRAS~16293-2422.

\begin{figure*}[h!]
  \centering
  \includegraphics[width=\columnwidth]{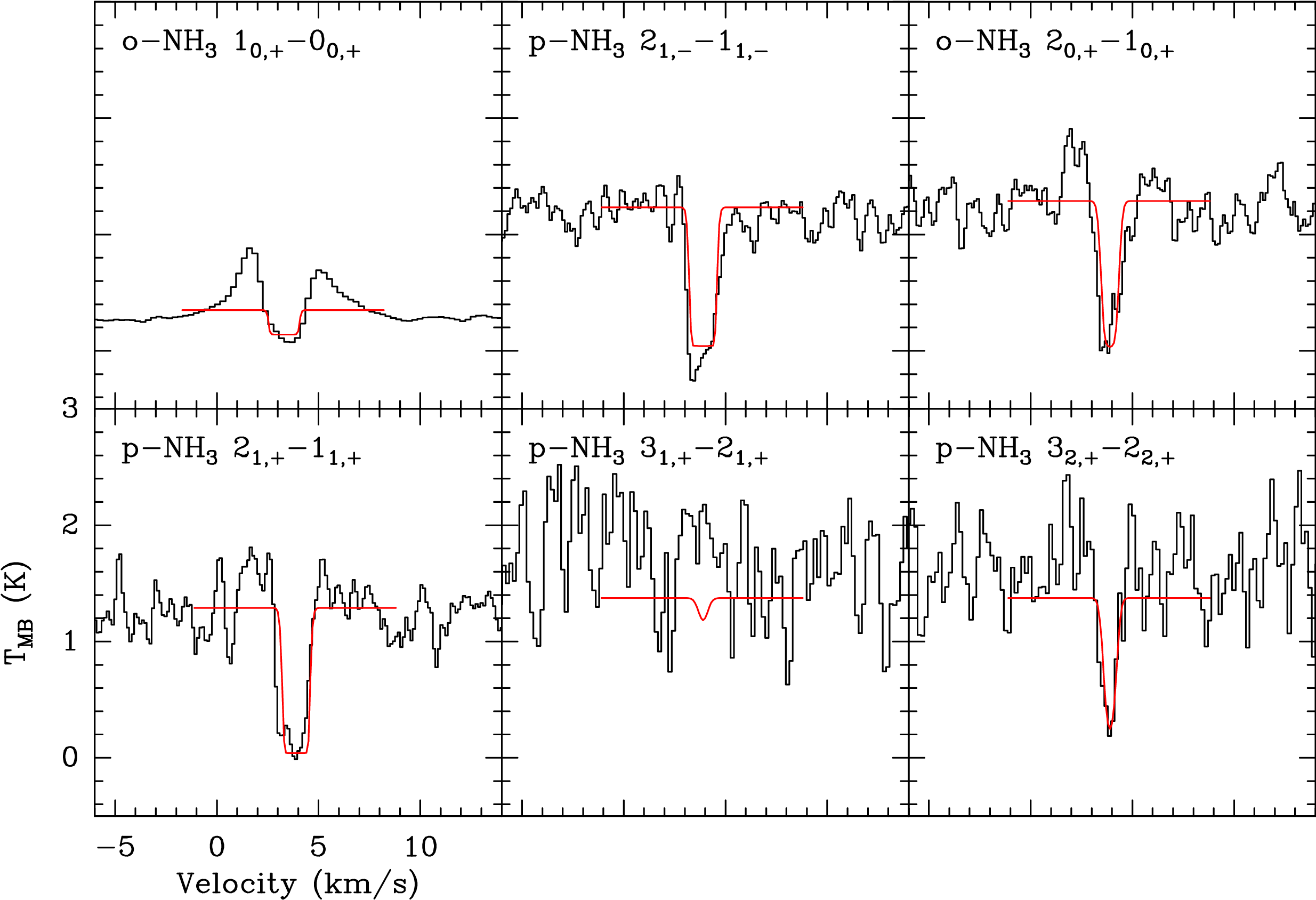}
  \caption{Comparison between the observed NH$_{3}$ spectrum (black)
    from \cite{hilyblant2010nh}) and the best-fit model (red). The $+$
    and $-$ signs in the transition labelling are the same as in
    \cite{maret2009}.}
  \label{fig:ratran-nh3}
\end{figure*}

\begin{figure*}[h!]
  \centering
  \includegraphics[width=\columnwidth]{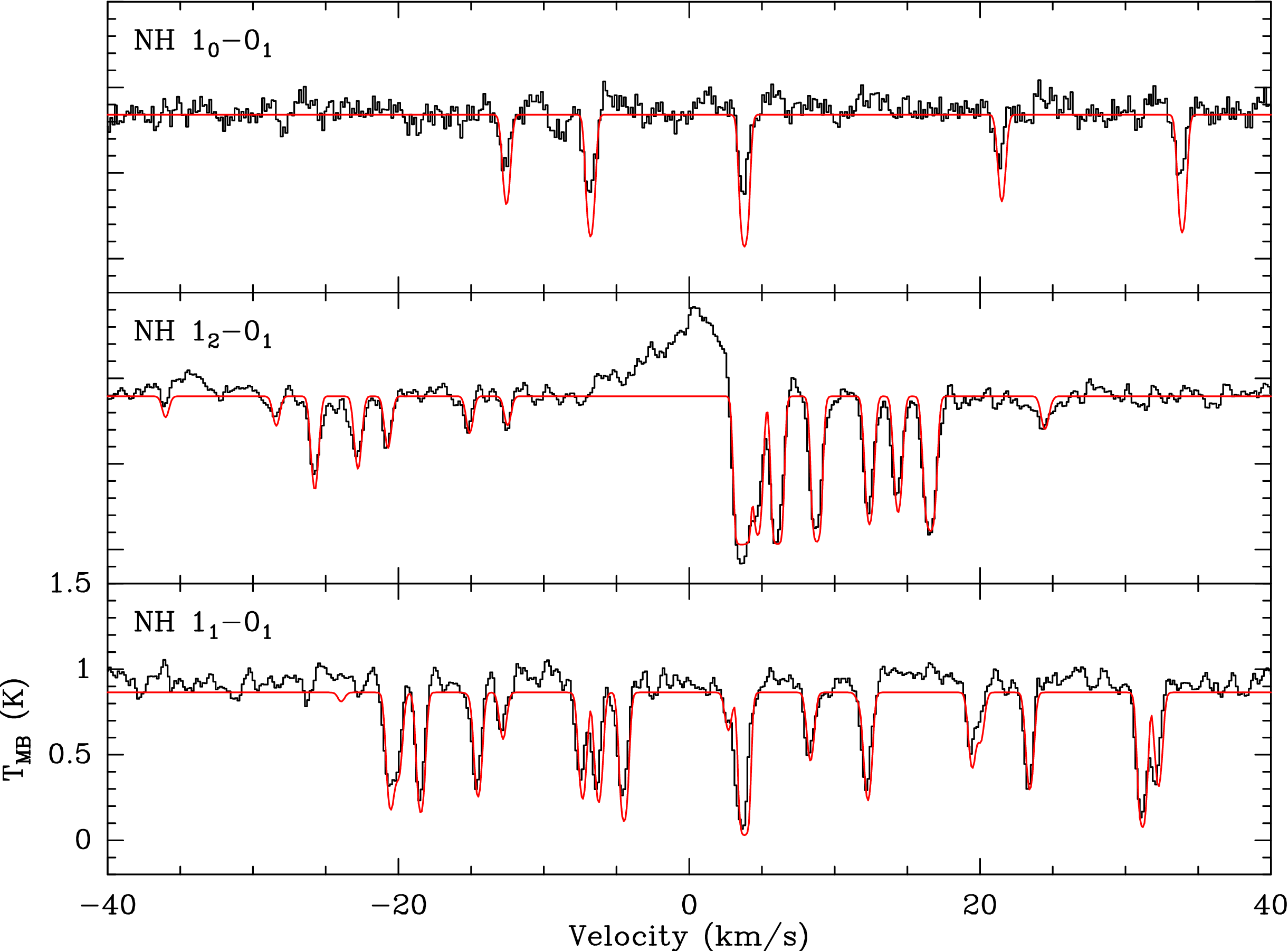}
  \caption{Same as in Fig.~\ref{fig:ratran-nh3} for the NH multiplets,
    using the same physical model as for \ce{NH3}, and adopting a
    single NH column density of 2.2$\pm$0.8\tdix{14}\cc.}
  \label{fig:ratran-nh}
\end{figure*}

An ensemble of solutions is then found for the ammonia column density
and the kinetic temperature, with a best agreement corresponding to
N(\ce{NH3})=1.4\tdix{15}\cc\ and $T=$11~K respectively (see
Fig.~\ref{fig:ratran-nh3}). This column density is a factor of 2.5
below the lower limit of \cite{hilyblant2010nh}. A cross-check of the
best agreement was performed by computing the emergent hyperfine
spectra of NH. To this aim, we used the \texttt{LIME} radiative
transfer code \citep{brinch2010}, which takes line blending into
account. The collisional rates for NH-\phh\ are scaled from the NH-He
rates of \cite{dumouchel2012} by applying the standard reduced mass
ratio of 1.33. The model that best reproduces the three hyperfine
multiplets of NH (see Fig.~\ref{fig:ratran-nh}) has a column density
of NH in the foreground layer of 2.0\tdix{14}\cc, in excellent
agreement with the determination of \cite{bacmann2010} based on the
``HFS'' method of the CLASS software. Therefore this simple 2-layers
model succesfully reproduces both the NH and NH$_{3}$ spectra. The
\ce{NH2}(o) column density was not re-analyzed owing to the lack of
collisional rates. However, the column density of \ce{NH2}(o) was
derived by \cite{hilyblant2010nh} using the same method and under the
same assumptions as those for NH by \cite{bacmann2010}, and is thus
expected to be reliable as well. In addition, here, we assumed an
$\opr(\ce{NH2})$ of 2 to estimate the total column density of
\ce{NH2}.

\clearpage

\section{Chemical updates}
\label{chem_updates}

%
%

\onecolumn
\small
%
%
%
  \renewcommand{\arraystretch}{1.2}
  \begin{longtable}{llclllcccccr}
    \caption{\label{tab:newin} New ion-neutral chemical reaction rates
      and ortho-para branching ratios.}\\
    \toprule
    \multicolumn{6}{c}{Chemical reactions \tablefootmark{a}} & $\alpha$
    & $\beta$ & $\gamma$ & $k$\tablefootmark{b} &References& NSBR\tablefootmark{c}\\
    & & & & & & & & &\cccs\\
    \endfirsthead
    \caption{continued.}\\
    \toprule
    \multicolumn{6}{c}{Chemical reactions \tablefootmark{a}} &
    $\alpha$ & $\beta$ & $\gamma$ & $k$\tablefootmark{b} &References& NSBR\tablefootmark{c}\\
    & & & & & & & & &\cccs \\
    \midrule
    \endhead
    \bottomrule
    \endfoot
    \midrule
    \multicolumn{12}{c}{Formation of H$_3^+$}\\
    \midrule
    H$_2^+$(o)  &H$_2$(o) & \ra & H$_3^+$(o) & H &&1.40(-9)&0.00&0.00&1.40(-9)&1&2/3\\
    H$_2^+$(o)  &H$_2$(o) & \ra & H$_3^+$(p) & H &&7.00(-10)&0.00&0.00&7.00(-10)&1&1/3\\
    H$_2^+$(o)  &H$_2$(p) & \ra & H$_3^+$(o) & H &&7.00(-10)&0.00&0.00&7.00(-10)&1&1/3\\
    H$_2^+$(o)  &H$_2$(p) & \ra & H$_3^+$(p) & H &&1.40(-9)&0.00&0.00&1.40(-9)&1&2/3\\
    H$_2^+$(p)  &H$_2$(o) & \ra & H$_3^+$(o) & H &&7.00(-10)&0.00&0.00&7.00(-10)&1&1/3\\
    H$_2^+$(p)  &H$_2$(o) & \ra & H$_3^+$(p) & H &&1.40(-9)&0.00&0.00&1.40(-9)&1&2/3\\
    H$_2^+$(p)  &H$_2$(p) & \ra & H$_3^+$(p) & H &&2.10(-9)&0.00&0.00&2.10(-9)&1&1\\
    \midrule
    \multicolumn{12}{c}{Main ortho-to-para conversion reactions of H$_2$}\\
    \midrule
  H$_3^+$(o) &H$_2$(o) & \ra & H$_3^+$(o)  & H$_2$(p) &&9.67(-11) & 0.00& -0.14 &9.81(-11)&2\\
  H$_3^+$(o) &H$_2$(o) & \ra & H$_3^+$(p)  & H$_2$(o) &&4.00(-10) & 0.00& -0.19 &4.08(-10)&2\\
  H$_3^+$(o) &H$_2$(o) & \ra & H$_3^+$(p)  & H$_2$(p) &&1.04(-10) & 0.00& -0.14 &1.04(-10)&2\\
  H$_3^+$(o) &H$_2$(p) & \ra & H$_3^+$(o)  & H$_2$(o) &&8.84(-10) & 0.00& 170 &3.66(-17)&2\\
  H$_3^+$(o) &H$_2$(p) & \ra & H$_3^+$(p)  & H$_2$(o) &&1.50(-9) & 0.00& 136.2 &1.82(-15)&2\\
  H$_3^+$(p) &H$_2$(o) & \ra & H$_3^+$(o)  & H$_2$(o) &&8.03(-10) & 0.00& 32.6 &3.08(-11)&2\\
  H$_3^+$(p) &H$_2$(o) & \ra & H$_3^+$(o)  & H$_2$(p) &&3.46(-10) & 0.00& -0.69 &3.71(-10)&2\\
  H$_3^+$(p) &H$_2$(o) & \ra & H$_3^+$(p)  & H$_2$(p) &&2.98(-10) & 0.00& -0.69 &3.19(-10)&2\\
  H$_3^+$(p) &H$_2$(p) & \ra & H$_3^+$(p)  & H$_2$(p) &&5.88(-10) & 0.00& 198.2 &1.45(-18)&2\\
  H$_3^+$(p) &H$_2$(p) & \ra & H$_3^+$(p)  & H$_2$(p) &&8.16(-10) & 0.00& 164.9 &5.63(-17)&2\\
  H$^+$ &H$_2$(o) & \ra & H$^+$ &H$_2$(p) & &1.82(-10)&0.13&-0.02&1.17(-10)&3\\ 
  H$^+$ &H$_2$(p) & \ra & H$^+$ &H$_2$(o) & &1.64(-9) &0.13 & 170.5& 4.15(-17)&3\\ 
  HCO$^+$ &H$_2$(o) & \ra & HCO$^+$ &H$_2$(p) & &1.27(-10)&0.00&0.00&1.27(-10)&4\\ 
  HCO$^+$ &H$_2$(p) & \ra & HCO$^+$ &H$_2$(o) & &1.14(-9) &0.00 & 170.5& 4.49(-17)&4\\ 
  \midrule
  \multicolumn{12}{c}{Nitrogen hydrides formation} \\
  \midrule
  N$^+$   & H$_2$(o)   & \ra & NH$^+$ & H && 4.20(-10) & -0.15 & 44.1 & 8.50(-12)& 5\\ 
  N$^+$   & H$_2$(p)   & \ra & NH$^+$ & H && 8.35(-10) & 0.00 & 168.5 & 4.02(-17)& 5 \\
  NH$^+$ & H$_2$(o) & \ra & NH$_2^+$(o) & H && 1.06(-9) & 0.00 & 0.00 & 1.06(-9) & 6&5/6\\ 
  NH$^+$ & H$_2$(o) & \ra & NH$_2^+$(p) & H && 2.13(-10) & 0.00 & 0.00 & 2.13(-10) & 6 & 1/6 \\ 
  NH$^+$ & H$_2$(p) & \ra & NH$_2^+$(o) & H && 6.38(-10) & 0.00 & 0.00 & 6.38(-10) & 6 & 1/2 \\
  NH$^+$ & H$_2$(p) & \ra & NH$_2^+$(p) & H && 6.38(-10) & 0.00 & 0.00 & 6.38(-10) & 6 & 1/2\\ 
  NH$_2^+$(o) & H$_2$(o) & \ra & NH$_3^+$(o) & H && 1.80(-10) & 0.00 & 0.00 &1.80(-10) & 6 & 2/3 \\  
  NH$_2^+$(o) & H$_2$(o) & \ra & NH$_3^+$(p) & H && 9.00(-11) & 0.00 & 0.00 &9.00(-11) & 6 & 1/3 \\  
  NH$_2^+$(o) & H$_2$(p) & \ra & NH$_3^+$(o) & H &&  9.00(-11) & 0.00 & 0.00 &9.00(-11) & 6 & 1/3 \\  
  NH$_2^+$(o) & H$_2$(p) & \ra & NH$_3^+$(p) & H && 1.80(-10) & 0.00 & 0.00 &1.80(-10) & 6 & 2/3 \\   
  NH$_2^+$(p) & H$_2$(o) & \ra & NH$_3^+$(o) & H &&  9.00(-11)  & 0.00 & 0.00 & 9.00(-11) & 6 & 1/3 \\   
  NH$_2^+$(p) & H$_2$(o) & \ra & NH$_3^+$(p) & H && 1.80(-10)  & 0.00 & 0.00 &1.80(-10) & 6 & 2/3 \\    
  NH$_2^+$(p) & H$_2$(p) & \ra & NH$_3^+$(p) & H && 2.70(-10) & 0.00 & 0.00 &2.70(-10) & 6 & 1 \\
  NH$_3^+$(o) & H$_2$(o) & \ra & NH$_4^+$(I=2) & H && 1.40(-12) & 0.00 & 0.00&1.40(-12) & 6& 7/12\\ 
  NH$_3^+$(o) & H$_2$(o) & \ra & NH$_4^+$(I=1) & H && 8.40(-13) & 0.00 & 0.00& 8.40(-13) & 6& 21/60 \\ 
  NH$_3^+$(o) & H$_2$(o) & \ra & NH$_4^+$(I=0) & H && 1.60(-13) & 0.00 & 0.00& 1.60(-13) & 6& 1/15 \\
  NH$_3^+$(o) & H$_2$(p) & \ra & NH$_4^+$(I=2) & H && 1.80(-12) & 0.00 & 0.00& 1.80(-12) & 6& 1/4 \\
  NH$_3^+$(o) & H$_2$(p) & \ra & NH$_4^+$(I=1) & H && 6.00(-13) & 0.00 & 0.00& 6.00(-13) & 6& 3/4 \\
  NH$_3^+$(o) & H$_2$(p) & \ra & NH$_4^+$(I=0) & H && 0.00 & 0.00 & 0.00& 0.00 & 6& 0 \\
  NH$_3^+$(p) & H$_2$(o) & \ra & NH$_4^+$(I=2) & H && 4.00(-13) & 0.00 & 0.00& 4.00(-13) & 6& 1/6 \\   
  NH$_3^+$(p) & H$_2$(o) & \ra & NH$_4^+$(I=1) & H && 1.68(-12)& 0.00 & 0.00& 1.68(-12) & 6& 21/30 \\ 
  NH$_3^+$(p) & H$_2$(o) & \ra & NH$_4^+$(I=0) & H && 3.20(-13) & 0.00 & 0.00& 3.20(-13) & 6& 2/15 \\ 
  NH$_3^+$(p) & H$_2$(p) & \ra & NH$_4^+$(I=2) & H && 0.00 & 0.00 & 0.00& 0.00 & 6& 0 \\   
  NH$_3^+$(p) & H$_2$(p) & \ra & NH$_4^+$(I=1) & H && 1.44(-12)& 0.00 & 0.00& 1.44(-12) & 6& 3/5\\ 
  NH$_3^+$(p) & H$_2$(p) & \ra & NH$_4^+$(I=0) & H && 9.60(-13) & 0.00 & 0.00& 9.60(-13) & 6& 2/5\\   
  \midrule
  \multicolumn{12}{c}{Destruction of Nitrogen hydrides by H$_3^+$, HCO$^+$ \& N$_2$H$^+$} \\
  \midrule
  H$_3^+$(p) & NH & \ra & NH$_2^+$(p) & H$_2$(p) && 1.63(-10) & 0.00 & 0.00 &  1.63(-10) & 7 & 1/8 \\
  H$_3^+$(p) & NH & \ra & NH$_2^+$(o) & H$_2$(p) && 3.25(-10) & 0.00 & 0.00 & 3.25(-10) & 7 & 1/4 \\
  H$_3^+$(p) & NH & \ra & NH$_2^+$(p) & H$_2$(o) && 3.25(-10) & 0.00 & 0.00 & 3.25(-10) & 7 & 1/4\\
  H$_3^+$(p) & NH & \ra & NH$_2^+$(o) & H$_2$(o) && 4.88(-10) & 0.00 & 0.00 & 4.88(-10) & 7 & 3/8\\
  H$_3^+$(p) & NH$_2$(p) & \ra & NH$_3^+$(p) & H$_2$(p) && 7.20(-10) & 0.00 & 0.00 & 7.20(-10) & 7 & 8/5/4\\
  H$_3^+$(p) & NH$_2$(p) & \ra & NH$_3^+$(p) & H$_2$(o) && 7.20(-10) & 0.00 & 0.00 & 7.20(-10) & 7 & 8/5/4\\
  H$_3^+$(p) & NH$_2$(p) & \ra & NH$_3^+$(o) & H$_2$(o) && 3.60(-10) & 0.00 & 0.00 & 3.60(-10) & 7 & 4/5/4\\
  
  H$_3^+$(p) & NH$_2$(o) & \ra & NH$_3^+$(p) & H$_2$(o) && 8.40(-10) & 0.00 & 0.00 & 8.40(-10) & 7 & 28/5/12\\
  H$_3^+$(p) & NH$_2$(o) & \ra & NH$_3^+$(o) & H$_2$(o) && 4.20(-10) & 0.00 & 0.00 & 4.20(-10) & 7 & 14/5/12\\
  H$_3^+$(p) & NH$_2$(o) & \ra & NH$_3^+$(p) & H$_2$(p) && 2.40(-10) & 0.00 & 0.00 & 2.40(-10) & 7 & 8/5/12\\
  H$_3^+$(p) & NH$_2$(o) & \ra & NH$_3^+$(o) &  H$_2$(p) && 3.00(-10) & 0.00 & 0.00 & 3.00(-10) & 7 & 2/12\\
  H$_3^+$(p) & NH$_3$(p) & \ra & NH$_4^+$(I=0) & H$_2$(p) && 9.10(-10) & 0.00 & 0.00 & 9.10(-10) & 8 &8/5/16 \\
  H$_3^+$(p) & NH$_3$(p) & \ra & NH$_4^+$(I=1) & H$_2$(p) && 2.28(-9)& 0.00 & 0.00 & 2.28(-9) &8 & 4/16 \\
  H$_3^+$(p) & NH$_3$(p) & \ra & NH$_4^+$(I=0) & H$_2$(o) && 1.52(-9) & 0.00 & 0.00 & 1.52(-9) & 8 & 8/3/16\\
  H$_3^+$(p) & NH$_3$(p) & \ra & NH$_4^+$(I=1) & H$_2$(o) && 3.64(-9) & 0.00 & 0.00 & 3.64(-9) & 8 & 32/5/16\\
  H$_3^+$(p) & NH$_3$(p) & \ra & NH$_4^+$(I=2) & H$_2$(o) && 7.58(-10) & 0.00 & 0.00 & 7.58(-10) & 8 & 4/3/16\\
  H$_3^+$(p) & NH$_3$(o) & \ra & NH$_4^+$(I=0) & H$_2$(o) && 7.58(-10) & 0.00 & 0.00 & 7.58(-10) & 8 & 4/3/16\\
  H$_3^+$(p) & NH$_3$(o) & \ra & NH$_4^+$(I=1) & H$_2$(o) && 4.55(-9) & 0.00 & 0.00 & 4.55(-9) & 8 & 8/16\\
  H$_3^+$(p) & NH$_3$(o) & \ra & NH$_4^+$(I=2) & H$_2$(o) && 1.52(-9) & 0.00 & 0.00 & 1.52(-9) & 8 & 8/3/16\\
  H$_3^+$(p) & NH$_3$(o) & \ra & NH$_4^+$(I=2) & H$_2$(p) && 1.14(-9) & 0.00 & 0.00 & 1.14(-9) &8 & 2/16 \\
  H$_3^+$(p) & NH$_3$(o) & \ra & NH$_4^+$(I=1) & H$_2$(p) && 1.14(-9) & 0.00 & 0.00 & 1.14(-9) & 8 & 2/16\\

  H$_3^+$(o)& NH & \ra & NH$_2^+$(p) & H$_2$(o) && 1.63(-10) & 0.00 & 0.00 & 1.63(-10) & 7 & 1/8\\
  H$_3^+$(o)& NH & \ra & NH$_2^+$(o) & H$_2$(o) && 9.75(-10) & 0.00 & 0.00 & 9.75(-10) & 7 & 6/8\\
  H$_3^+$(o)& NH & \ra & NH$_2^+$(o) & H$_2$(p) && 1.63(-10) & 0.00 & 0.00 & 1.63(-10) & 7 & 1/8\\
  H$_3^+$(o)& NH$_2$(p) & \ra & NH$_3^+$(o) & H$_2$(p) && 4.50(-10) & 0.00 & 0.00 & 4.50(-10) & 7 & 1/4\\
  H$_3^+$(o)& NH$_2$(p) & \ra & NH$_3^+$(p) & H$_2$(o) && 9.00(-10) & 0.00 & 0.00 & 9.00(-10) & 7 & 2/4\\
  H$_3^+$(o)& NH$_2$(p) & \ra & NH$_3^+$(o) & H$_2$(o) && 4.50(-10) & 0.00 & 0.00 & 4.50(-10) & 7 & 1/4\\

  H$_3^+$(o)& NH$_2$(o) & \ra & NH$_3^+$(p) & H$_2$(p) && 1.20(-10) & 0.00 & 0.00 & 1.20(-10) & 7 & 4/5/12\\
  H$_3^+$(o)& NH$_2$(o) & \ra & NH$_3^+$(p) & H$_2$(o) && 4.20(-10) & 0.00 & 0.00 & 4.20(-10) & 7 & 14/5/12\\
  H$_3^+$(o)& NH$_2$(o) & \ra & NH$_3^+$(o) & H$_2$(p) && 1.50(-10) & 0.00 & 0.00 & 1.50(-10) & 7 & 1/12\\
  H$_3^+$(o)& NH$_2$(o) & \ra & NH$_3^+$(o) & H$_2$(o) && 1.11(-9) & 0.00 & 0.00 & 1.11(-9) & 7 & 37/5/12\\
  H$_3^+$(o)& NH$_3$(p) & \ra & NH$_4^+$(I=1) & H$_2$(p) && 1.14(-9) & 0.00 & 0.00 & 1.14(-9) & 8 & 2/16\\
  H$_3^+$(o)& NH$_3$(p) & \ra & NH$_4^+$(I=2) & H$_2$(p) && 1.14(-9) & 0.00 & 0.00 & 1.14(-9) & 8 & 2/16\\
  H$_3^+$(o)& NH$_3$(p) & \ra & NH$_4^+$(I=0) & H$_2$(o) && 7.58(-10) & 0.00 & 0.00 & 7.58(-10) & 8 & 4/3/16\\
  H$_3^+$(o)& NH$_3$(p) & \ra & NH$_4^+$(I=1) & H$_2$(o) && 4.55(-9) & 0.00 & 0.00 & 4.55(-9) & 8 & 8/16\\
  H$_3^+$(o)& NH$_3$(p) & \ra & NH$_4^+$(I=2) & H$_2$(o) && 1.52(-9) & 0.00 & 0.00 & 1.52(-9) & 8 & 8/3/16\\
  H$_3^+$(o)& NH$_3$(o) & \ra & NH$_4^+$(I=0) & H$_2$(p) && 2.28(-10) & 0.00 & 0.00 & 2.28(-10) & 8 & 2/5/16\\
  H$_3^+$(o)& NH$_3$(o) & \ra & NH$_4^+$(I=1) & H$_2$(p) && 5.69(-10) & 0.00 & 0.00 & 5.69(-10) & 8 & 1/16\\
  H$_3^+$(o)& NH$_3$(o) & \ra & NH$_4^+$(I=2) & H$_2$(p) && 5.69(-10) & 0.00 & 0.00 & 5.69(-10) & 8 & 1/16\\
  H$_3^+$(o)& NH$_3$(o) & \ra & NH$_4^+$(I=0) & H$_2$(o) && 3.79(-10) & 0.00 & 0.00 & 3.79(-10) & 8 & 2/3/16\\
  H$_3^+$(o)& NH$_3$(o) & \ra & NH$_4^+$(I=1) & H$_2$(o) && 2.62(-9) & 0.00 & 0.00 & 2.62(-9) & 8 & 23/5/16\\
  H$_3^+$(o)& NH$_3$(o) & \ra & NH$_4^+$(I=2) & H$_2$(o) && 4.74(-9)& 0.00 & 0.00 & 4.74(-9) & 8 & 25/3/16\\
  HCO$^+$& NH & \ra & NH$_2^+$(p) & CO && 1.60(-10) & 0.00 & 0.00&1.60(-10) & 7 & 1/4\\
  HCO$^+$& NH & \ra & NH$_2^+$(o) & CO && 4.80(-10) & 0.00 &   0.00&4.80(-10)& 7& 3/4 \\
  HCO$^+$& NH$_2$(p) & \ra & NH$_3^+$(p) & CO && 8.90(-10) & 0.00 & 0.00&8.90(-10) & 7 & 1/1\\
  HCO$^+$& NH$_2$(o) & \ra & NH$_3^+$(p) & CO && 3.00(-10) & 0.00 &   0.00&3.00(-10)& 7& 1/3 \\
  HCO$^+$& NH$_2$(o) & \ra & NH$_3^+$(o) & CO && 5.90(-10) & 0.00 &   0.00& 5.90(-10) & 7 & 2/3 \\
  HCO$^+$& NH$_3$(p) & \ra & NH$_4^+$(I=0) & CO && 4.80(-10) &  0.00&0.00&4.80(-10) & 6 &1/4 \\
  HCO$^+$& NH$_3$(p) & \ra & NH$_4^+$(I=1) & CO && 1.40(-9) &  0.00 & 0.00 & 1.40(-9) & 6 &3/4 \\
  HCO$^+$& NH$_3$(o)& \ra & NH$_4^+$(I=1) & CO && 7.10(-10) &  0.00 & 0.00 &7.10(-10)& 6 &3/8 \\
  HCO$^+$& NH$_3$(o) & \ra & NH$_4^+$(I=2) & CO && 1.20(-9) &  0.00 & 0.00 &1.20(-9)& 6 &5/8 \\
  N$_2$H$^+$ & NH$_3$(p) & \ra & NH$_4^+$(I=0) & N$_2$ && 5.75(-10) & 0.00 & 0.00 & 5.75(-10) & 6 & 1/4\\
  N$_2$H$^+$ & NH$_3$(p) & \ra & NH$_4^+$(I=1) & N$_2$ && 1.73(-9) & 0.00 & 0.00 & 1.73(-9) & 6 & 3/4\\
  N$_2$H$^+$ & NH$_3$(o) & \ra & NH$_4^+$(I=2) & N$_2$ && 8.63(-10) & 0.00 & 0.00 & 8.63(-10) & 6 &  3/8\\
  N$_2$H$^+$ & NH$_3$(o) & \ra & NH$_4^+$(I=1) & N$_2$ && 1.43(-9) & 0.00 & 0.00 & 1.43(-9) & 6 &  5/8\\
  \midrule
  \multicolumn{12}{c}{\ce{H3+ + O}}\\
  \midrule H$_3^+$(o) & O &\ra& OH$^+$ & H$_2$(o) & & 7.98(-10)
  &-0.156 &1.41 &
  1.18(-9)& 9&1\\
  H$_3^+$(o) & O &\ra& H$_2$O$^+$ & H & & 3.42(-10) &-0.156 &1.41 &
  5.05(-10)& 9\\
  H$_3^+$(p) & O &\ra& OH$^+$ & H$_2$(o) & & 3.99(-10) &-0.156 &1.41 &
  5.89(-10)&9&1/2\\
  H$_3^+$(p) & O &\ra& OH$^+$ & H$_2$(p) & & 3.99(-10) &-0.156 &1.41 &
  5.89(-10)& 9&1/2\\
  H$_3^+$(p) & O &\ra& H$_2$O$^+$ & H & & 3.42(-10) &-0.156 &1.41 &
  5.05(-10)& 9\\
\end{longtable}
\tablefoot{%
  Numbers in parentheses are power of 10.%
  \tablefoottext{a}{o, and p stand for ortho, and para
    states respectively. As a spherical top with four identical
    protons, the ammonium ion exists in three nuclear spin states
    noted as in \cite{faure2013}:
    para (I = 0), meta (I = 2), and ortho (I = 1). We note that the meta and ortho species are
    inverted in \cite{rist2013}.}  \tablefoottext{b}{Rates
    $k=\alpha\,(T/300)^\beta\,\exp[-\gamma/T]$ have been computed
    for a 10~K temperature.}
  \tablefoottext{c}{NSBR stands for nuclear-spin branching ratio. These were combined with the overall rate coefficients taken from the cited references, e.g. \cite{prasad1980}. The integer ratios like e.g. 28/5/12 are normalized NSBR and stand for (28/5)/12 i.e. 28/60.}  
  \tablefoottext{d}{\cite{wakelam2012}}}
\tablebib{
  \tiny
  (1) Langevin rate: 2.10$\times10^{-9}$ \cccs; 
  (2) \citet{hugo2009}; 
  (3) \citet{honvault2011}; 
  (4) Langevin rate: 1.52$\times10^{-9}$ \cccs; 
  (5) \citet{dislaire2012}; 
  (6) \cite{anicich1986}; 
  (7)  \citet{prasad1980};
  (8) \citet{marquette1989};
  (9) datasheet by  Ian Smith from KIDA\tablefootmark{d}.
}

\twocolumn
%
\onecolumn
\small
\begin{longtable}{llclllcccccc}
  \caption{\label{tab:newdr} New dissociative recombination reaction rates and branching
    ratios.}\\
  \toprule
  \multicolumn{6}{c}{Reactions} & $\alpha$ &
  $\beta$ & $\gamma$ & $k$\tablefootmark{b} &References& NSBR\tablefootmark{c}\\
  & & & & & & & & &\cccs \\
  \endfirsthead
  \caption{continued.}\\
  \toprule
  \multicolumn{6}{c}{Reactions} & $\alpha$ &
  $\beta$ & $\gamma$ & $k$\tablefootmark{b} &References& NSBR\tablefootmark{c}\\
  & & & & & & & & &\cccs \\
  \midrule
  \endhead
  \bottomrule
  \endfoot
    \midrule
    \multicolumn{6}{l}{DR of \ce{H$_3^+$}} \\
    \midrule
    H$_3^+$(o) & e$^-$ & \ra & H$_2$(o) & H & & 2.51(-8) &0.16&-1.01& 1.61(-8) & 1,2& 1\\
    H$_3^+$(o) & e$^-$ & \ra & H & H & H& 4.87(-8) &0.16 &-1.01& 3.13(-8)& 1,2\\
    H$_3^+$(p) & e$^-$ & \ra & H$_2$(o)  & H & & 0.92(-8) &-0.73 &0.98& 9.94e(-8) & 1,2 & 1/2\\ 
    H$_3^+$(p) & e$^-$ & \ra & H$_2$(p)  & H & & 0.92(-8) &-0.73 &0.98& 9.94e(-8) & 1,2 & 1/2\\ 
    H$_3^+$(p)  & e$^-$ & \ra & H & H & H& 3.56(-8) &-0.73 & 0.98 &3.86(-7) & 1,2\\
    \midrule
    \multicolumn{6}{l}{DR of Nitrogen hydrides} \\
    \midrule
    N$_2$H$^+$ & \ec & \ra & N$_2$ & H && 2.77(-7) & -0.50 & 0.00 & 3.43(-6) & 3\\ 
    N$_2$H$^+$ & \ec & \ra & NH    & N && 2.09(-8) & -0.50 & 0.00 & 2.59(-7) & 3\\
    NH$^+_2$(o) & \ec & \ra & NH & H &   & 1.17(-7) & -0.50 & 0.00 & 6.41(-7) & 4,5\\
    NH$^+_2$(p) & \ec & \ra & NH & H &  & 1.17(-7) & -0.50 & 0.00 & 6.41(-7) &4,5\\
    NH$^+_2$(o) & \ec & \ra & N  & H & H & 1.71(-7) & -0.50 & 0.00 & 9.37(-7) & 4,5\\
    NH$^+_2$(p) & \ec & \ra & N  & H & H & 1.71(-7) & -0.50 & 0.00 & 9.37(-7) & 4,5\\
    NH$^+_2$(o) & \ec & \ra & N  & H$_2$(o) &  & 1.20(-8) & -0.50 & 0.00 &6.57(-8)& 4,5\\
    NH$^+_2$(p) & \ec & \ra & N  & H$_2$(p) &  & 1.20(-8) & -0.50 & 0.00 &6.57(-8)& 4,5\\
    NH$^+_3$(p) & \ec & \ra & NH$_2$(p) & H & & 7.75(-8) & -0.50 & 0.00 & 4.25(-7)& 6 & 1/2\\ 
    NH$^+_3$(p) & \ec & \ra & NH$_2$(o) & H & & 7.75(-8) & -0.50 & 0.00 & 4.25(-7)& 6 & 1/2\\ 
    NH$^+_3$(o) & \ec & \ra & NH$_2$(o) & H & & 1.55(-7) & -0.50 & 0.00 & 8.49(-7) & 6 & 1\\  
    NH$^+_3$(p) & \ec & \ra & NH & H & H & 1.55(-7) & -0.50 & 0.00 &8.49(-7) & 6\\  
    NH$^+_3$(o) & \ec & \ra & NH & H & H & 1.55(-7) & -0.50 & 0.00 & 8.49(-7)& 6\\  
    NH$_4^+$ (I=2) & \ec & \ra & NH$_2$(o) & H & H  &1.22(-7) &  -0.60 & 0.00 &  9.39(-7) & 7 & 1\\ 
    NH$_4^+$ (I=1) & \ec & \ra & NH$_2$(o) & H & H  & 8.07(-8)  & -0.60 & 0.00 & 6.21(-7) & 7 &  2/3\\ 
    NH$_4^+$ (I=1) & \ec & \ra & NH$_2$(p) & H & H  & 4.03(-8)  & -0.60 & 0.00 & 3.10(-7) & 7 &  1/3\\  
    NH$_4^+$ (I=0) & \ec & \ra & NH$_2$(o) & H & H  & 6.11(-8) & -0.60 & 0.00 & 4.70(-7) & 7 &  1/2\\
    NH$_4^+$ (I=0) & \ec & \ra & NH$_2$(p) & H & H  & 6.11(-8) & -0.60 & 0.00 & 4.70(-7) & 7 &  1/2\\
    NH$_4^+$ (I=2) & \ec & \ra & NH$_2$(o) & H$_2$(o) &  & 1.88(-8) & -0.60 & 0.00 &1.45(-7)& 7 &  1\\  
    NH$_4^+$ (I=1) & \ec & \ra & NH$_2$(o) & H$_2$(o) &  & 6.27(-9) & -0.60 & 0.00 & 4.83(-8)& 7 &  1/3\\   
    NH$_4^+$ (I=1) & \ec & \ra & NH$_2$(o) & H$_2$(p) &  & 6.27(-9) & -0.60 & 0.00 & 4.83(-8)& 7 &  1/3\\    
    NH$_4^+$ (I=1) & \ec & \ra & NH$_2$(p) & H$_2$(o) &  & 6.27(-9) & -0.60 & 0.00 & 4.83(-8)& 7 &  1/3\\    
    NH$_4^+$ (I=0) & \ec & \ra & NH$_2$(o) & H$_2$(o) &  & 9.40(-9) & -0.60 & 0.00 0& 7.23(-8)& 7 &  1/2\\      
    NH$_4^+$ (I=0) & \ec & \ra & NH$_2$(p) & H$_2$(p) &  & 9.40(-9) & -0.60 & 0.00 & 7.23(-8)& 7 &  1/2\\      
    NH$_4^+$ (I=2) & \ec & \ra & NH$_3$(o) & H &  & 8.00(-7) & -0.60 & 0.00 & 6.16(-6)& 7 & 1\\
    NH$_4^+$ (I=1) & \ec & \ra & NH$_3$(o) & H &  & 2.66(-7) & -0.60 & 0.00 &2.05(-6)& 7 & 1/3\\
    NH$_4^+$ (I=1) & \ec & \ra & NH$_3$(p) & H &  & 5.33(-7) &-0.60 & 0.00 &4.10(-6)& 7 & 2/3\\
    NH$_4^+$ (I=0) & \ec & \ra & NH$_3$(p) & H & & 8.00(-7) & -0.60 & 0.00 &6.16(-6)& 7 & 1\\
    \midrule
    \multicolumn{10}{l}{DR of H$_3$O$^+$} \\
    \midrule
    H$_3$O$^+$  &e$^-$  &\ra& OH & H$_2$(o) & &3.00(-8) & -0.50 &  0.00 & 1.64(-7) & 6 & 1/2\\
    H$_3$O$^+$  &e$^-$  &\ra& OH & H$_2$(p) & &3.00(-8) & -0.50 &  0.00 & 1.64(-7) & 6 & 1/2\\
    H$_3$O$^+$  &e$^-$  &\ra& OH & H & H& 2.60(-7)&-0.50&0.00&1.42(-6)& 6 &\\
    H$_3$O$^+$  &e$^-$  &\ra& H$_2$O & H & &1.10(-7)&-0.50&0.00&6.03(-7)& 6\\
    H$_3$O$^+$  &e$^-$  &\ra& H$_2$(o) & H & O& 2.80(-9) &-0.50&0.00&1.53(-8) & 6 & 1/2\\
    H$_3$O$^+$  &e$^-$  &\ra& H$_2$(p) & H & O& 2.80(-9) &-0.50&0.00&1.53(-8) & 6 & 1/2\\
  \end{longtable}
  \tablefoot{Numbers in parentheses are power of 10.
    \tablefoottext{a}{DR stands for dissociative recombination.}
    \tablefoottext{b}{Rates $k=\alpha\,(T/300)^\beta\,\exp(-\gamma/T)$
      have been computed for a 10~K temperature.}
    \tablefoottext{c}{NSBR stands for nuclear-spin branching ratio. These were combined with the overall rate coefficients taken from the cited references, e.g. \cite{dossantos2007}.}
  }
  \tablebib{
    (1) \citet{dossantos2007}; 
    (2) \citet{mccall2004}; 
    (3) \citet{vigren2012}; 
    (4) \citet{mitchell1990}; 
    (5) \citet{thomas2005}; 
    (6) \citet{jensen2000};
    (7) \citet{ojekull2004}.
  }

\twocolumn

%
%
\begin{table*}[t]
  \centering
  \tiny
  \caption{Neutral-neutral chemical reaction rates
    and branching ratios considered}
  \begin{tabular}{llclllccccl}
    \toprule
    \multicolumn{6}{c}{Chemical reactions} & $\alpha$ &
    $\beta$ & $\gamma$ & $k$\tablefootmark{a} &References\\
    & & & & & & & & &\cccs \\
    \midrule
    \multicolumn{6}{c}{N to N$_2$ conversion}\\
    \midrule
    N   & OH   & \ra & NO & H  && 8.9(-11)& 0.20 & 0.00  & 4.5(-11)&
    datasheet by Bergeat from KIDA\tablefootmark{b}\\
    N   & NO   & \ra & N$_2$ & O && 7.2(-11)& 0.44 & 12.7& 4.6(-12)&
    fit of calculations from \cite{jorfi2009b}\\
    N   &CN   & \ra & N$_2$ & C  &&8.8(-11)&0.42 & 0.00 & 2.1(-11)&\cite{daranlot2012}\\
    N   &CH   & \ra & CN & H &&1.7(-10) & 0.18 & 0.00 & 9.0(-11)& datatsheet by Smith \& Loison from KIDA\tablefootmark{b}\\
    C   &NO   & \ra & CN & O &&6.0(-11) & -0.16 & 0.00 & 1.0(-10)& \cite{chastaing2000,bergeat1999}\\
     C  &NO   & \ra & CO & N &&9.0(-11) & -0.16 & 0.00 & 1.6(-10)& \cite{chastaing2000,bergeat1999}\\
   \midrule
    \multicolumn{6}{c}{O to O$_2$ conversion} &\\
    \midrule
    O  &OH  &\ra& O$_2$ & H & & 4.0(-11)  & 0.00 & 0.00 & 4.0(-11)&
    datasheet by Loison et al. from KIDA\tablefootmark{b}\\
    \bottomrule
  \end{tabular}
  \tablefoot{Numbers in parentheses are powers of 10.
    \tablefoottext{a}{Rates of the form $k=\alpha\,(T/300)^\beta\,\exp(-\gamma/T)$
      have been computed at 10~K.}
    \tablefoottext{b}{\cite{wakelam2012}.}
  }
  \label{tab:newnn}
\end{table*}

\end{document}